\documentclass[lineno]{jfm}
\usepackage{graphicx}
\usepackage{newtxtext}
\usepackage{newtxmath}
\usepackage{natbib}
\usepackage{hyperref}
\hypersetup{
	colorlinks = true,
	urlcolor   = blue,
	citecolor  = blue,
	linkcolor  = blue,
	filecolor  = blue,
}

\newcommand{\RomanNumeralCaps}[1]
\linenumbers

\usepackage{float}
\usepackage{caption}
\usepackage{subcaption}
\usepackage{amssymb}
\usepackage{booktabs}
\usepackage{multirow}
\usepackage[font=normalsize,justification=justified]{caption}
\usepackage{bm}
\usepackage{comment}
\usepackage{soul}
\usepackage{xcolor}
\usepackage{comment}
\title{Clapping propulsion and thin vortex rings: a computational study of vortex dynamics, energy equivalence, and core potential energy}
\author{Suyog Mahulkar\aff{1}
	\corresp{\email{suyogm@iisc.ac.in}},
	Jaywant Arakeri\aff{1}}

\affiliation{\aff{1}Indian Institute of Science, Banglore, India}

\begin{document}
\maketitle
\begin{abstract}
We report a computational study of clapping propulsion using two thin rigid plates forming a $60^\circ$ interplate cavity that generates a thrust-producing jet during closure. Plate kinematics are prescribed from experiments for two cases: dynamic and stationary, with forward motion constrained in the latter. The computations show that interplate pressure is higher in the stationary case compared to that in the dynamic case, resulting in differences in the thrust produced and in the evolution of wake vortices, with the stationary case forming triangular and $\Omega$-shaped loops, while the dynamic case forms an elliptical loop for each plate. We examine the energy budget in the post-clapping phase, when the vortices are fully formed. The energy consists of kinetic energy and a component associated with vortex formation, whose sum approximately matches the work done on the fluid. This extra term, which we call the core potential energy, is found to be equal to the integral of pressure over the core volume of the vortex. This component is also checked using separate axisymmetric vortex ring simulations, where the kinetic energy is about 60\% of the injected slug energy, and the remaining part is the core potential energy. Sullivan \textit{et al.}\cite{Sullivan08} had commented on this deficit for vortex rings and hypothesized the existence of a potential energy associated with the vortex structure.
\end{abstract}


\section{Introduction}
 In aquatic habitats, animals such as squid and jellyfish commonly use pulse jet propulsion. A review by Gemmell {\it et al.}\cite{Gemmell21} discusses several other aquatic invertebrates that employ this mode of propulsion. Interest in pulse-jet propulsion has also grown in recent years, reflected in the development of bio-inspired aquatic robots such as Robosquid (Nichols, Moslemi and Krueger\cite{robosquid08}), CALAMAR-E (Krieg and Mohseni\cite{Krieg08}), and biohybrid robots (Xu and Dabiri\cite{xu20}). These systems are inspired by the jet production mechanism observed in jellyfish and squid. In jellyfish, contraction of the umbrella cavity generates thrust, and the wake evolves in the form of a starting vortex ring (Dabiri {\it et al.}\cite{Dabiri05},\cite{Dabiri06}). Squid operate similarly, although their wake patterns may vary, showing either isolated vortex rings or rings with a trailing jet (Bartol {\it et al.}\cite{Bartol2D09},\cite{Bartol3D16}). Pulse-jet production is also observed in butterflies during the clapping phase of wing motion (Brodsky\cite{Brodsky91} and Johansson and Henningsson\cite{Johanassan21}). This clapping motion provides a simple way to generate pulse jets. D. Kim {\it et al.}\cite{Kim13} conducted the first controlled study on two clapping plates submerged in a quiescent fluid to examine thrust generation and the associated flow field.\par
 
\begin{figure}
	   	\centering\
	   	\begin{subfigure}[b]{0.6\textwidth}
	   		\includegraphics[width=\textwidth]{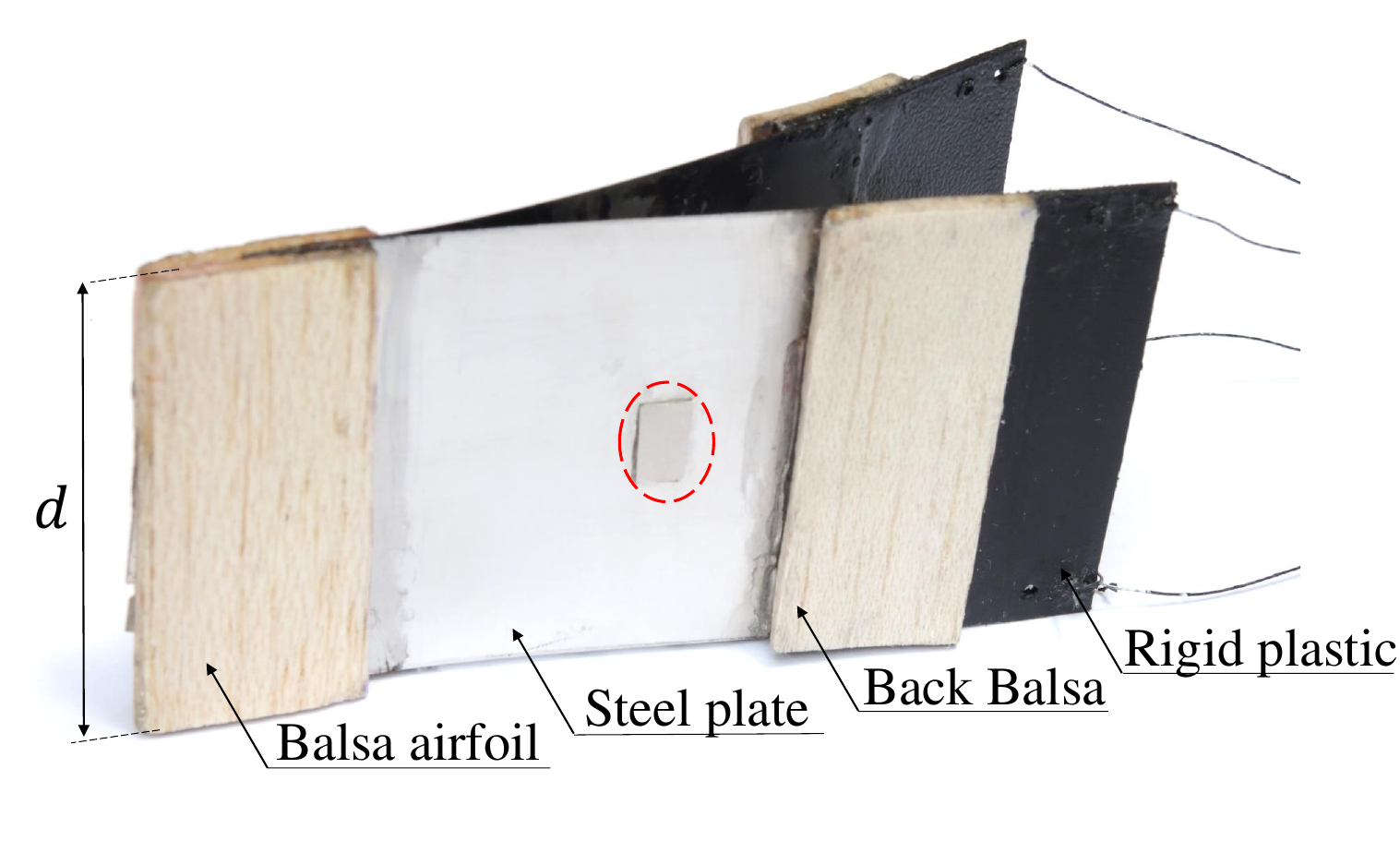}
	   		\label{fig:C_ExpApparatus}
	   	\end{subfigure}
	   	\caption{A perspective view of the clapping body showing both plates joined at the front. Each plate consists of a thin steel plate that provides the spring-back action required for the clapping motion, while balsa wood is used to achieve buoyancy balance. A balsa airfoil is attached at the leading edge of each steel plate, and a rigid plastic plate with a balsa piece is attached at its trailing edge. The component dimensions are: steel plate – 40 mm length, 0.14 mm thickness; rigid plastic plate – 30 mm length, 0.7 mm thickness; back balsa piece – 18 mm length, 2 mm thickness; front balsa airfoil – 19 mm chord length, 2 mm thickness. The placement of these components is designed to avoid any unbalanced torque. A small steel mass (red circle) is used to adjust the buoyancy of the body in water. A fishing thread of diameter 0.25 mm is tied at the top and bottom corners of the trailing edges to form a loop. The total body length is $L = 89$ mm and the depth is $d = 45$ mm.}\label{fig:C_ClapBody}	
   \end{figure}

 To examine how body motion influences pulse jet production, we conducted experiments (Mahulkar and Arakeri \cite{Mahulkar23}) using a freely moving clapping body consisting of two plates hinged at one end with an effective torsional spring and initially pulled apart at the other end (see figure \ref{fig:C_ClapBody}). Upon release, the plates perform a clapping motion and thereby eject fluid from the interplate cavity, generating a thrust-producing jet that propels the body forward. The construction details are provided in figure \ref{fig:C_ClapBody} and in Mahulkar and Arakeri \cite{Mahulkar23}. Previous studies on pulse jet formation typically consider stationary bodies (Kim {\it et al.}\cite{Kim13}, Martin {\it et al.}\cite{Martin17}, Das {\it et al.}\cite{Das13}, \cite{Das18}), whereas our subsequent study examined the differences between a stationary clapping body and one allowed to move freely forward (Mahulkar and Arakeri \cite{Mahulkar24}), termed stationary and dynamic, respectively. The experiments revealed that clapping was much slower for the stationary case, leading to distinctly different wake-vortex structures between the two cases. This naturally leads to a key question: how constraining the body motion alters the pressure field, and how this leads to differences in vortex formation and subsequent evolution. To address these aspects, we perform viscous computations of the clapping motion to obtain three-dimensional flow field information, which was not available from the experimental studies. In these computations, the interplate cavity is modeled primarily using two thin rigid plates that follow experimentally obtained motions of the clapping plates.\par
 
 In this paper, we also examine the energy associated with the formation of vortices in the clapping wake. For the vortex ring case, such an energy was originally hypothesized by Sullivan \textit{et al.}\cite{Sullivan08}. In their experiments, they observed that the kinetic energy of the ring is substantially lower than that of the injected fluid slug, while their momenta are nearly equal. They proposed that this energy deficit corresponds to some potential energy associated with the vortex bubble. To investigate this further, we perform axisymmetric simulations of vortex ring formation by varying the piston velocity program and the stroke-to-diameter ratio. The energy understanding gained from the vortex ring is then examined for the clapping wake. \par

 The paper is organized as follows: \S \ref{sec:C_CompModel} presents the computational model for the clapping body. \S \ref{sec:C_FlowField} describes the flow and pressure fields, along with the three-dimensional evolution of vorticity, including comparisons with experimental observations. \S \ref{sec:C_ThrustCoeff} discusses the variation in thrust coefficient. \S \ref{sec:C_WakeEnergy_Clapping} examines the energy budget of the clapping wake, based on the axisymmetric vortex ring analysis presented in \S \ref{sec:C_WakeEnergy_Ring}. Finally, the concluding remarks are presented in \S \ref{sec:C_conclusion}.

\section{Computational modeling}
\label{sec:C_CompModel}

	\begin{figure}
			\centering\
			\begin{subfigure}[b]{0.4\textwidth}
				\includegraphics[width=\textwidth]{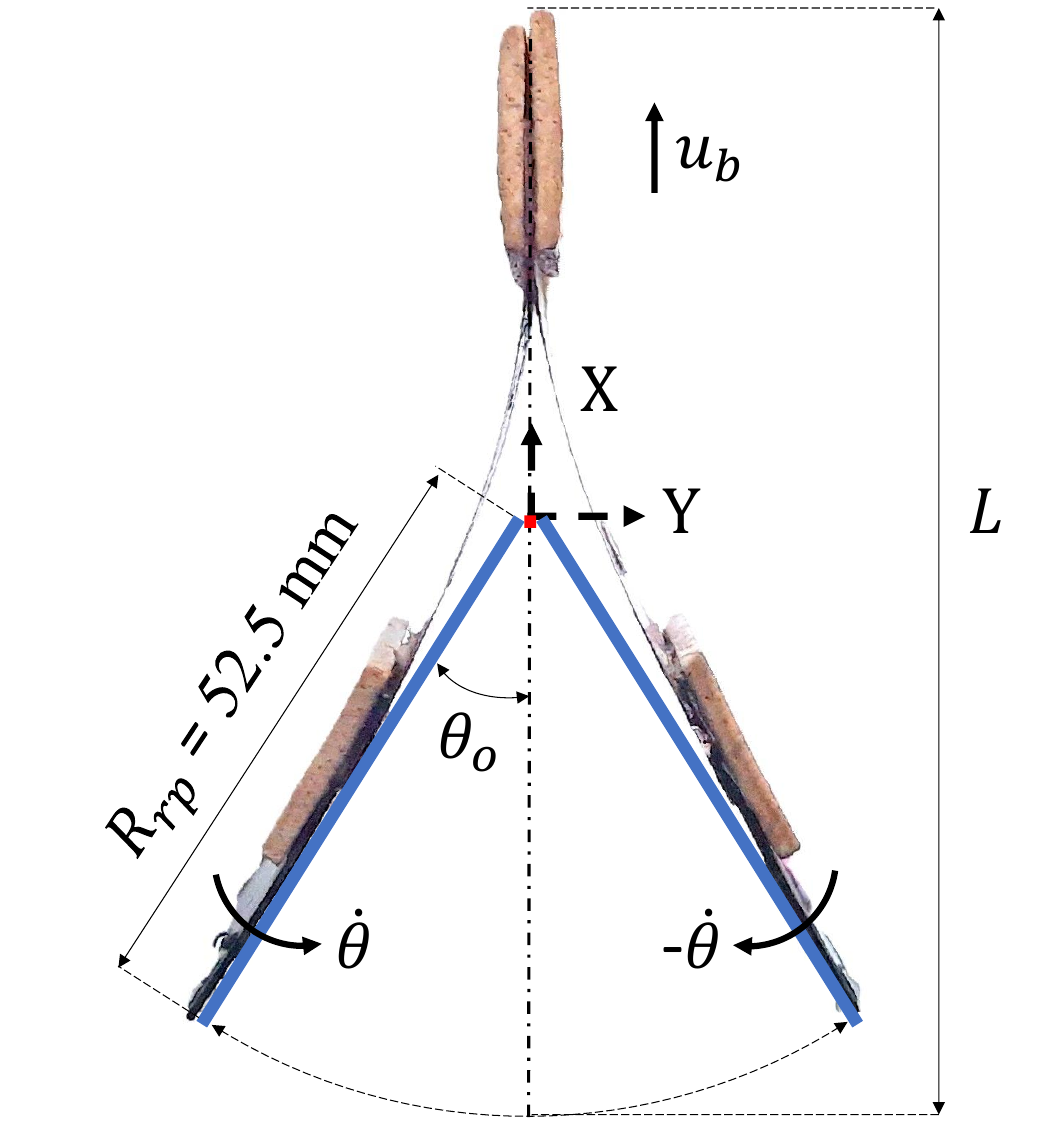}
				\vspace{02mm}
				\caption{}
				\label{fig:C_ExpApparatus_TV}
			\end{subfigure}
			\centering\
			\begin{subfigure}[b]{0.58\textwidth}
				\includegraphics[width=\textwidth]{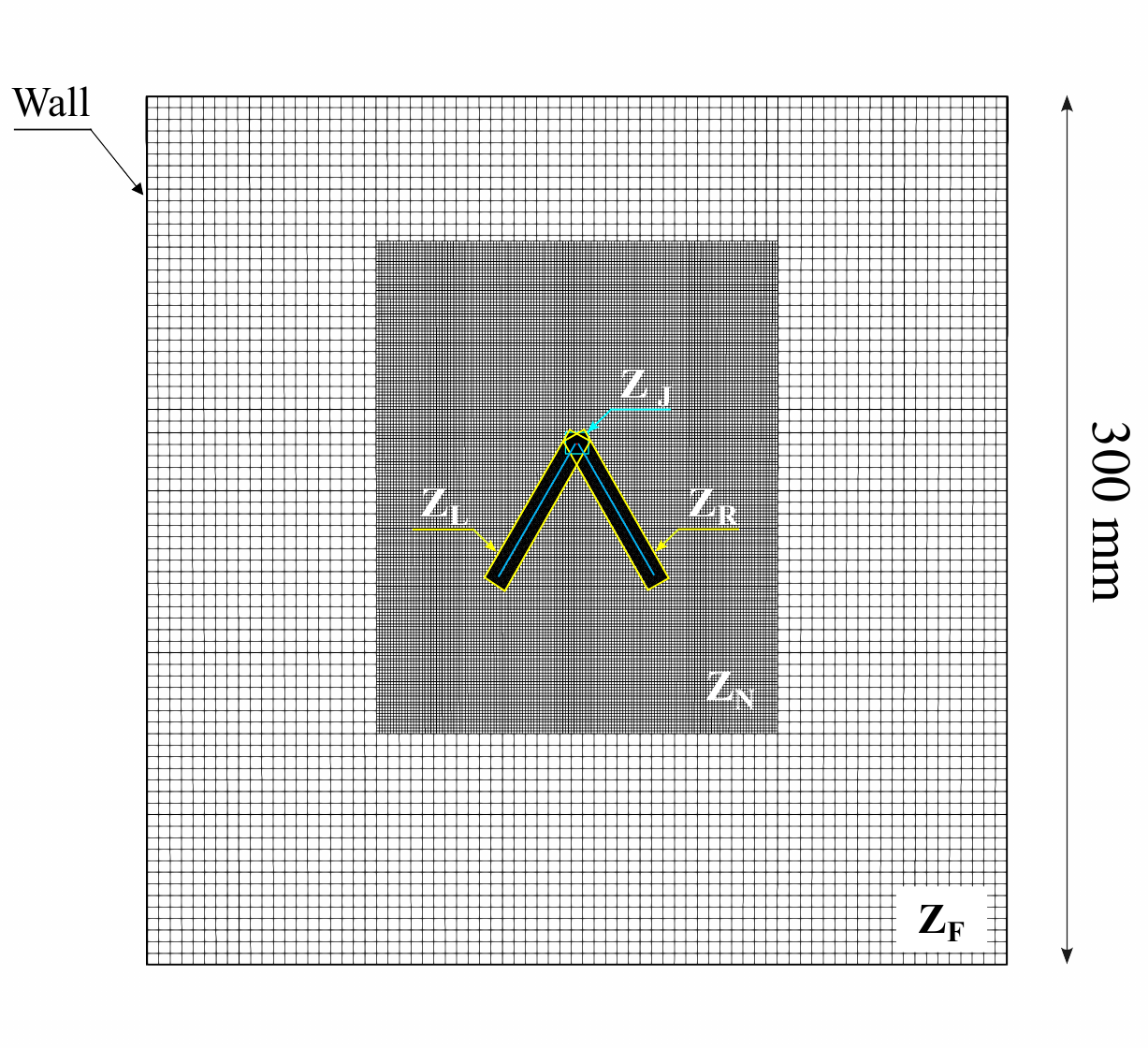}
				\caption{}
			\end{subfigure}
			\caption{(a) Computational model of the clapping body consisting of two rigid plates (52.5 mm length, 0.7 mm thickness) shown in blue and an interconnecting joint plate (0.7 mm length, 0.7 mm thickness) shown in red. The model is superimposed on the top view of the actual clapping body. The plates in the computational model are tangential to the rear portion of the rigid plastic plates in the clapping body (see figure \ref{fig:C_ClapBody}). In both stationary and dynamic cases, equal and opposite angular velocity, $\dot{\theta}$, is prescribed. In dynamic cases, an additional translational velocity, $u_b$, is prescribed. The inertial coordinate system aligns with the initial position of the joint plate. (b) Top view of the 3D overset mesh domain showing the background mesh zone, $\mathrm{Z_F}$, and the component mesh zones: the body-enveloping zones $\mathrm{Z_L}$, $\mathrm{Z_R}$, and $\mathrm{Z_J}$ surrounding the left, right, and joint plates, and the near-body zone $\mathrm{Z_N}$. The mesh resolution is finest (0.35 mm) in $\mathrm{Z_L}$, $\mathrm{Z_R}$, and $\mathrm{Z_J}$, intermediate (1 mm) in $\mathrm{Z_N}$, and coarsest (4 mm) in $\mathrm{Z_F}$.}\label{fig:C_CompModelling}	
	\end{figure}

We model the closing motion of the interplate cavity using clapping plate kinematics extracted from experiments (Mahulkar and Arakeri \cite{Mahulkar24}). Three rigid plates are used to construct a simplified configuration of the interplate cavity. As shown in figure \ref{fig:C_CompModelling}(a), two longer rigid plates represent the clapping plates, while a shorter plate placed between them acts as the joint plate. The origin of the inertial coordinate system coincides with the centroid of the joint plate at the initial time. The X direction is aligned with the direction of forward translation, Y is perpendicular to it, and Z is along the body depth direction. The longer rigid plates have length $R_{rp}=52.5$ mm and thickness $t_{rp}=0.7$ mm, while the joint plate has a length equal to $t_{rp}$. This three-plate representation captures most of the clapping cavity, although the portion of the body ahead of the joint plate is not considered. Computations are performed for stationary and dynamic cases with a constant initial interplate cavity angle of $2\theta_o = 60^\circ$. For both cases, two body depths are considered: $d=45$ mm and $d=89$ mm, corresponding to nondimensional depths $d^*(=d/L)$ of 0.5 and 1.0, obtained by normalizing the body depth by the total body length.\par

\begin{figure}
	\centering\
	\begin{subfigure}[b]{0.40\textwidth}
		\includegraphics[width=\textwidth]{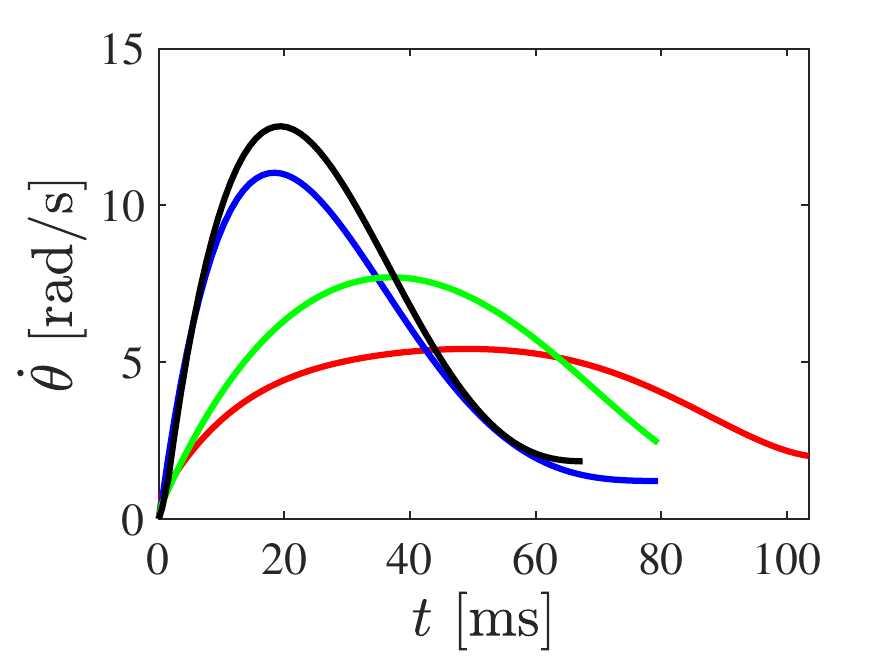}
		\caption{}
	\end{subfigure}\hspace{5mm}
	\begin{subfigure}[b]{0.40\textwidth}
		\includegraphics[width=\textwidth]{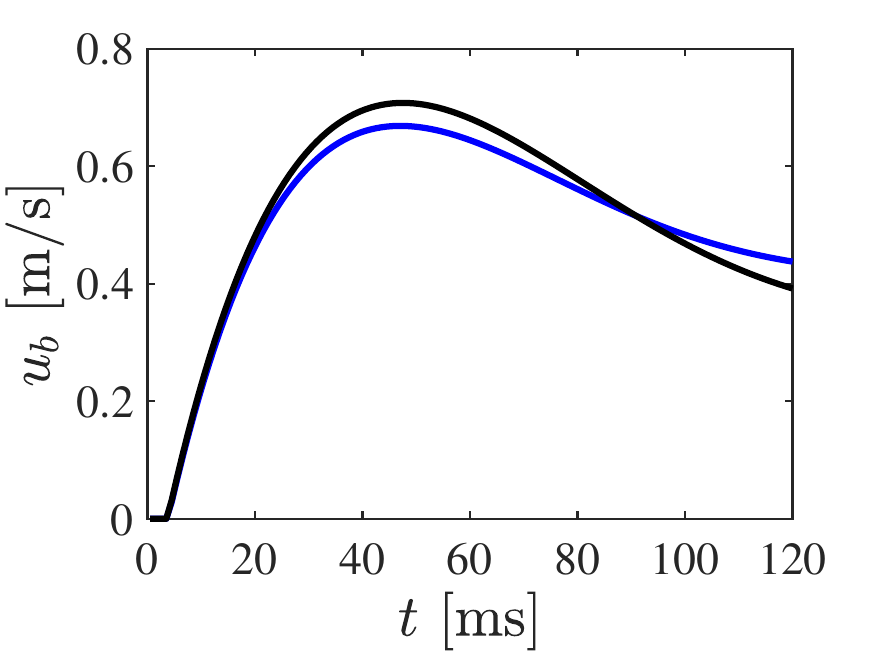}
		\caption{}
	\end{subfigure}
	\begin{subfigure}[b]{0.40\textwidth}
		\includegraphics[width=\textwidth]{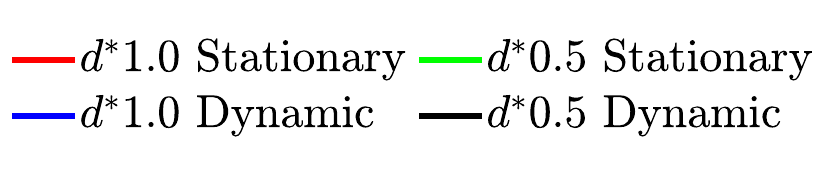}
	\end{subfigure}
	\caption{Experimentally obtained velocities used as input for computation: time variation of (a) the angular velocity of the clapping plates, $\dot{\theta}$, and (b) the translational velocity of the clapping body, $u_b$.}\label{fig:C_VeloInput_Stat_Dyn}	
\end{figure}

The clapping motion is prescribed using experimentally obtained plate kinematics. The angular velocity of the plates, $\dot{\theta}$, is extracted from experiments using image analysis. For both stationary and dynamic cases, $\dot{\theta}$ increases with time, reaches a maximum value $\dot{\theta}_m$, and subsequently decreases (figure \ref{fig:C_VeloInput_Stat_Dyn}(a)). The $\dot{\theta}$ data is used up to the clapping duration $t_c$, when the interplate cavity is nearly closed ($\theta \approx 5^\circ$); the remaining closure from $5^\circ$ to $0^\circ$ is ignored as it occurs very slowly. The center of rotation of each plate coincides with the corresponding edge of the joint plate. The dynamic cases show higher values of $\dot{\theta}_m$ (table \ref{tab:C_BodyDyn_gamma}), indicating faster clapping than the stationary cases. In dynamic cases, the translational velocity of the body, $u_b$, obtained from experiments, is also prescribed (figure \ref{fig:C_VeloInput_Stat_Dyn}(b)). During the initial $\approx 4$ ms, the body remains nearly stationary, after which it accelerates to a maximum velocity of approximately $0.7 \ \mathrm{m/s}$ at around 50 ms and then decelerates due to the drag force. During the clapping duration ($t \leq t_c$), the body travels nearly three-fourths of its length. In both cases, $\dot{\theta}(t)$ and $-\dot{\theta}(t)$ are prescribed for the left and right plates, respectively, while in dynamic cases the translational velocity $u_b(t)$ is applied to all plates. \par 

\begin{table}
	\centering
	\small
	\setlength{\tabcolsep}{5pt}
	
	\begin{tabular}{ccccccccc}
		\toprule
		\multirow{2}{*}{\text{Case}} 
		& \multirow{2}{*}{$d^*$} 
		& \multicolumn{3}{c}{\text{Kinematics}} 
		& \multicolumn{4}{c}{\text{Starting-vortex circulation}} \\
		\cmidrule(lr){3-5} \cmidrule(lr){6-9}
		& 
		& $t_c$ 
		& $\dot{\theta}_m$ 
		& $t_{\dot{\theta}_m}$ 
		& \multicolumn{2}{c}{$\Gamma_m$ [cm$^2$/s]} 
		& \multicolumn{2}{c}{$t_{\Gamma_m}$ [ms]} \\
		\cmidrule(lr){6-7} \cmidrule(lr){8-9}
		& & [ms] & [rad/s] & [ms] & CFD & EXP & CFD & EXP \\
		\midrule
		\multirow{2}{*}{\text{Stationary}} 
		& 0.5 & 80  & 7.7  & 37 & 192.3 & 156.9 & 68.5 & 70.5 \\
		& 1.0 & 104 & 5.4  & 49 & 211.4 & 196.2 & 88.5 & 94.5 \\
		\midrule
		\multirow{2}{*}{\text{Dynamic}} 
		& 0.5 & 68  & 12.5 & 20 & 158.7 & 138.8 & 33.5 & 33.5 \\
		& 1.0 & 80  & 11 & 19 & 163.1 & 138.9 & 31.5 & 32.5 \\
		\bottomrule
	\end{tabular}%
	
	\caption{Clapping kinematics and starting-vortex circulation for stationary and dynamic cases of $d^* = 0.5$ and $1.0$. The kinematic quantities include the clapping time period $t_c$, maximum angular velocity $\dot{\theta}_m$, and the time when it reaches its maximum, $t_{\dot{\theta}_m}$. The vortex-circulation data include the maximum starting-vortex circulation, $\Gamma_m$, obtained from computational (CFD) and experimental (EXP) data, and the corresponding time, $t_{\Gamma_m}$.}
	\label{tab:C_BodyDyn_gamma}%
\end{table}%
The meshing of the fluid domain is challenging because it must accommodate both the forward motion of the body and the strong contraction of the interplate cavity. To address this, an overset mesh technique is used. As shown in figure \ref{fig:C_CompModelling}(b), the domain consists of a background mesh zone $\mathrm{Z_F}$ and overlapping component meshes. The background mesh has dimensions $30\times30\times30 \ \mathrm{cm}^3$, providing a sufficiently large domain relative to the body; wall boundary conditions are applied on its outer surfaces. The component meshes consist of two regions. The first is the body-enveloping region, composed of three cuboid meshes surrounding the left plate, right plate, and joint plate ($\mathrm{Z_L}$, $\mathrm{Z_R}$, and $\mathrm{Z_J}$). Here, wall boundary conditions are applied on the plate surfaces and overset boundary conditions on the outer boundaries (figure \ref{fig:C_CompModelling}(b): $\mathrm{Z_L}$ and $\mathrm{Z_R}$ outer boundaries shown in yellow, plate surfaces marked by the thick blue line; $\mathrm{Z_J}$ outer and inner boundaries shown in cyan and red). The second region is the near-body zone $\mathrm{Z_N}$, located between the background mesh and the body-enveloping meshes. In the overset arrangement, the finest resolution is used in the body-enveloping zones, an intermediate resolution in $\mathrm{Z_N}$, and a coarser resolution in $\mathrm{Z_F}$. After the overset mesh operation, only a few orphan cells appear near the edges of $\mathrm{Z_J}$. However, no fluid leakage into the interplate cavity is observed during the clapping motion, confirming proper cavity modeling. \par

The governing equations for conservation of mass and momentum are solved in an inertial reference frame using the commercial solver Ansys Fluent 2020 R1. Previous studies by David {\it et al.}\cite{JeemReeves18} and Lee {\it et al.}\cite{Lee12} demonstrate the solver's reliability for flows involving rotating and translating plates. In the present simulations, the gradient is computed using the least-squares cell-based scheme, pressure is discretized using a second-order scheme, and momentum using a second-order upwind scheme. Pressure–velocity coupling is achieved using a coupled algorithm, and temporal discretization is performed using a first-order implicit scheme. A time step of $0.25$ ms is used for all cases, maintaining the Courant–Friedrichs–Lewy number below $0.5$. Water is used as the working fluid with density $998.2\,\mathrm{kg/m^3}$ and viscosity $0.001\,\mathrm{kg/(m\,s)}$. A residual convergence criterion of $10^{-4}$ is enforced at each time step. During post-processing in ParaView, ‘solve’ and ‘donor’ cells are retained while ‘receptor’, ‘dead’, and ‘orphan’ cells are filtered out. \par

Grid independence is examined for the dynamic case of $d^*=0.5$ using three mesh configurations: coarse (1,480,352 cells), medium (1,998,328 cells), and fine (3,982,656 cells). All meshes consist of hexahedral cells. The comparison metric is the cavity exit velocity $\overline{u}_e$, averaged across the opening width on the XY plane (Z=0). Over the clapping period, the difference in $\overline{u}_e$ between the coarse and fine meshes remains within $\pm13\%$ of the fine mesh value, while the difference between the medium and fine meshes remains within $\pm6\%$ of the fine mesh value. Based on this comparison, the fine mesh is selected for subsequent simulations. In this configuration the cell sizes are $4$ mm in $\mathrm{Z_F}$, $1$ mm in $\mathrm{Z_N}$, and $0.35$ mm in $\mathrm{Z_L}$, $\mathrm{Z_R}$, and $\mathrm{Z_J}$. This mesh variation is constrained by the available computational resources. \par

In dynamic cases simulations are performed over $132$ ms, during which the body completes the clapping motion and travels approximately three-quarters of its length. Extending the simulation further would require enlarging the near-body zone $\mathrm{Z_N}$, which exceeds the available computational resources. For stationary cases, where the body does not translate, the simulation time is extended to $250$ ms while maintaining the same mesh count. In this case, the hinge location is shifted closer to the upstream boundary of $\mathrm{Z_N}$. This adjustment creates a larger fine-mesh region in the wake, allowing the wake evolution to be captured over a longer duration. To prevent intersection of the body-enveloping meshes $\mathrm{Z_L}$ and $\mathrm{Z_R}$ during cavity closure, the clapping motion in both stationary and dynamic cases is simulated only until $\theta \approx 5^\circ$.

\section{The flow field}
\label{sec:C_FlowField}
In this section, we present the flow fields associated with the clapping of the plates for both the stationary and dynamic cases. Wherever relevant, experimental and computational data are presented together, which allows comparison and shows how well the computations reproduce the experimental observations.

\subsection{Flow in the central plane}
\label{sec:C_Validation}
	\begin{figure}
	\centering\
	\begin{subfigure}[b]{0.32\textwidth}
		\includegraphics[width=\textwidth]{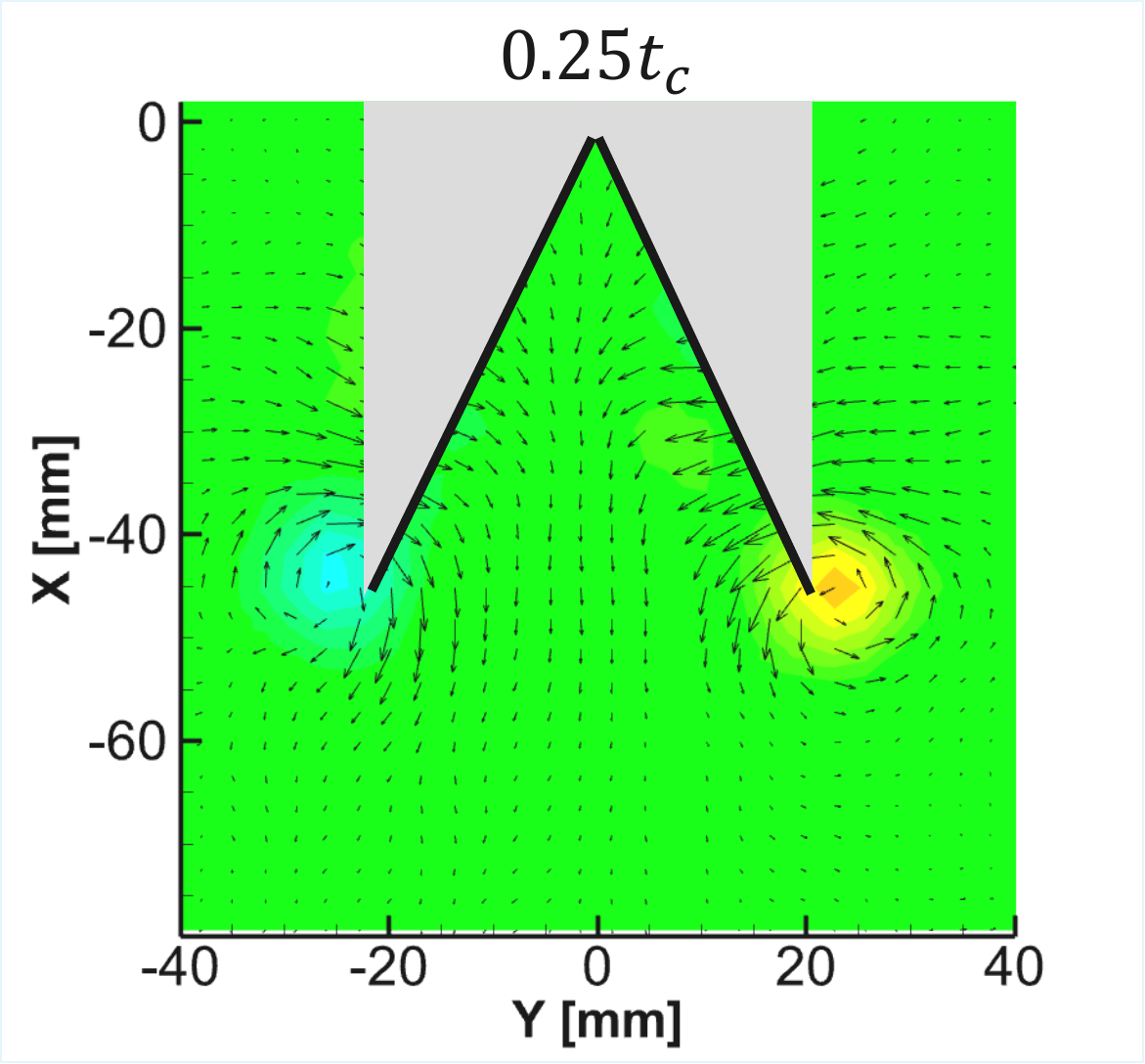}
		\caption{EXP: Stat}
	\end{subfigure}\hspace{05mm}
	\begin{subfigure}[b]{0.32\textwidth}
		\includegraphics[width=\textwidth]{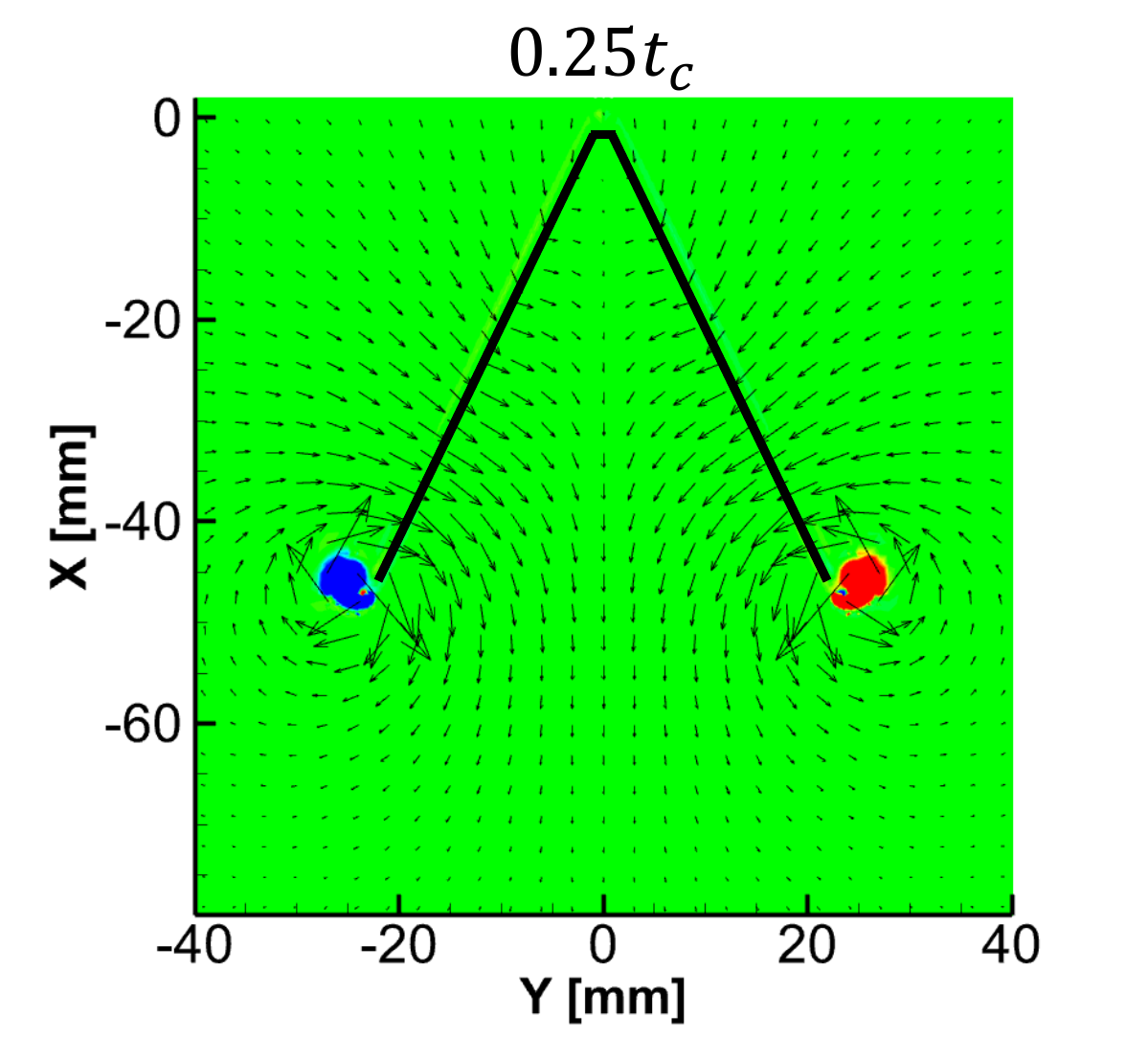}
		\caption{CFD: Stat}
	\end{subfigure}\vspace{01mm}
	\begin{subfigure}[b]{0.32\textwidth}
		\includegraphics[width=\textwidth]{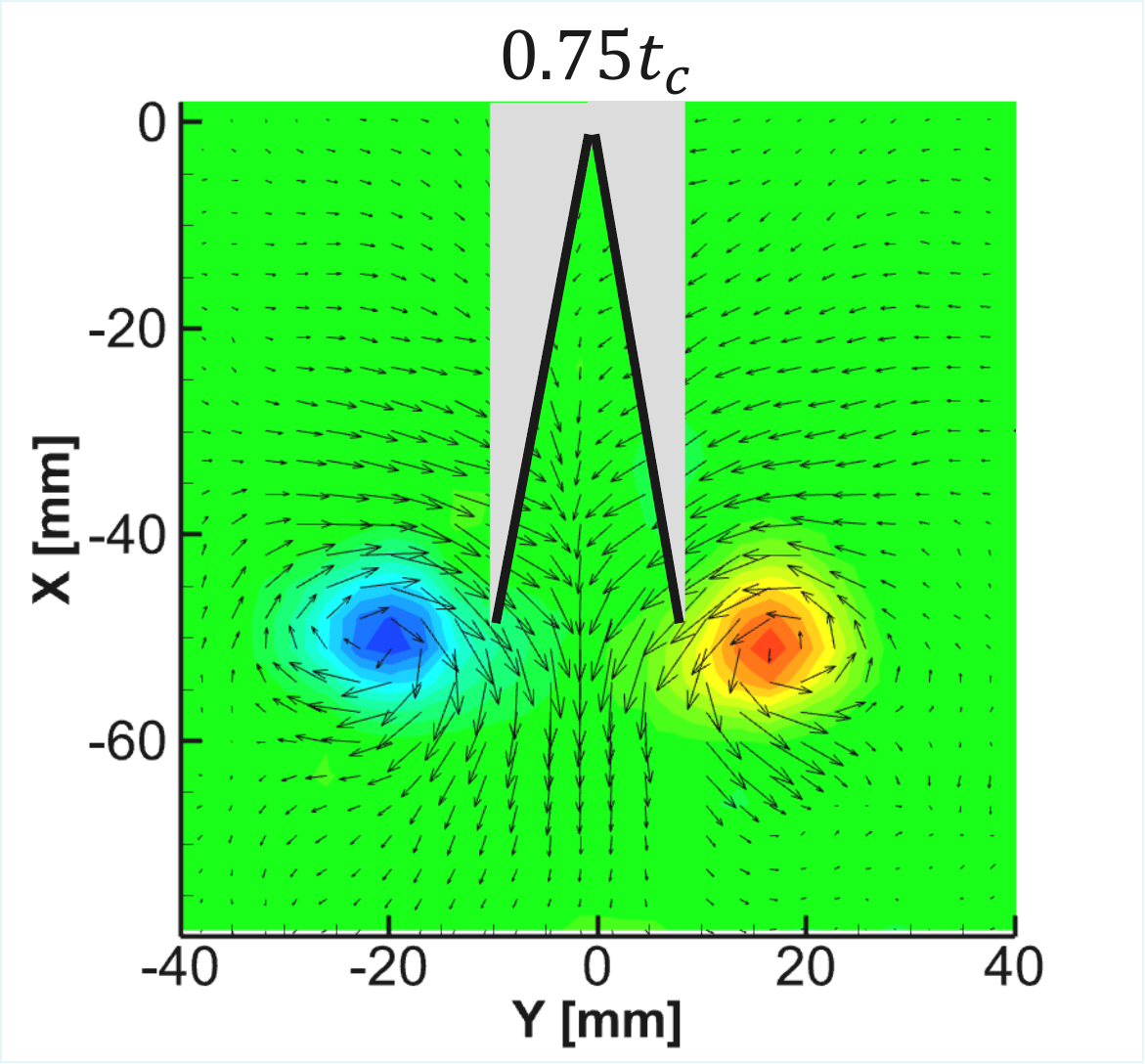}
		\caption{EXP: Stat}
	\end{subfigure}\hspace{05mm}
	\begin{subfigure}[b]{0.32\textwidth}
		\includegraphics[width=\textwidth]{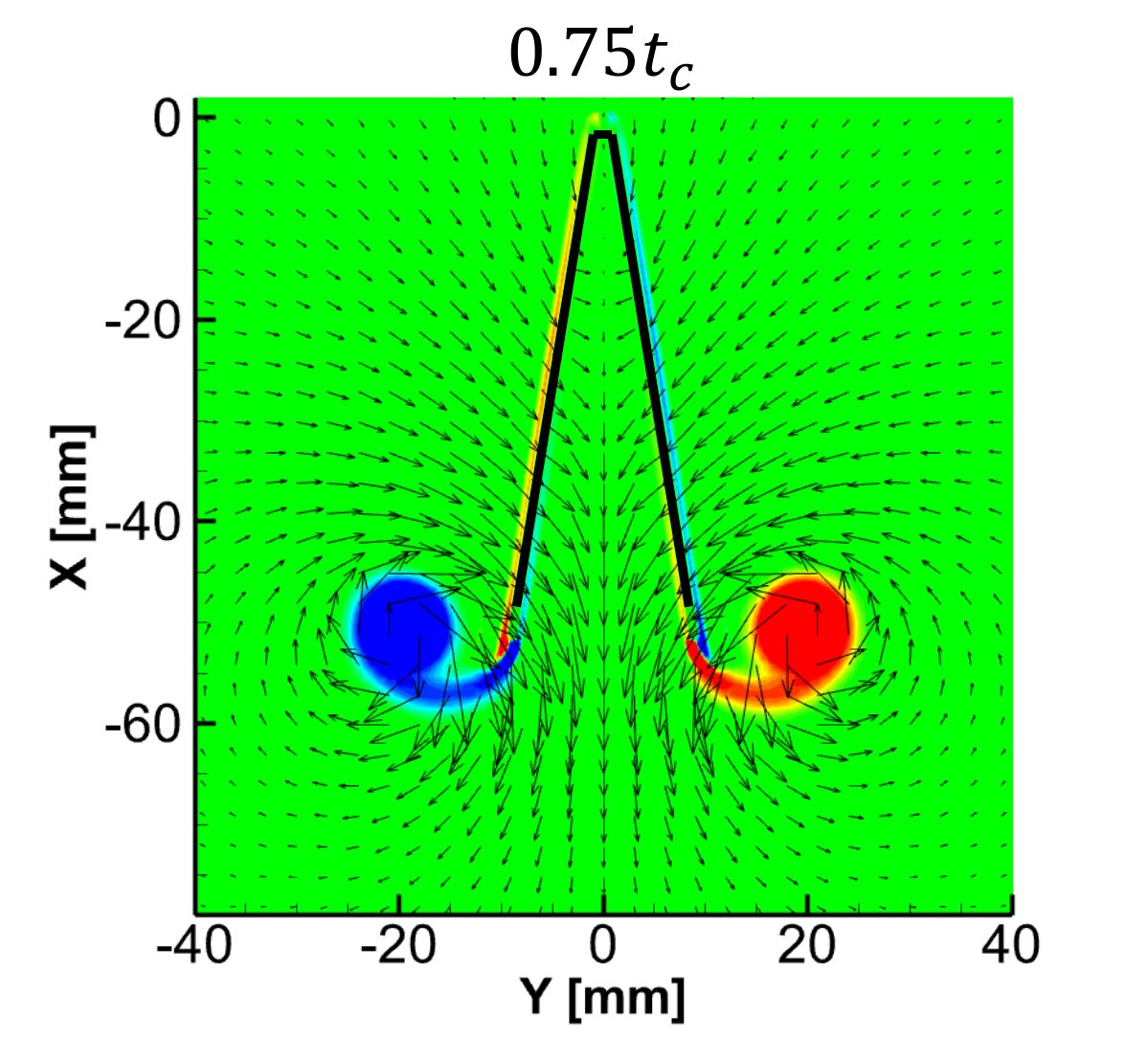}
		\caption{CFD: Stat}
	\end{subfigure}\vspace{01mm}
	\begin{subfigure}[b]{0.32\textwidth}
		\includegraphics[width=\textwidth]{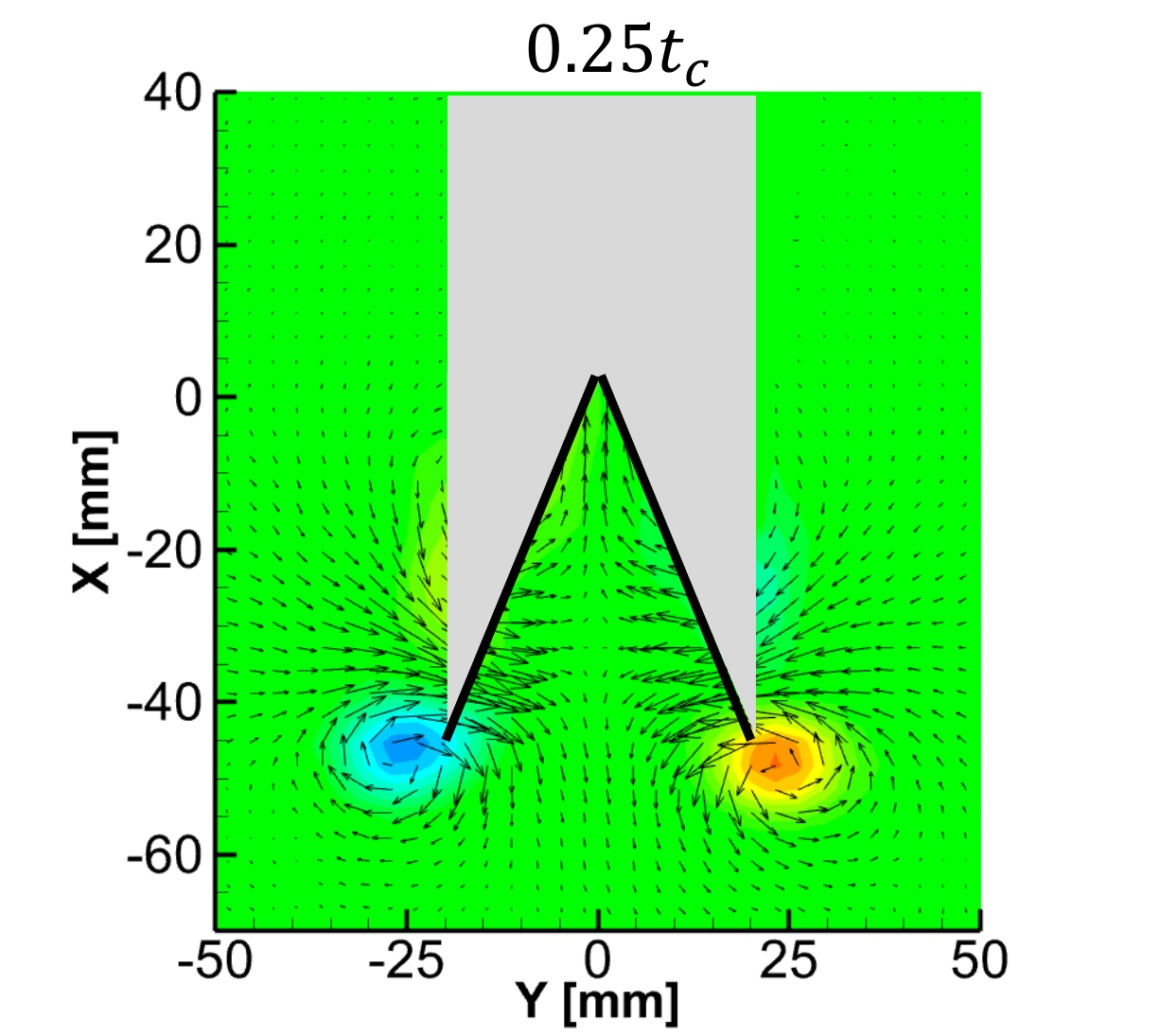}
		\caption{EXP: Dyn}
	\end{subfigure}\hspace{05mm}
	\begin{subfigure}[b]{0.32\textwidth}
		\includegraphics[width=\textwidth]{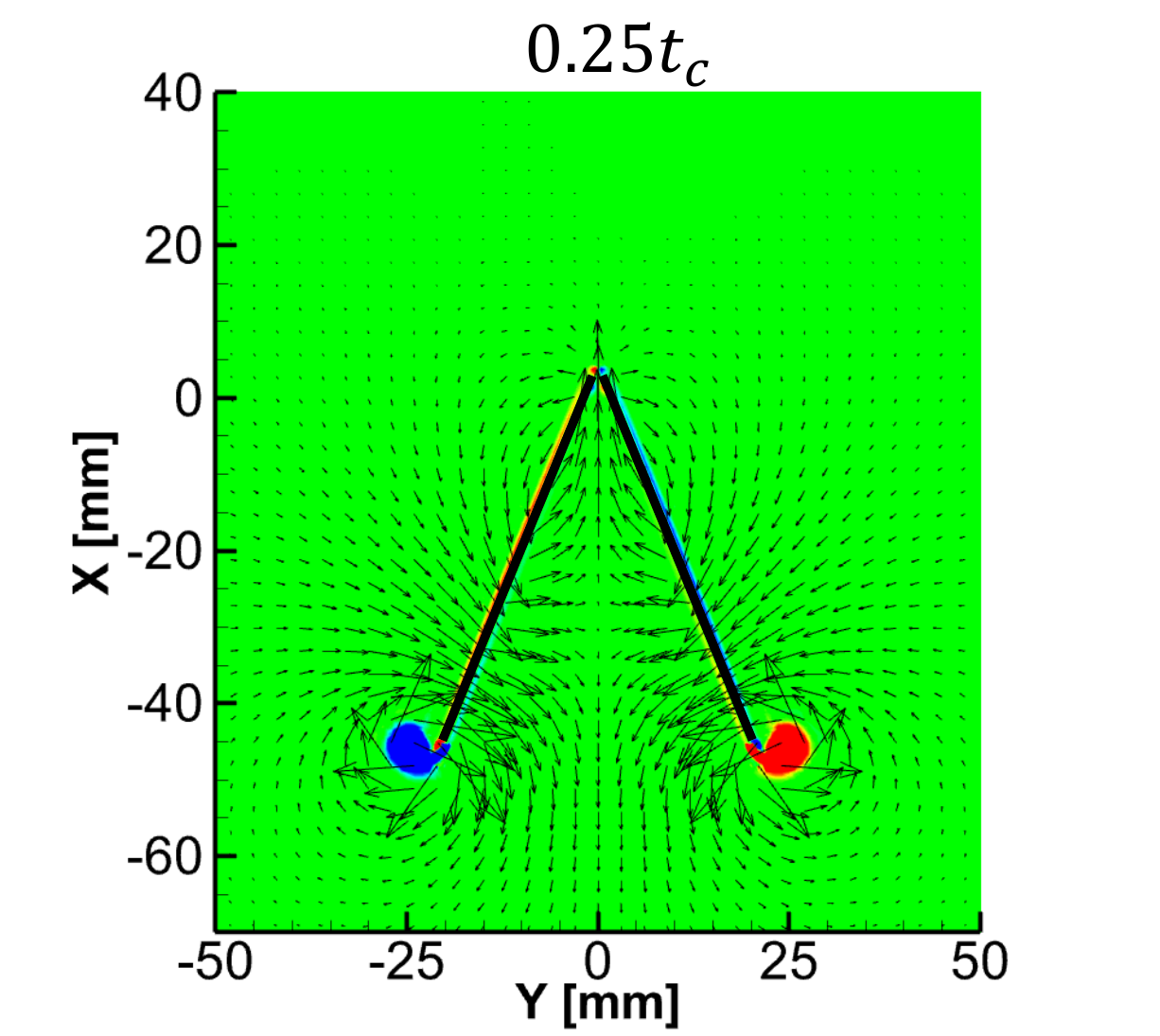}
		\caption{CFD: Dyn}
	\end{subfigure}\vspace{01mm}
	\begin{subfigure}[b]{0.32\textwidth}
		\includegraphics[width=\textwidth]{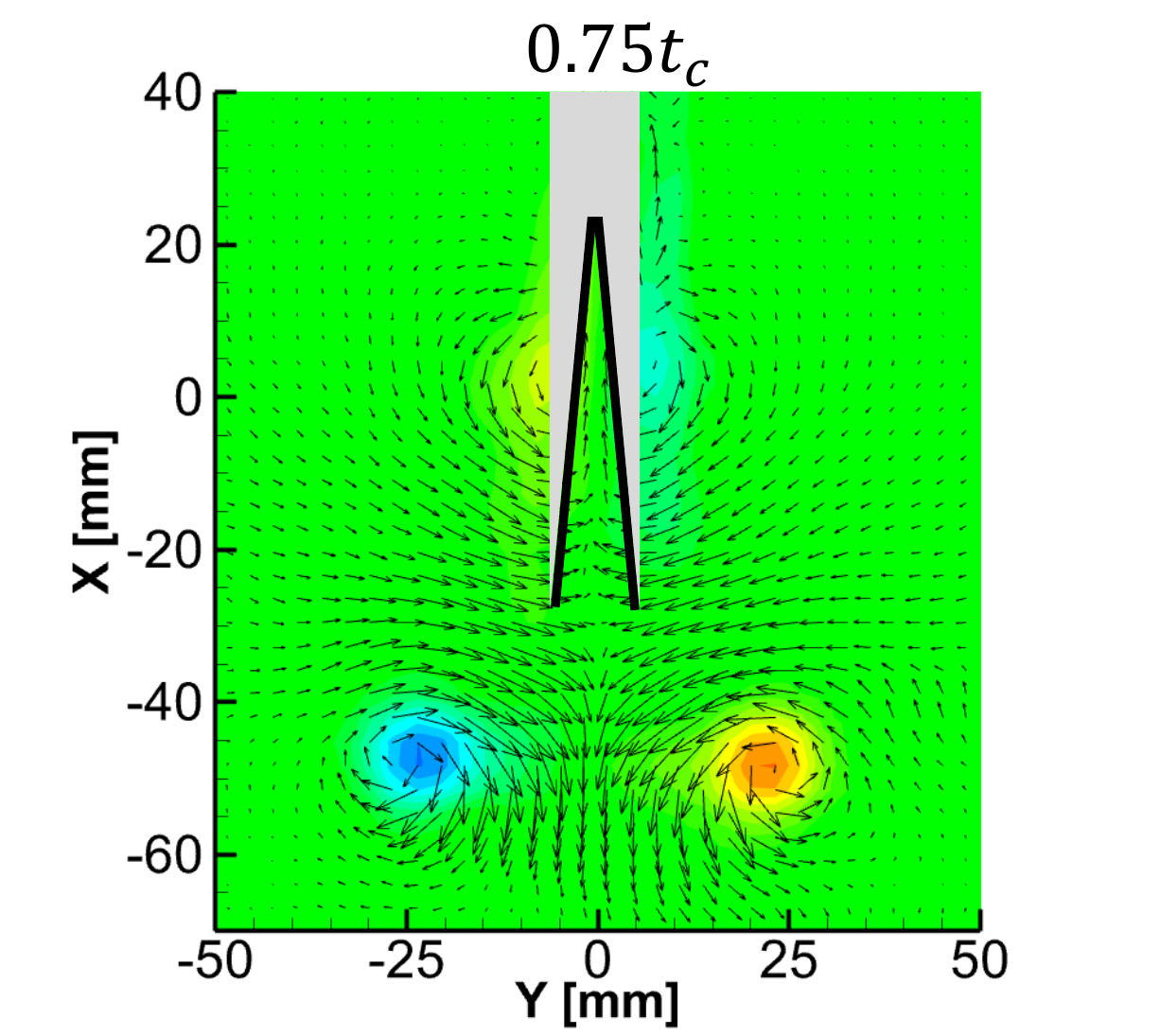}
		\caption{EXP: Dyn}
	\end{subfigure}\hspace{05mm}
	\begin{subfigure}[b]{0.32\textwidth}
		\includegraphics[width=\textwidth]{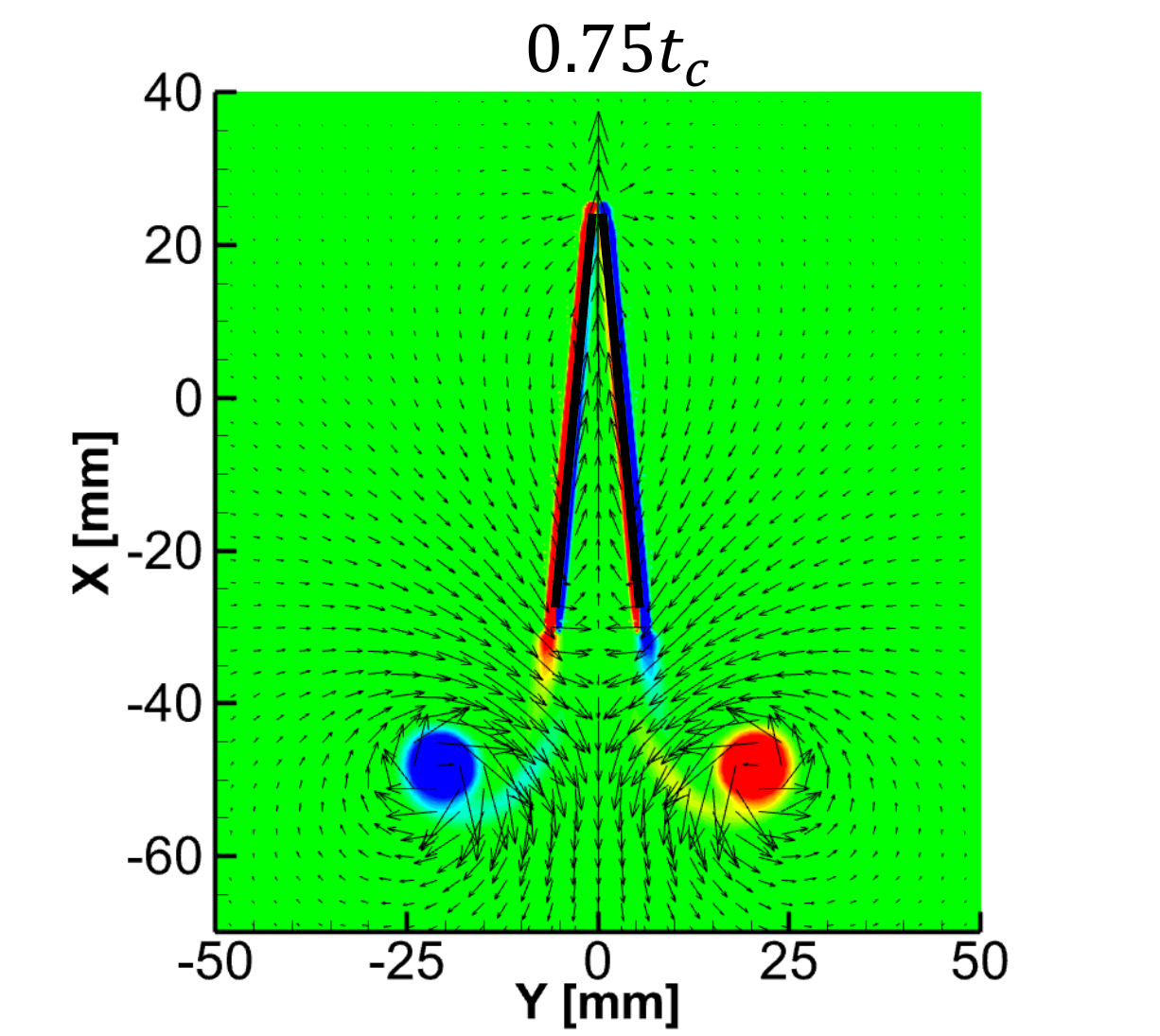}
		\caption{CFD: Dyn}
	\end{subfigure}
	\begin{subfigure}[b]{0.40\textwidth}
		\includegraphics[width=\textwidth]{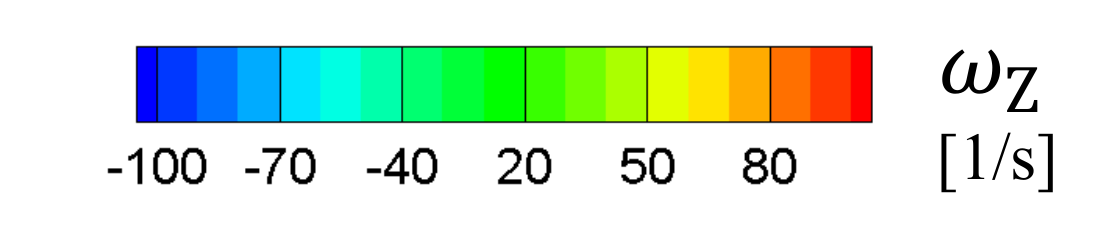}
	\end{subfigure}
	\caption{Comparison of experimental (EXP) and computational (CFD) flow fields for stationary (Stat) and dynamic (Dyn) cases for $d^* = 0.5$. The flow fields, plotted at $t = 0.25 t_c$ (a, b, e, and f) and $t = 0.75 t_c$ (c, d, g, and h), show velocity vectors and the Z-component of vorticity, $\omega_Z$, on the XY plane at $Z = 0$.}\label{fig:C_VortZ_AR200_Dyn_Stat}	
\end{figure}

\begin{figure}
	\centering\
	\begin{subfigure}[b]{0.40\textwidth}
		\includegraphics[width=\textwidth]{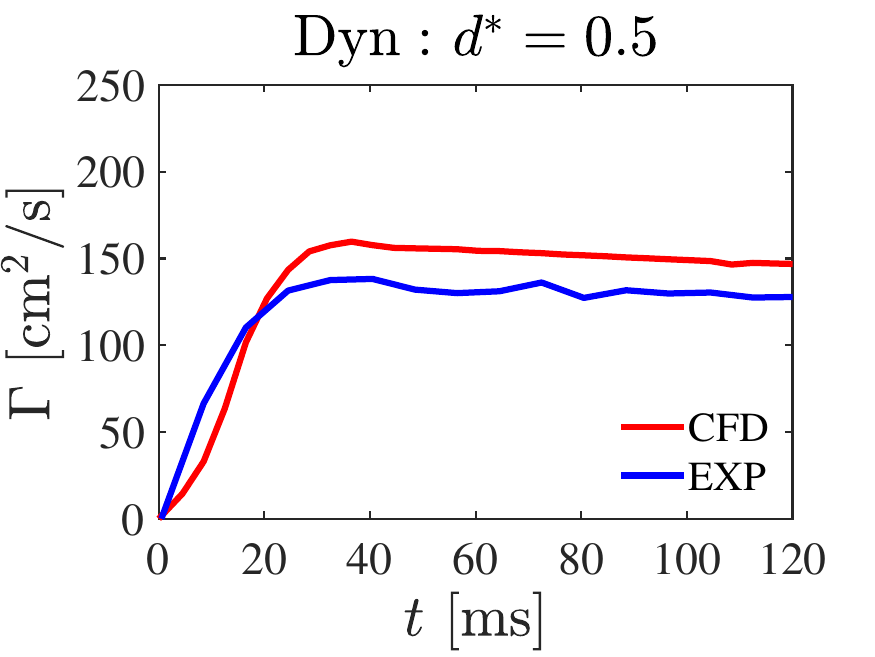}
		\caption{}
	\end{subfigure}\hspace{5mm}
	\begin{subfigure}[b]{0.40\textwidth}
		\includegraphics[width=\textwidth]{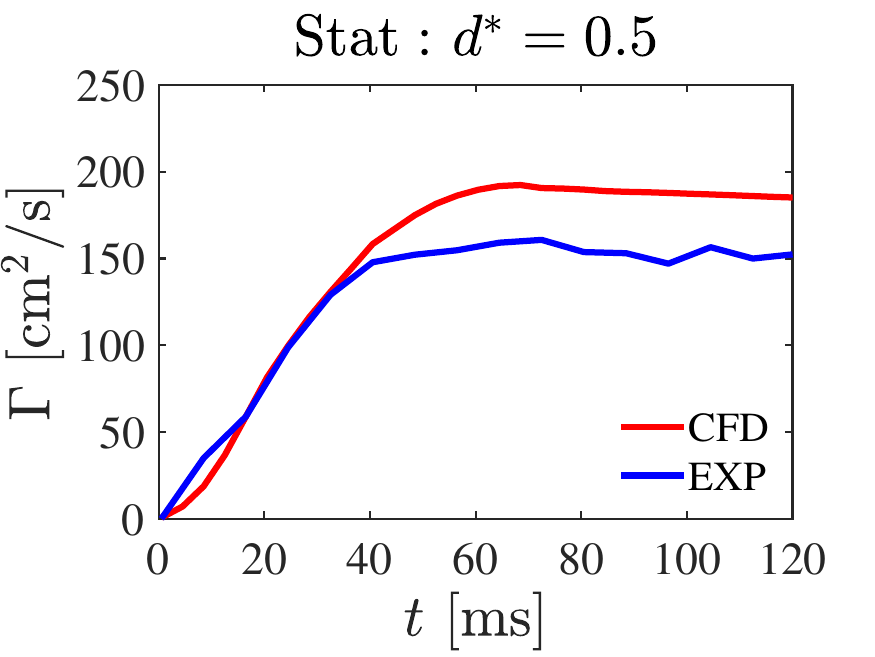}
		\caption{}
	\end{subfigure}\vspace{05mm}
	\begin{subfigure}[b]{0.40\textwidth}
		\includegraphics[width=\textwidth]{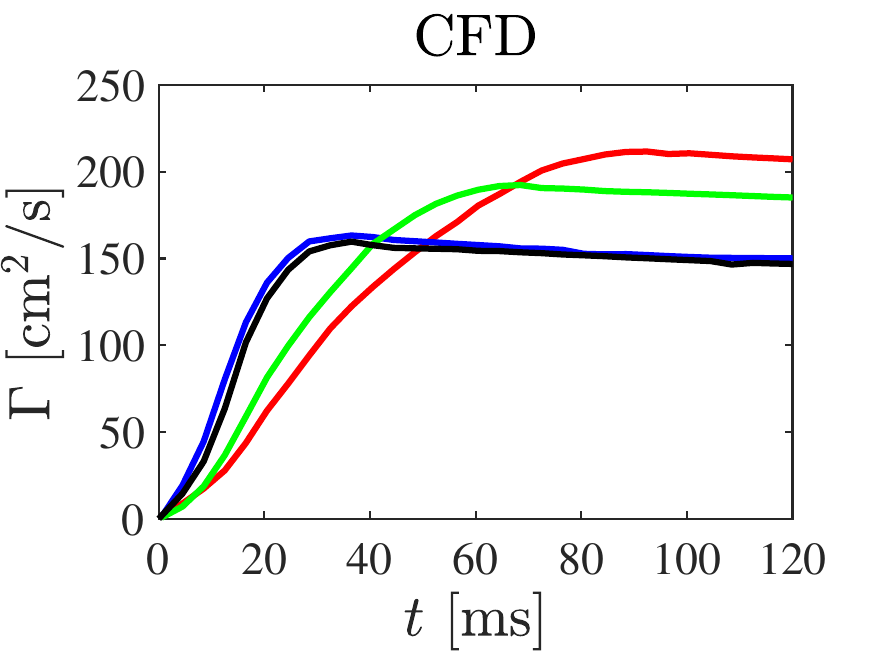}
		\caption{}
	\end{subfigure}\hspace{5mm}
	\begin{subfigure}[b]{0.40\textwidth}
		\includegraphics[width=\textwidth]{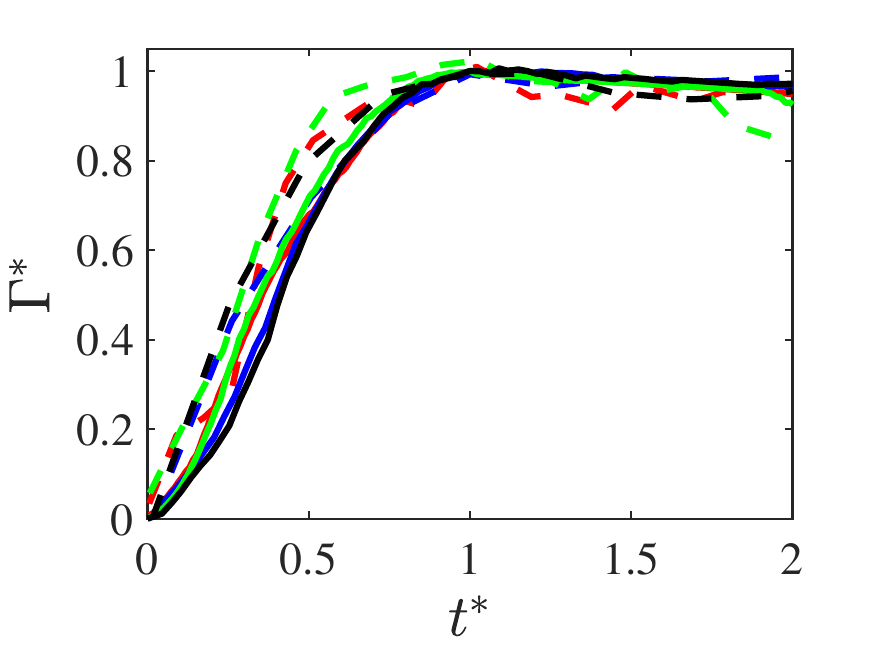}
		\caption{}
	\end{subfigure}
	\begin{subfigure}[b]{0.80\textwidth}
		\includegraphics[width=\textwidth]{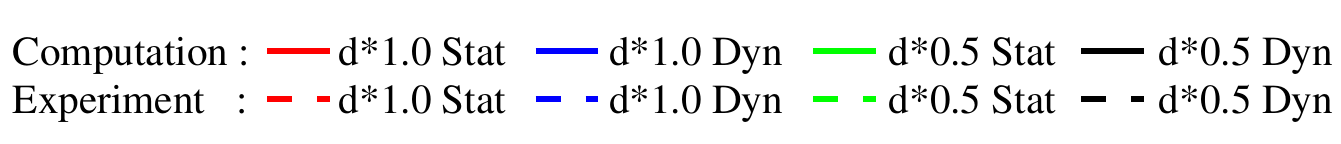}
	\end{subfigure}
	\caption{Evolution of starting vortex circulation, $ \Gamma $, with time, $ t $: (a) and (b) compare the computationally obtained values (CFD), shown in red, with the experimentally obtained values (EXP), shown in blue, for the dynamic (Dyn) and stationary (Stat) cases of $ d^* = 0.5 $; (c) compares the evolution of $ \Gamma $ obtained from computations for stationary and dynamic cases for $ d^* = 0.5 $ and $ d^* = 1.0 $; (d) shows a plot of nondimensional circulation, $ \Gamma^* (=\Gamma / \Gamma_m) $, versus nondimensional time, $ t^* (= t / t_{\Gamma_m}) $. Here, the maximum circulation, $ \Gamma_m $, and the time to reach maximum circulation, $ t_{\Gamma_m} $ (listed in table~\ref{tab:C_BodyDyn_gamma}), are used to normalize $ \Gamma $ and $ t $, respectively. Data from experiments and computations for both $ d^* $ values in stationary and dynamic cases are compared, with the corresponding legends provided below the plot.}\label{fig:C_Gamma}	
\end{figure}

	In both the stationary and dynamic cases, the clapping motion initiated by cutting the thread leads to the formation of vortices along the trailing and side edges of the plates, which eventually form vortex loops shed into the wake. Figure \ref{fig:C_VortZ_AR200_Dyn_Stat} shows the velocity vectors and vorticity contours in the horizontal (XY) plane at two instants, $0.25t_c$ and $0.75t_c$, for the stationary (a, b, c, and d) and dynamic (e, f, g, and h) cases of $d^* = 0.5$. The corresponding interplate cavity angles are approximately $51^\circ$ and $19^\circ$ for the stationary case, and $45^\circ$ and $11^\circ$ for the dynamic case, respectively. Both experimental and computational data are shown in the figure. The flow field is coloured by the Z-component of vorticity, $\omega_Z$, where red–blue patches indicate the starting vortices. The black lines show the rear part of the clapping plates, and the gray region on their back side indicates the shadow region in the PIV. Further details of the PIV setup can be found in Mahulkar and Arakeri \cite{Mahulkar24}.\par
	
	The flow fields in the stationary case during the initial phase of the clapping motion ($ t = 0.25t_c $) show that the fluid jet begins to eject from the opening of the cavity (see figure \ref{fig:C_VortZ_AR200_Dyn_Stat}(a, b)), while starting vortices begin to form along the trailing edges of the plates. At the later time ($ t = 0.75t_c $), a higher jet velocity is observed and the starting vortices have grown in size. Two important differences are observed between the flow fields on the XY plane for the stationary and dynamic cases. In the dynamic case, only part of the fluid is ejected, while the rest is carried forward with the body, as seen in figures \ref{fig:C_VortZ_AR200_Dyn_Stat}(e, f, g, and h). A second difference is that the starting vortices are left behind as the body moves forward (figures \ref{fig:C_VortZ_AR200_Dyn_Stat}(g, h)). Note that these vortices represent a cross-section of the three-dimensional vortex loop around the plate edges; their evolution is discussed in \S\ref{sec:C_3DWakeVortices}. Further, flow fields on the XZ plane reveal the development of stronger sideways flow in the stationary cases, which is discussed in the next subsection (figure \ref{fig:C_sideways_flow_field}). The flow field description presented above for $ d^* = 0.5 $ remains largely valid for $ d^* = 1.0 $.\par
	
	In general, the computational flow fields match well with the experimentally obtained ones, except for slightly higher vorticity levels in the starting vortices in the computations, indicated by darker red and blue colours. In addition, vortex sheets downstream of the trailing edges are clearly visible in the computed flow fields but not in the experimental data. \par

	The strength of the vortex is quantified by the circulation, $ \Gamma \ (= \int \omega_Z \ dA_{cr}) $, calculated using the Z-component of vorticity, $\omega_Z$, in the XY plane ($Z=0$) over the vortex core area, $ A_{cr} $. The vortex core is identified using the criterion: $ \omega_Z \geq 5\% \ \omega_{Z|\mathrm{max}} $. The time evolution of circulation in the starting vortices in experiments and computations is shown in figures \ref{fig:C_Gamma}(a, b) for the dynamic and stationary cases, respectively, for the $ d^* = 0.5 $ body. In both cases, $ \Gamma $ initially increases before reaching a nearly steady value, with slightly higher circulation levels in the computations; the probable cause of this difference is discussed later in the section. Figure \ref{fig:C_Gamma}(c) compares the circulation evolution for dynamic and stationary cases for bodies with $ d^* = 0.5 $ and $1.0$. Two important conclusions can be drawn: the maximum steady circulation value, $ \Gamma_m $, is higher for stationary cases and increases with $ d^* $, while in dynamic cases $ \Gamma_m $ is lower and relatively constant with $ d^* $ (see table \ref{tab:C_BodyDyn_gamma}). The lower circulation in dynamic cases occurs because the starting vortices detach earlier from the forward-moving plates (figures \ref{fig:C_VortZ_AR200_Dyn_Stat}(g, h)), and vorticity is therefore fed into them for a shorter duration; in contrast, in stationary cases the starting vortices remain close to the plates, where vorticity is fed for a longer time. Further, the increase in $ \Gamma_m $ with $ d^* $ in stationary bodies is primarily due to reduced sideways flow as body height (or $ d^* $) increases, which leads to higher flow velocity in the wake behind the body and hence higher circulation. In contrast, in the dynamic cases negligible sideways flow results in nearly constant $ \Gamma_m $ over $ d^* $ variations. These aspects were discussed in detail in our previous experimental study (Mahulkar and Arakeri,\cite{Mahulkar24}). Furthermore, the time, $ t_{\Gamma_m} $, at which the starting vortex reaches its maximum value shows a close match between computational and experimental data (see table \ref{tab:C_BodyDyn_gamma}), and its variation with $ d^* $ follows the same trend as that of $ \Gamma_m $ for stationary and dynamic cases. Nondimensional circulation, $ \Gamma^* (= \Gamma / \Gamma_m) $, when plotted against nondimensional time, $ t^* (= t / t_{\Gamma_m}) $, shows a collapse of curves for both $ d^* $ values in dynamic and stationary cases for both computational and experimental data (see figure \ref{fig:C_Gamma}(d)), showing that the overall nature of circulation evolution remains the same across the parametric space. \par

	We comment on the higher vorticity and circulation levels observed in the computations than in the experiments. Unlike the flat-plate geometry used in the computations, each clapping plate in the experiments has two balsa pieces ($\approx$ 2 mm thick): an aerofoil-shaped balsa at the leading edge and a rectangular balsa at the rear end of the plate. Flow over the leading balsa sheds vorticity that is visible on the outer side of the plates (figure \ref{fig:C_VortZ_AR200_Dyn_Stat}(g)). This outer-side vorticity is oppositely oriented and weaker than the vorticity shed from the inner side facing the interplate cavity. Near the trailing edge, interaction between the oppositely shed layers weakens the inner-side vorticity, which rolls up into a starting vortex with lower circulation. This reduction in circulation is negligible in the computations because of the flat-plate geometry.
	
\subsection{Three-dimensional pressure fields}
	\label{sec:C_Pressure distribution}
		
	\begin{figure}
		\centering\
		\begin{subfigure}[b]{0.40\textwidth}
			\includegraphics[width=\textwidth]{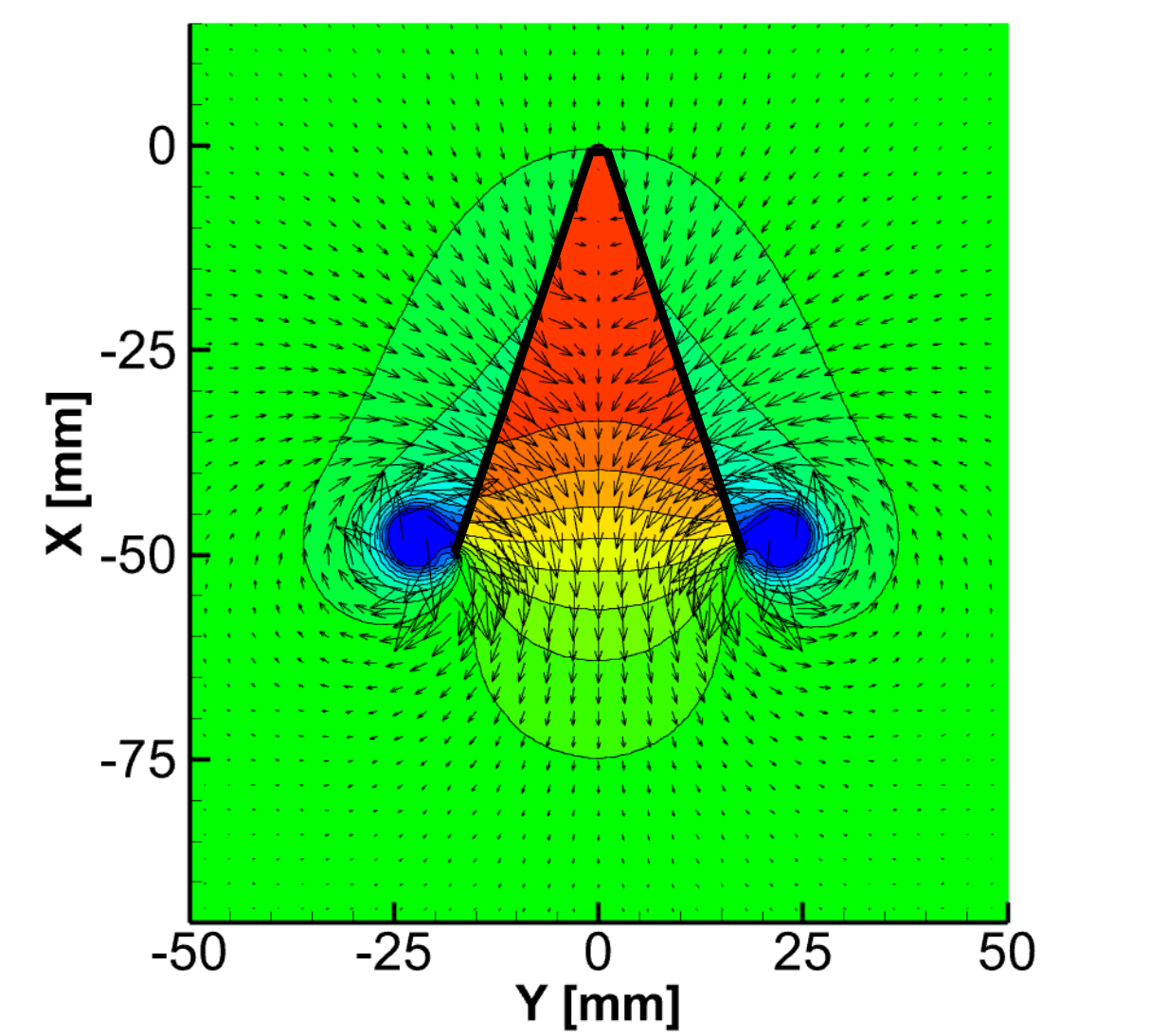}
			\caption{Stat ($t \approx 0.5 t_c$)}
		\end{subfigure}\hspace{05mm}
		\begin{subfigure}[b]{0.40\textwidth}
			\includegraphics[width=\textwidth]{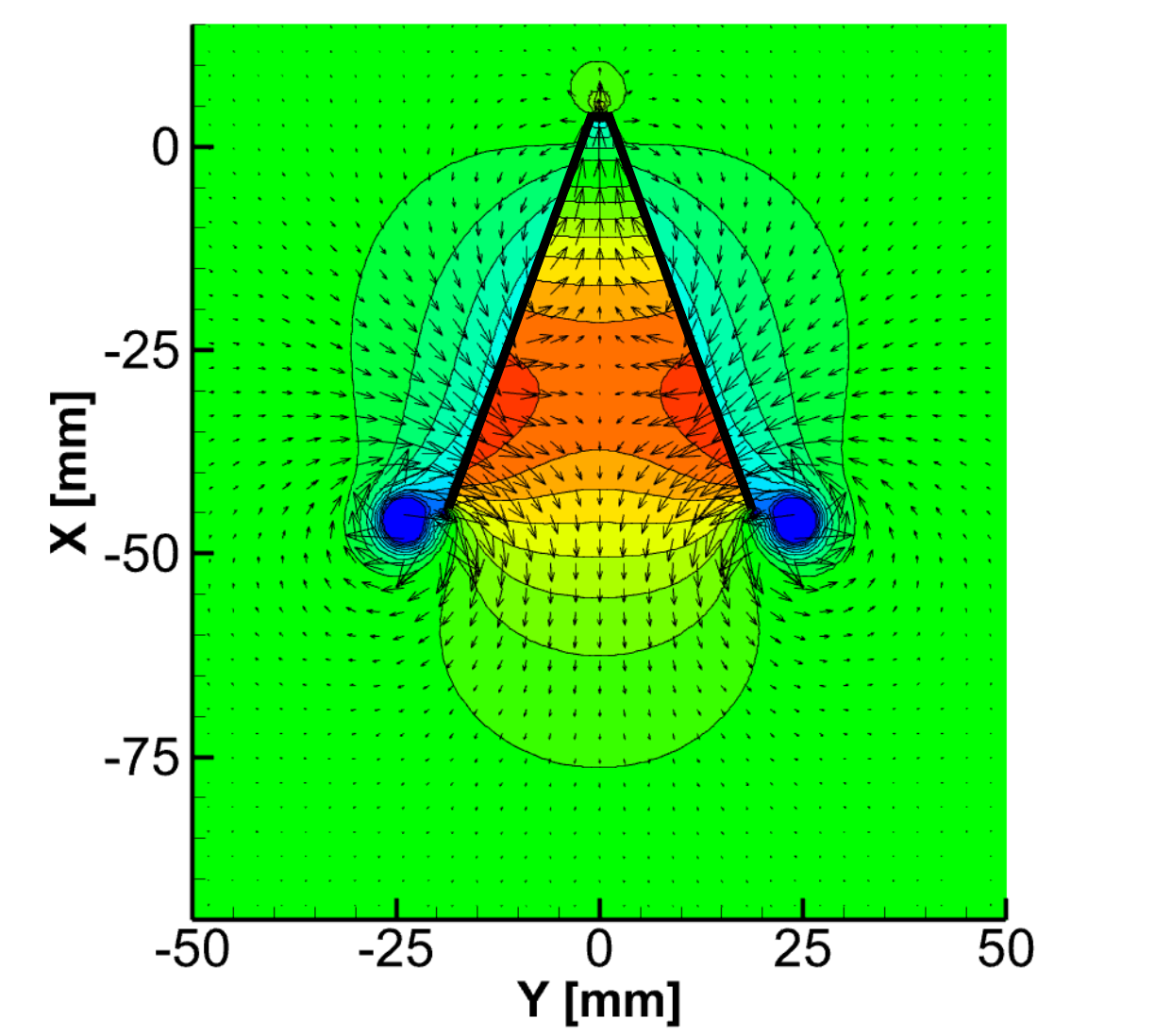}
			\caption{Dyn ($t \approx 0.3 t_c$)}
		\end{subfigure}
		
		\begin{subfigure}[b]{0.40\textwidth}
			\includegraphics[width=\textwidth]{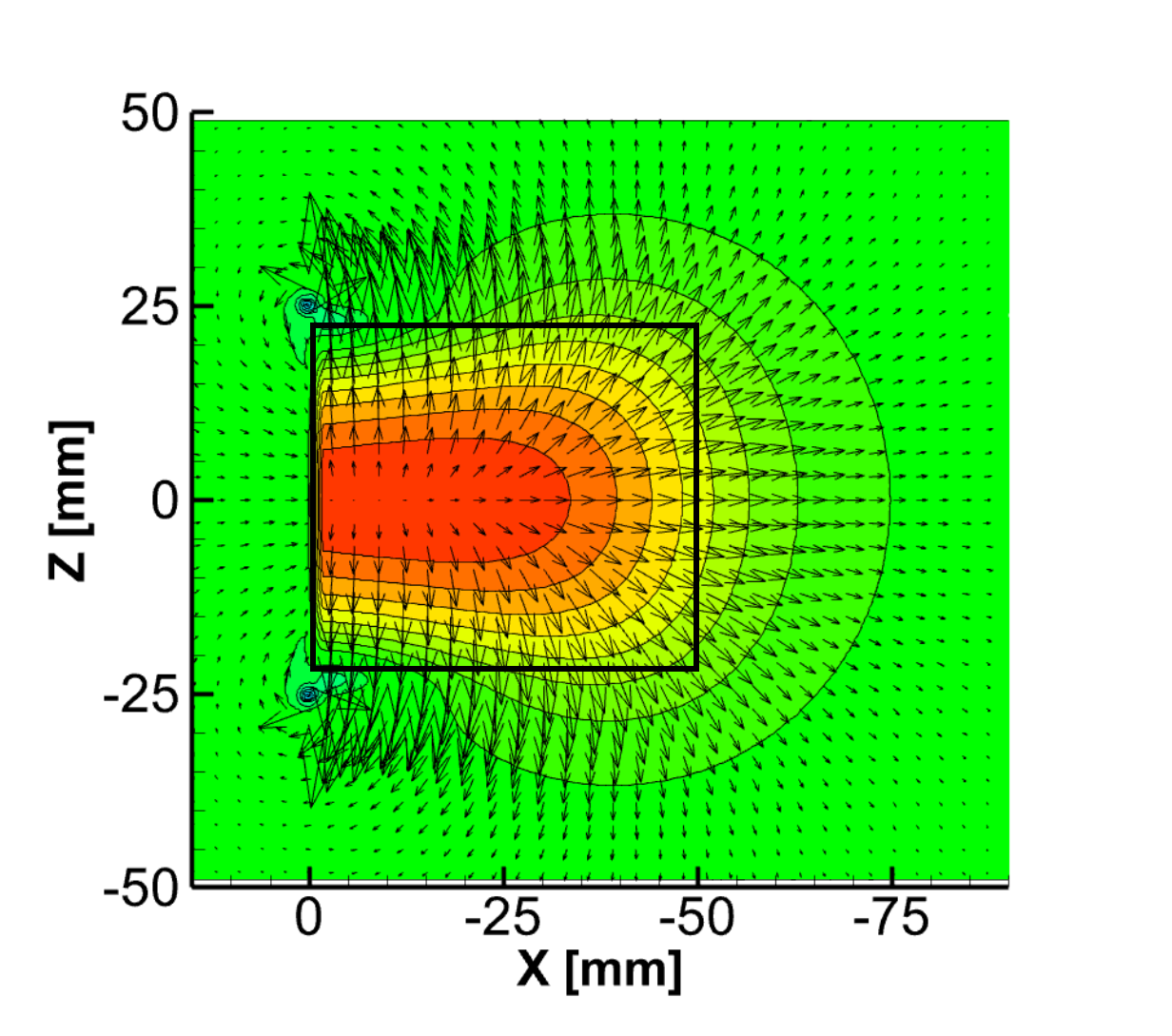}
			\caption{Stat ($t \approx 0.5 t_c$)}
		\end{subfigure}\hspace{05mm}
		\begin{subfigure}[b]{0.40\textwidth}
			\includegraphics[width=\textwidth]{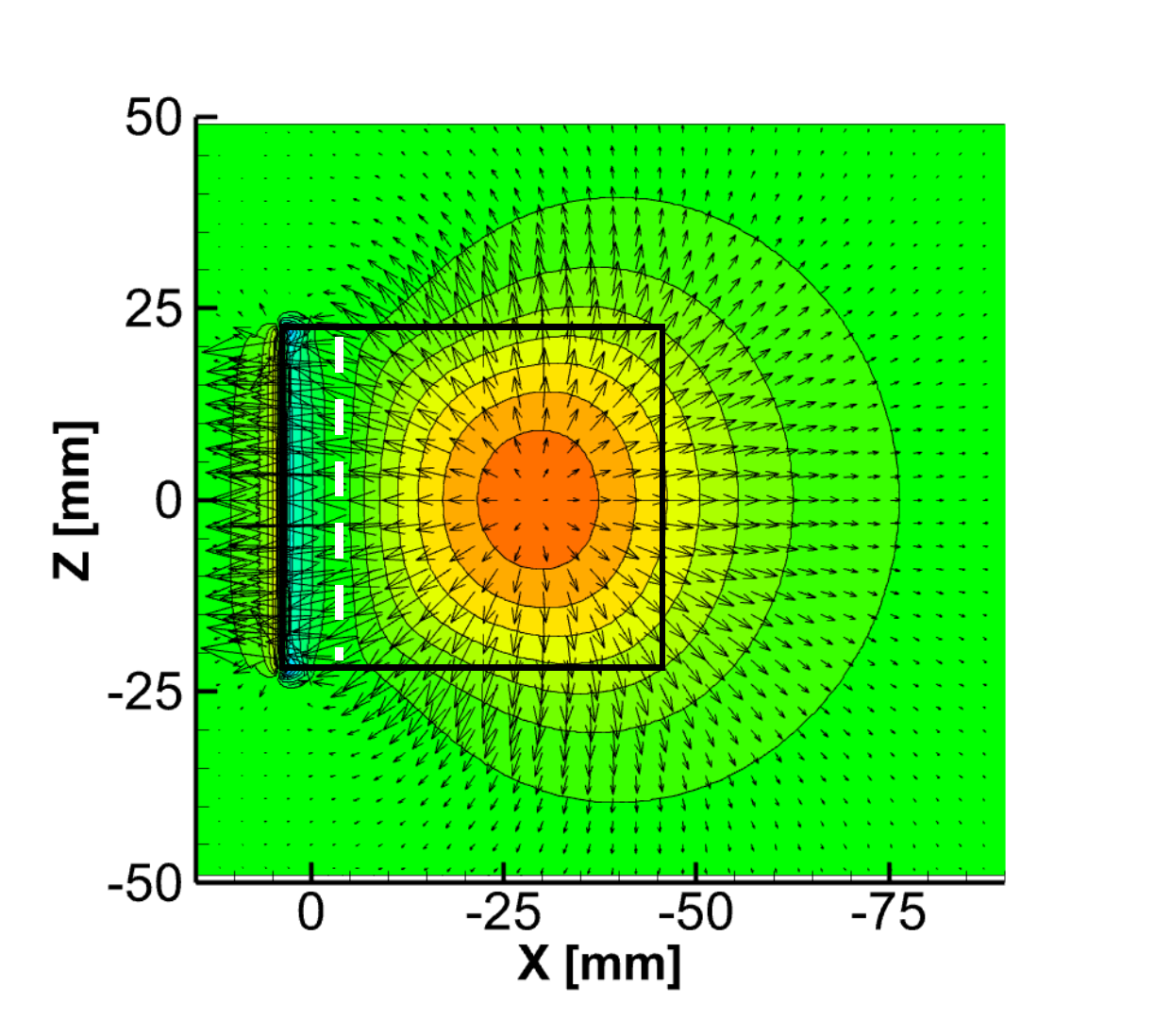}
			\caption{Dyn ($t \approx 0.3 t_c$)}
		\end{subfigure}
		\begin{subfigure}[b]{0.50\textwidth}
			\includegraphics[width=\textwidth]{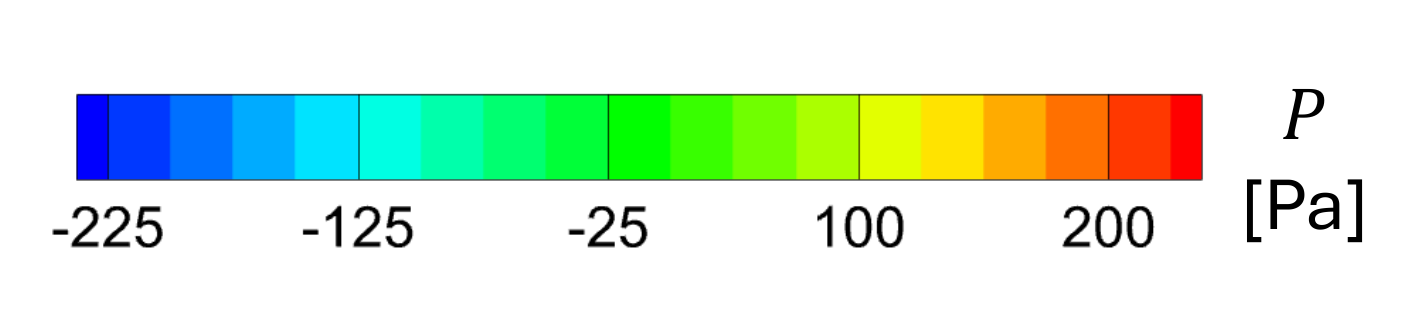}
		\end{subfigure}
		
		\caption{ Static pressure (gauge) field and velocity vectors at the time $t_{\dot{\theta}m}$, when the angular velocity of the plate reaches its maximum value (table \ref{tab:C_BodyDyn_gamma}). The flow field is plotted on the XY plane (Z=0) for (a) the stationary case (Stat) and (b) the dynamic case (Dyn) for $d^* = 0.5$, and on the XZ plane (Y=0) for (c) the stationary case and (d) the dynamic case for the same $d^*$ value. In (a) and (b), the top view of the clapping body is represented by two black lines, while in (c) and (d), the side view projection of the clapping plate is shown as a rectangle. In (d), the white dashed line marks the transition zone, where the pressure is negative (blue) on the left side and positive (red) on the right side. This zone is observed only in the dynamic case.}\label{fig:C_Pr_Stat_dyn}	
	\end{figure}
		We now give the pressure fields and velocity vectors obtained from the CFD simulations in two perpendicular planes in figure \ref{fig:C_Pr_Stat_dyn} for the stationary (a,c) and dynamic (b,d) cases of $d^* = 0.5$, at the time instant when the angular velocity of the clapping plates reaches its maximum, $t_{\dot{\theta}m}$ (table \ref{tab:C_BodyDyn_gamma}); the corresponding interplate angles, $2\theta$, are approximately $37^\circ$ and $40^\circ$, respectively. We first discuss the pressure field generated by the stationary clapping body. Expectedly, the shrinking of the interplate cavity raises the pressure inside the cavity, while the rotation of the plates produces negative pressure, with respect to ambient, along their outer surfaces. Figure \ref{fig:C_Pr_Stat_dyn}(a) shows this gauge pressure distribution on the XY plane ($Z=0$), with positive pressure (red) inside the cavity and negative pressure (blue) on the back side of the plates. The rotation of both plates forms starting vortices near the trailing edge, which show low-pressure values in their core regions. In the XZ-plane (figure~\ref{fig:C_Pr_Stat_dyn}(c)), the pressure reduces from a maximum in the mid-region towards the three edges. In addition to the flow from the cavity trailing edge, significant sideways flow occurs across the top and bottom edges. The above description corresponds to the effective clapping phase. However, in the final phase of clapping, a \textit{pressure crossover} is observed at $t_{\pm p} \approx 65$ ms, when negative pressure develops inside the closing cavity and positive pressure develops outside (not shown). In this phase ($ t \geq t_{\pm p} $), the spring force driving the plate motion has considerably reduced, while the fluid continues to move out of the cavity. We quantify the effect of pressure changes on the body (both plates) using the thrust coefficient, which is discussed later in \S\ref{sec:C_ThrustCoeff}. The pressure field description remains the same for the stationary body of $d^* = 1.0$, except for a slight delay in pressure crossover ($t_{\pm p} \approx 85$ ms), attributed to the delayed closing of the clapping motion ($t_c = 104$ ms, see table \ref{tab:C_BodyDyn_gamma}).\par
	
		For the dynamic case of $d^*=0.5$, during the first 4 ms, the plates undergo only clapping motion without translation. During this time, the pressure distributions are similar to those in the stationary case. After 4 ms, as the body starts moving forward, the pressure field becomes very different from that in the stationary case (figure~\ref{fig:C_Pr_Stat_dyn}(b,d)). The positive pressure zone within the cavity moves towards the trailing edge, while negative pressure exists close to the hinge region, in contrast to the stationary case, which shows only positive pressure in the cavity. In figure~\ref{fig:C_Pr_Stat_dyn}(d), the transition region from negative to positive pressure is indicated by a white dashed line, where the pressure is nearly ambient. The positive pressure distribution in the cavity is nearly circular, with outward flow in all directions. As in the stationary case, the pressure on the outer surfaces of the two plates is negative, except near the hinge region, where a `stagnation' region exists. Due to the body's forward motion, the sideways flow is significantly reduced compared to the stationary case. Towards the end of the clapping phase, a pressure crossover also occurs in the dynamic case at $t_{\pm p} \approx 40$ ms, when the pressure inside the cavity becomes completely negative (not shown). At this phase, the clapping cavity is nearly closed. Further, in contrast to the stationary case, where the pressure field eventually reaches ambient, the dynamic case exhibits a second pressure crossover around 60 ms (not shown). Here, the fluid trapped in the very narrow interplate cavity is dragged along with the purely translating body, resulting in low positive pressure inside the cavity ($\sim 10$ Pa) and low negative pressure of similar magnitude outside, particularly toward the trailing edges. The pressure description at the hinge region remains unchanged. This second pressure crossover can be understood from the motion of the trapped fluid in the narrow translating cavity. As the cavity narrows, the trapped fluid tends to accelerate relative to the cavity, while confinement by the plates limits this motion and produces a weak pressure buildup inside. Near the trailing edges, flow separation and wake formation produce a localized region of negative pressure relative to ambient. In this phase, the thrust coefficient is nearly zero (see \S\ref{sec:C_ThrustCoeff}). Crucially, a second pressure crossover cannot occur if the interplate cavity fully closes. For the dynamic case of $d^*=1.0$, this pressure field description remains the same.  \par

		An approximate analysis based on the unsteady Bernoulli equation is presented in Appendix~\ref{sec:Pr_ub_dot}. It shows that, in the dynamic cases, forward translation reduces the pressure difference across the rotating plates during the initial stage of motion, consistent with the lower pressure levels observed in these cases.		

\subsection{Evolution of wake vorticity structures}
	\label{sec:C_3DWakeVortices}

	An overall understanding of the flow field evolution can be obtained by examining how the vortex structures form and evolve during the clapping motion. In our previous experimental study comparing stationary and freely moving clapping bodies (Mahulkar and Arakeri \cite{Mahulkar24}), an approximate three-dimensional structure of the wake vortices was reconstructed using PIV data from two perpendicular planes. The computational data presented here provide a more detailed description of this vortex evolution. \par
	
	\subsubsection{Stationary cases}
	\begin{figure}
	\centering\
	\begin{subfigure}[b]{0.23\textwidth}				     \framebox{\includegraphics[width=\textwidth]{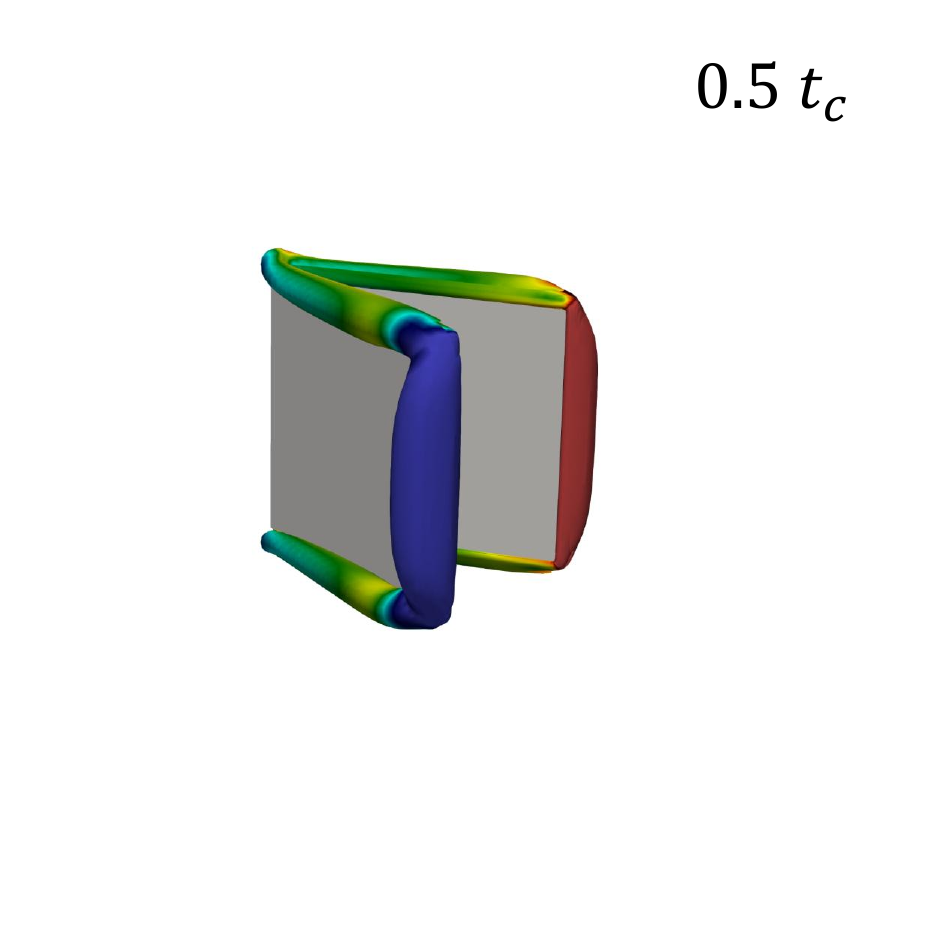}}
		\caption{}
	\end{subfigure}\hspace{20mm}
	\begin{subfigure}[b]{0.23\textwidth}
		\framebox{\includegraphics[width=\textwidth]{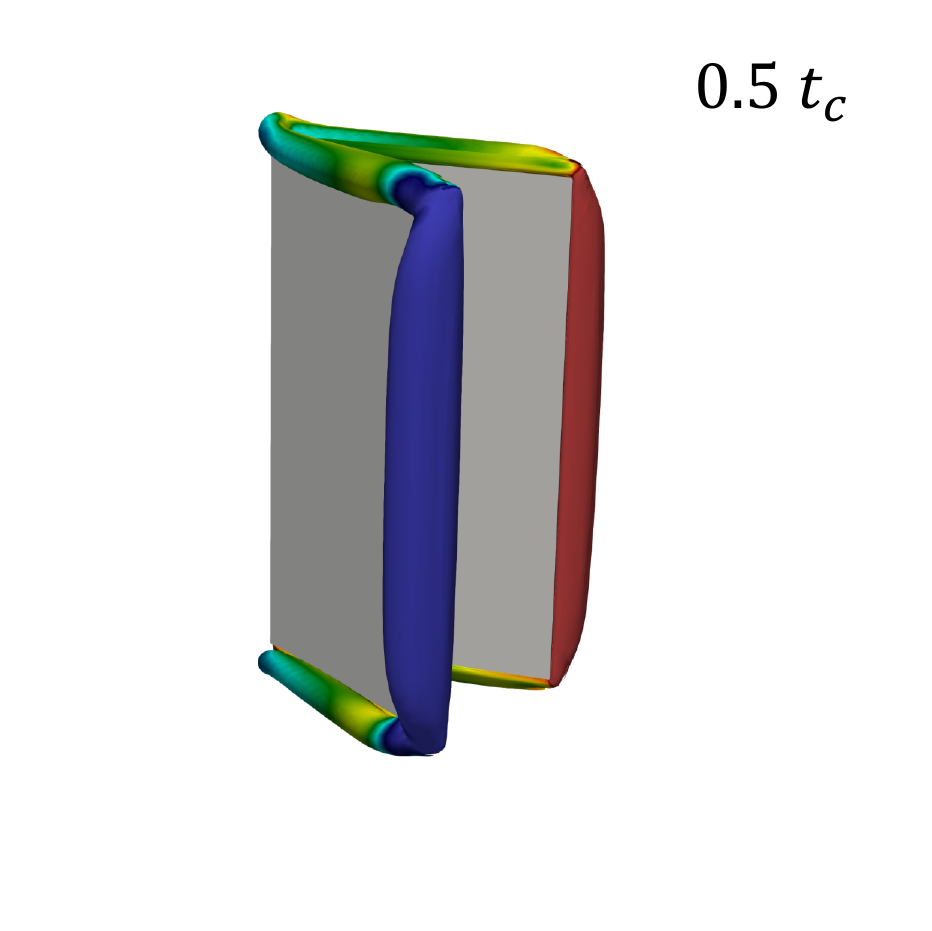}}
		\caption{}
	\end{subfigure}
	
	\begin{subfigure}[b]{0.23\textwidth}								\framebox{\includegraphics[width=\textwidth]{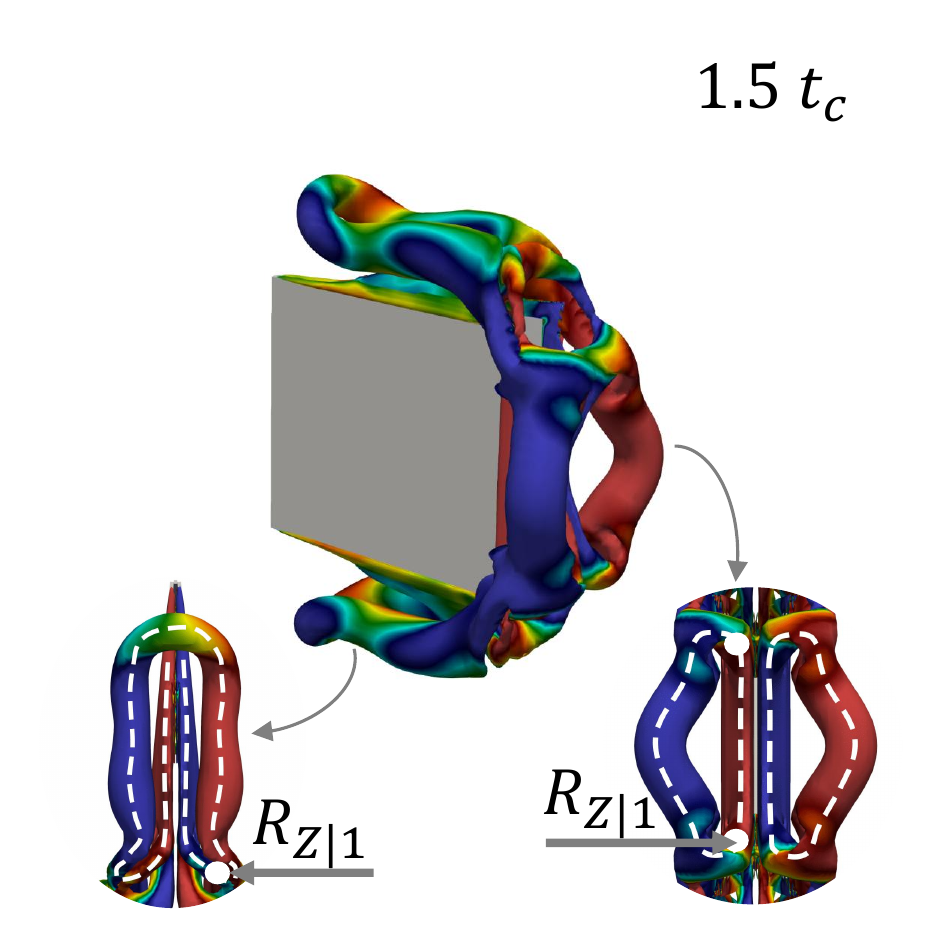}}
		\caption{}
	\end{subfigure}\hspace{20mm}
	\begin{subfigure}[b]{0.23\textwidth}
		\framebox{\includegraphics[width=\textwidth]{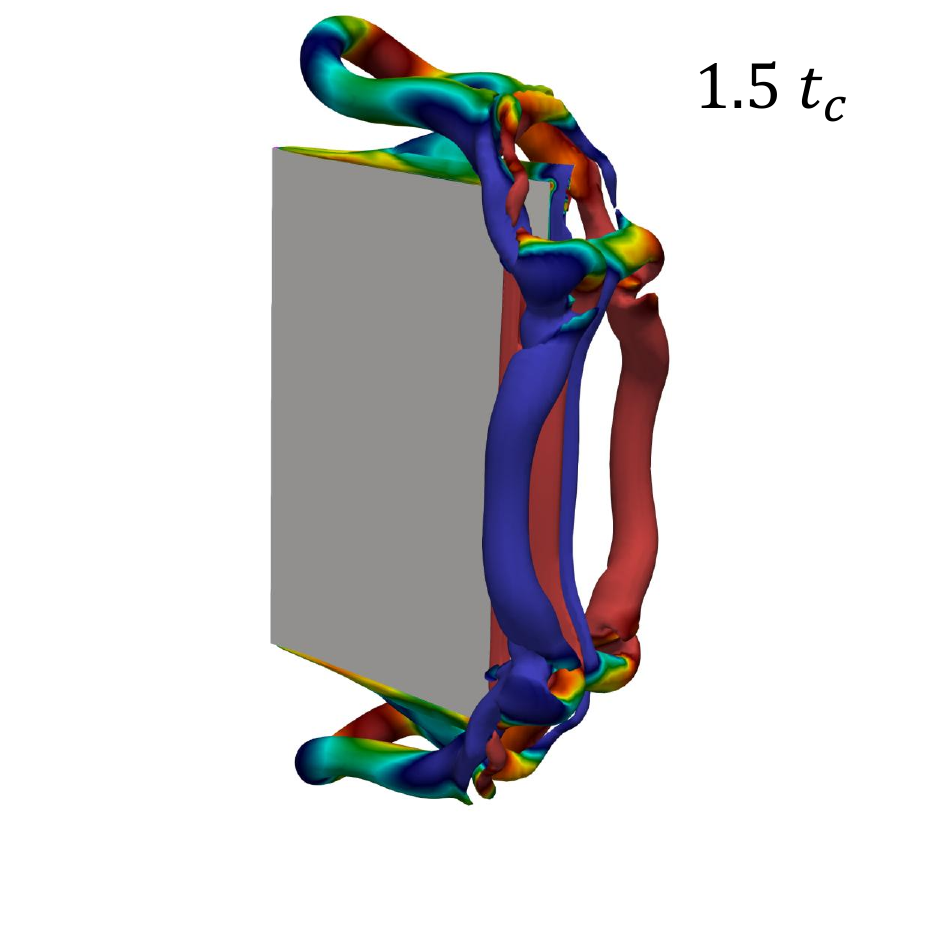}}
		\caption{}
	\end{subfigure}
	
	\begin{subfigure}[b]{0.23\textwidth}
		\framebox{\includegraphics[width=\textwidth]{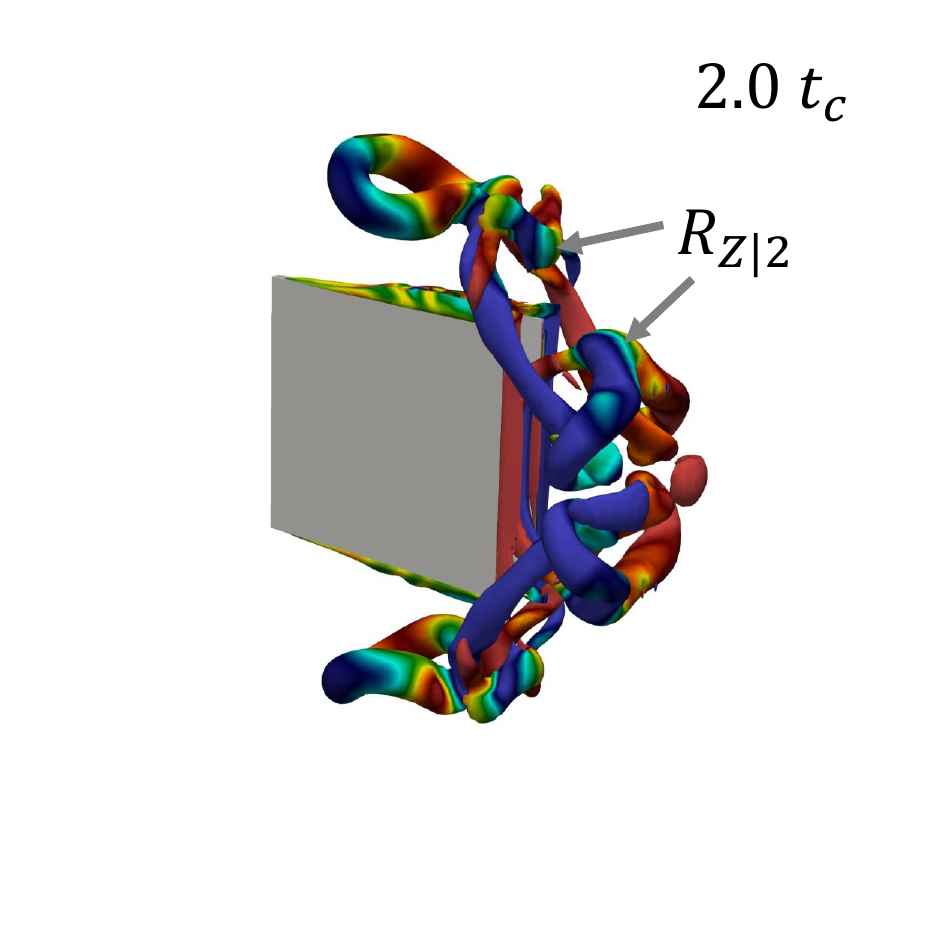}}
		\caption{}
	\end{subfigure}\hspace{20mm}
	\begin{subfigure}[b]{0.23\textwidth}
		\framebox{\includegraphics[width=\textwidth]{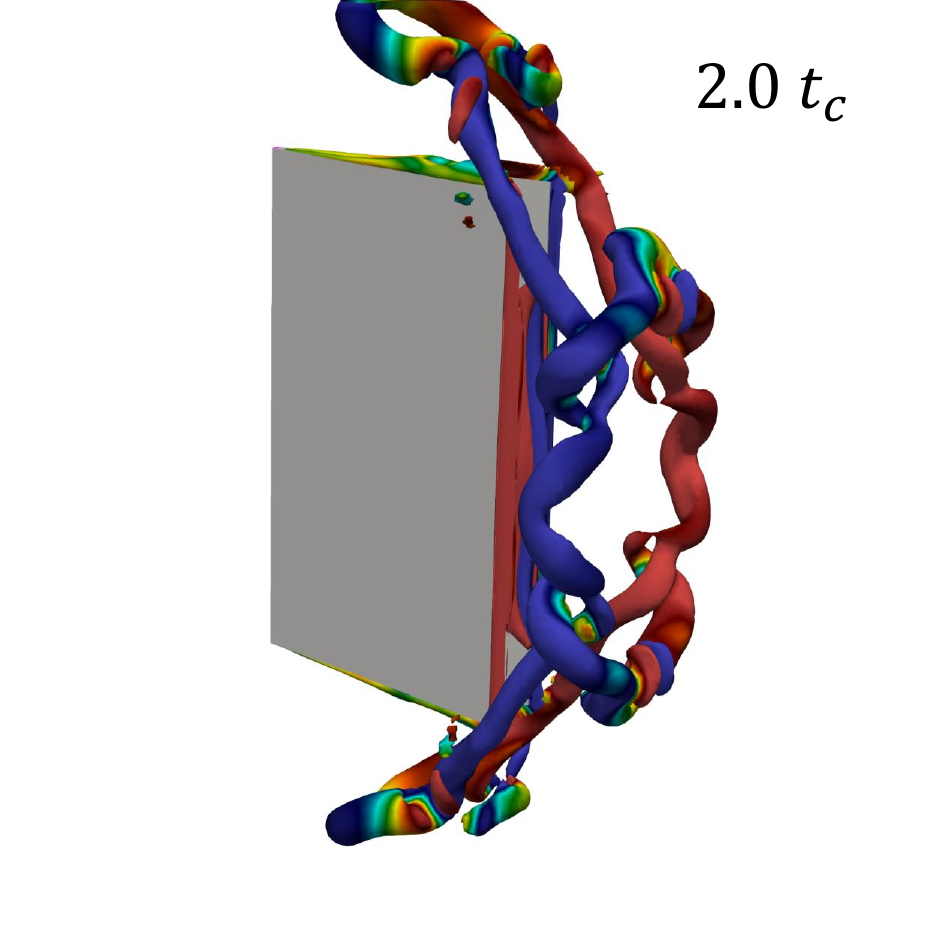}}
		\caption{}
	\end{subfigure}
	
	\begin{subfigure}[b]{0.23\textwidth}
		\framebox{\includegraphics[width=\textwidth]{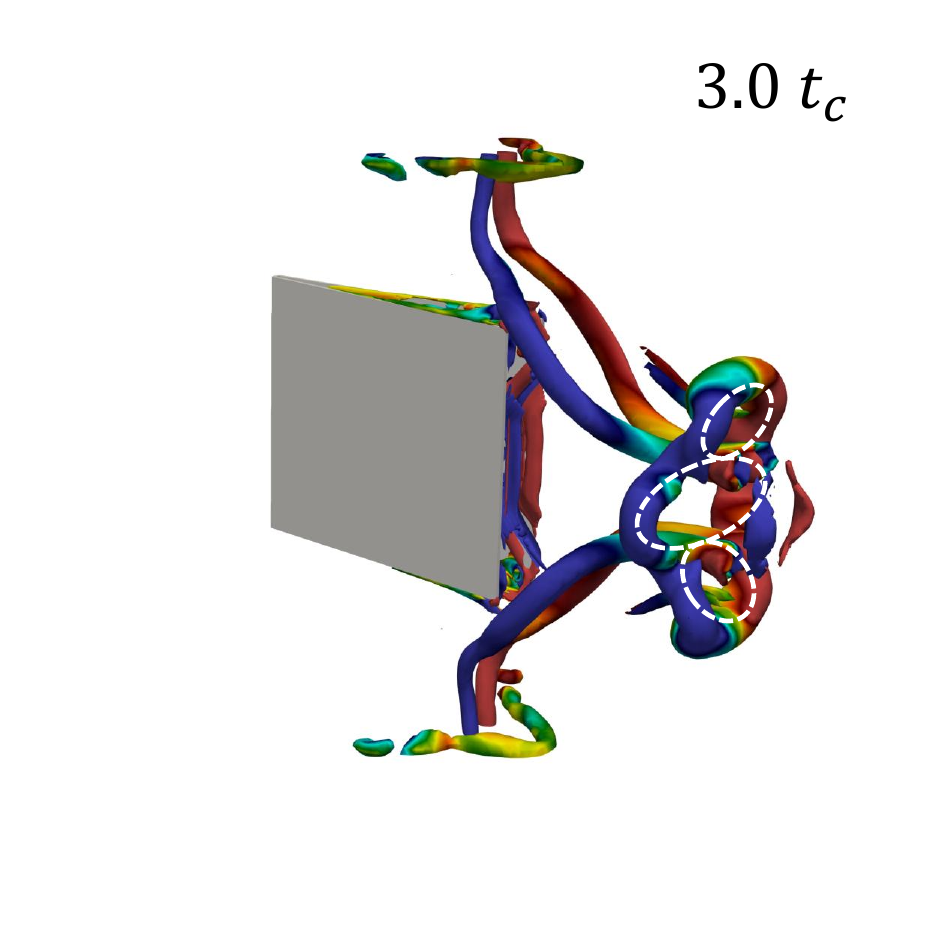}}
		\caption{}
	\end{subfigure}\hspace{20mm}
	\begin{subfigure}[b]{0.23\textwidth}				\framebox{\includegraphics[width=\textwidth]{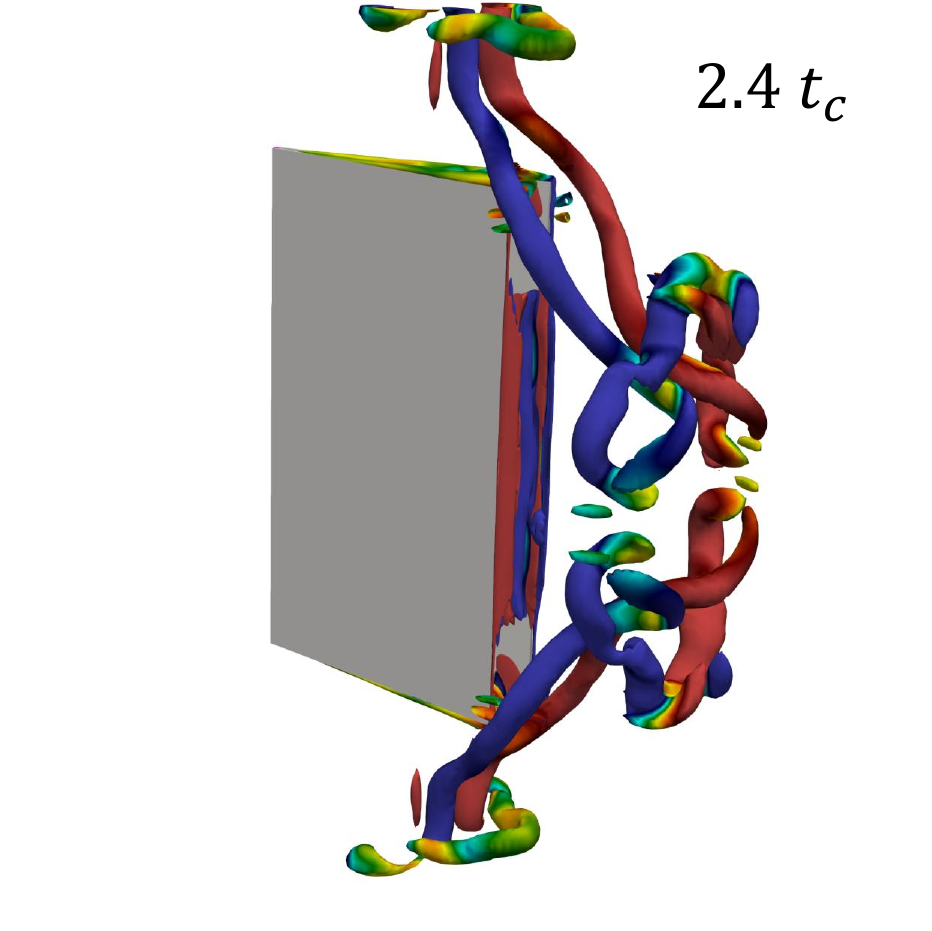}}
		\caption{}
	\end{subfigure}
	\begin{subfigure}[b]{0.60\textwidth}
		\includegraphics[width=\textwidth]{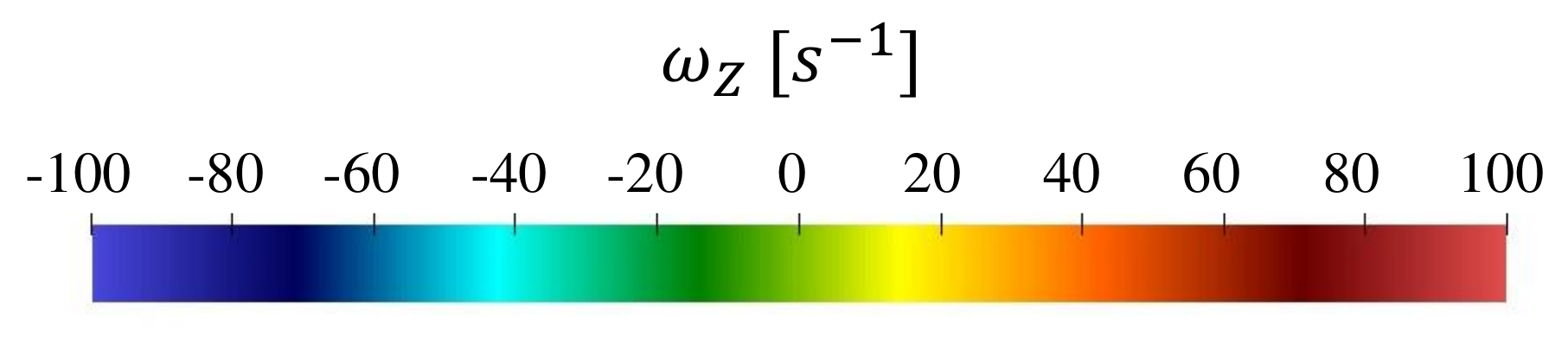}
	\end{subfigure}
	\caption{Perspective view of the vorticity iso-surface at a magnitude of 130 s$^{-1}$, colored with the Z-component of vorticity, $\omega_Z$, for the stationary case with $d^* = 0.5$ (a, c, e, and g) and $d^* = 1.0$ (b, d, f, and h). The inset figure at the bottom right of (c) shows the vortex structure spanning the depth, viewed from behind and colored with $\omega_Z$. The inset figure at the bottom left of (c) shows the top view of the vortex structure colored with streamwise vorticity, $\omega_X$, where the color legend is the same as for $\omega_Z$, shown at the bottom. In both inset figures, the white-dashed line identifies the vortex loop, and the white circle marks the primary reconnection zone, $R_{Z|1}$. The secondary reconnection zone, $R_{Z|2}$, is marked in (e).}\label{fig:C_3D_IsoVort_Stat}	
\end{figure}

\begin{figure}
	\centering\
	\begin{subfigure}[b]{0.23\textwidth}				\framebox{\includegraphics[width=\textwidth]{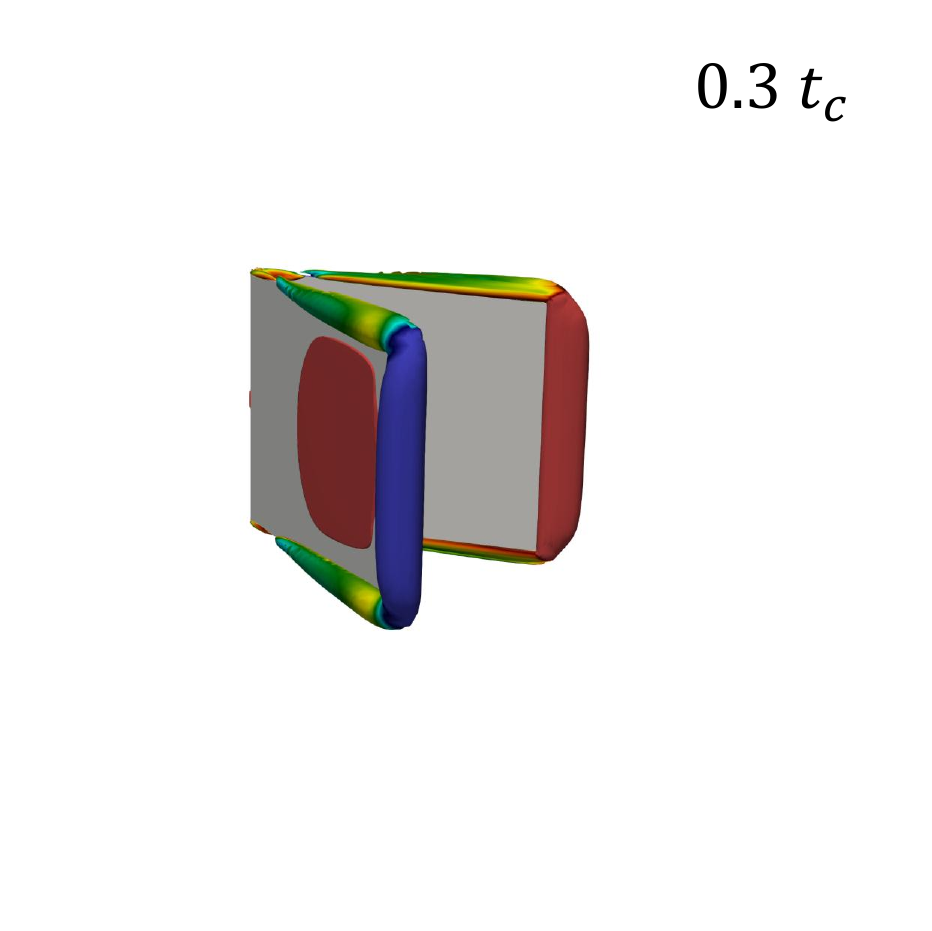}}
		\caption{}
	\end{subfigure}\hspace{20mm}
	\begin{subfigure}[b]{0.23\textwidth}
		\framebox{\includegraphics[width=\textwidth]{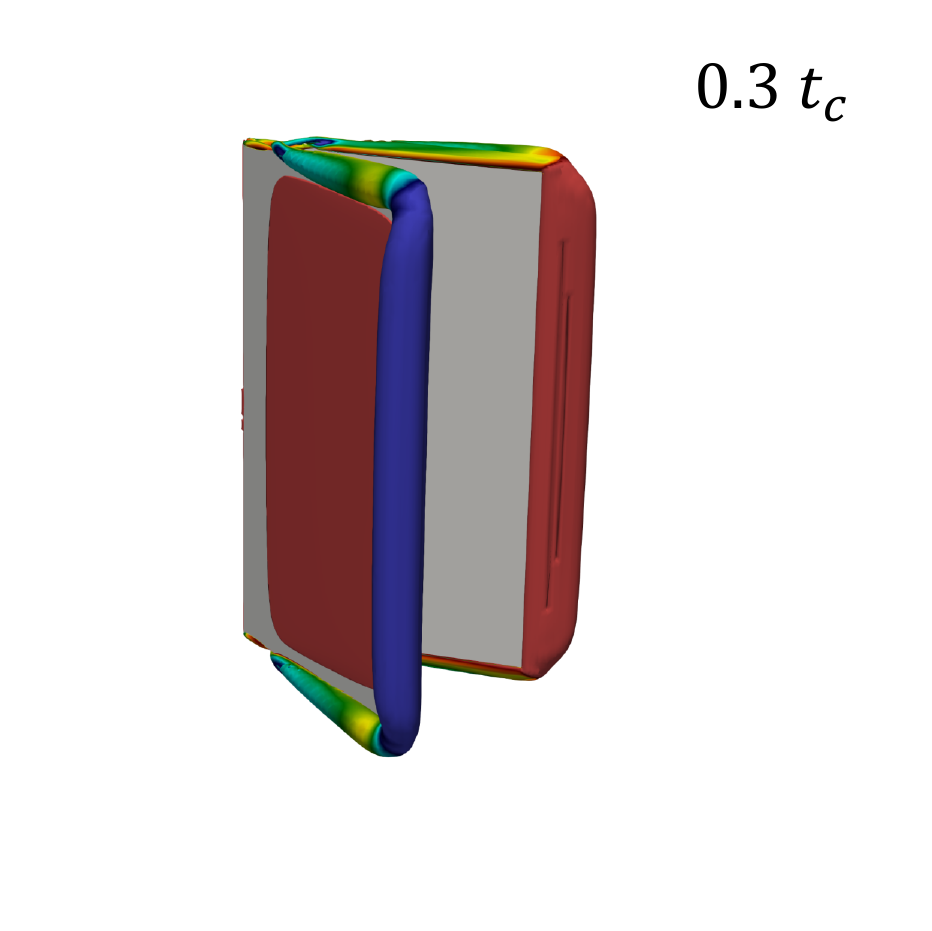}}
		\caption{}
	\end{subfigure}
	
	\begin{subfigure}[b]{0.23\textwidth}
		\framebox{\includegraphics[width=\textwidth]{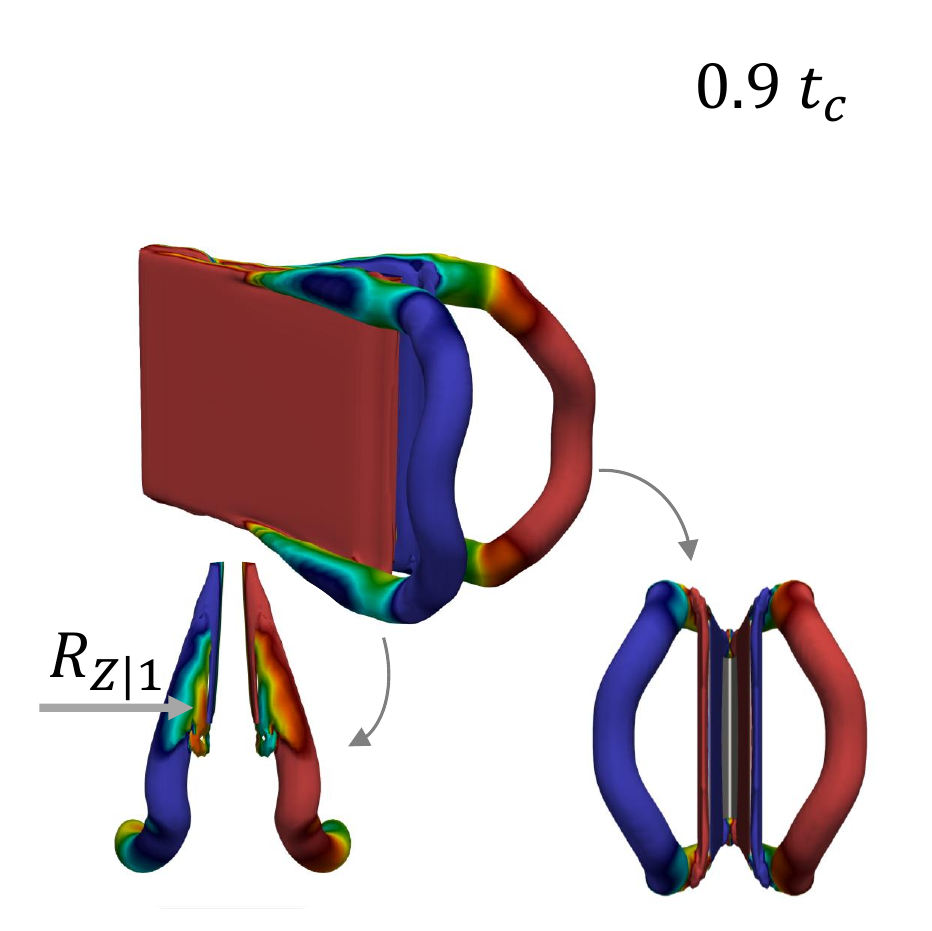}}
		\caption{}
	\end{subfigure}\hspace{20mm}
	\begin{subfigure}[b]{0.23\textwidth}
		\framebox{\includegraphics[width=\textwidth]{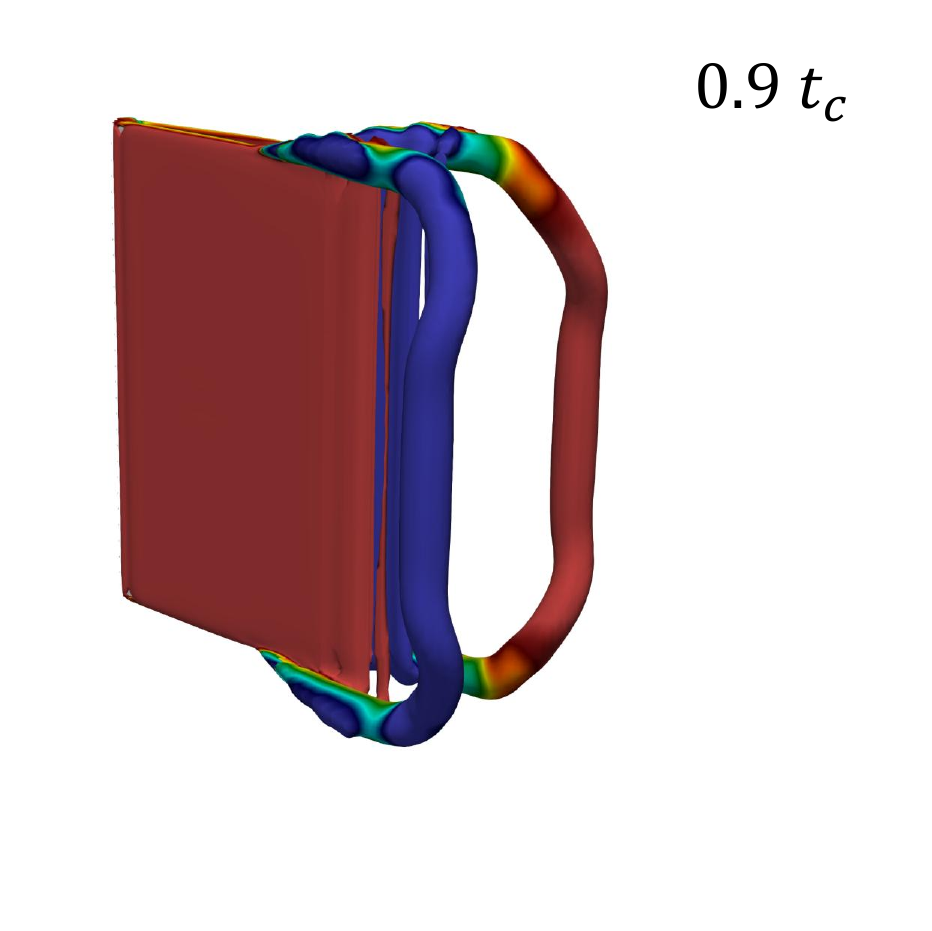}}
		\caption{}
	\end{subfigure}
	
	\begin{subfigure}[b]{0.23\textwidth}
		\framebox{\includegraphics[width=\textwidth]{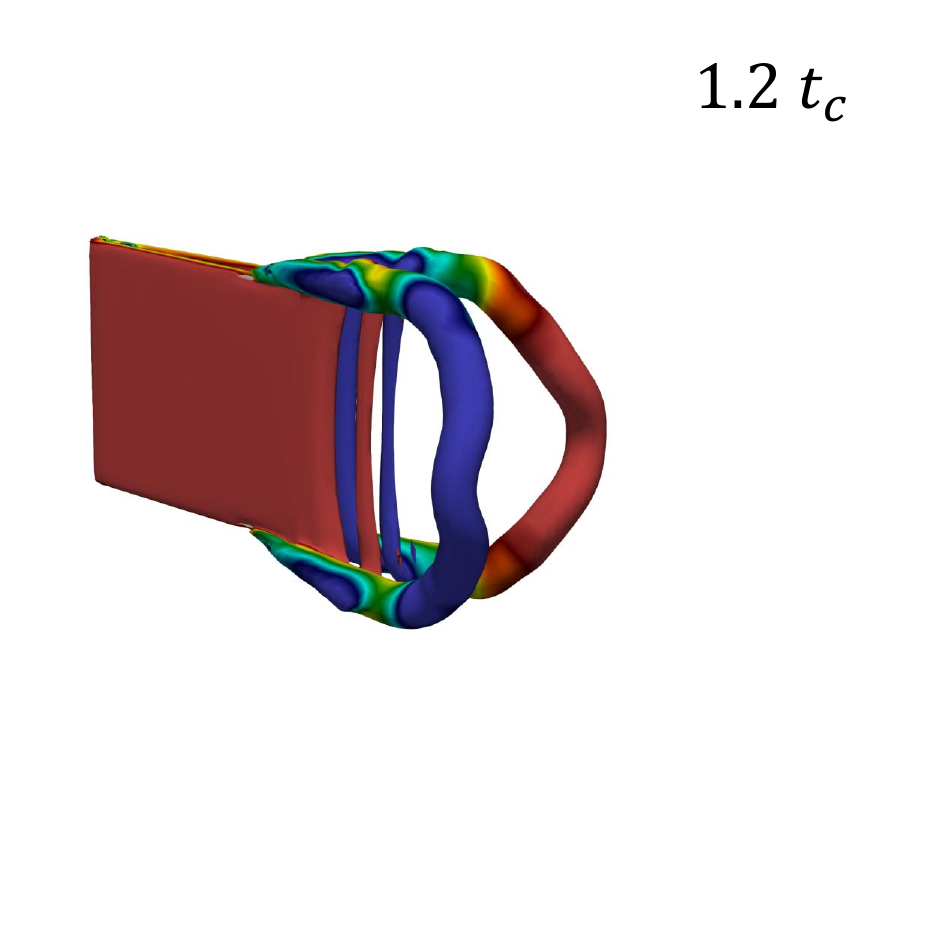}}
		\caption{}
	\end{subfigure}\hspace{20mm}
	\begin{subfigure}[b]{0.23\textwidth}
		\framebox{\includegraphics[width=\textwidth]{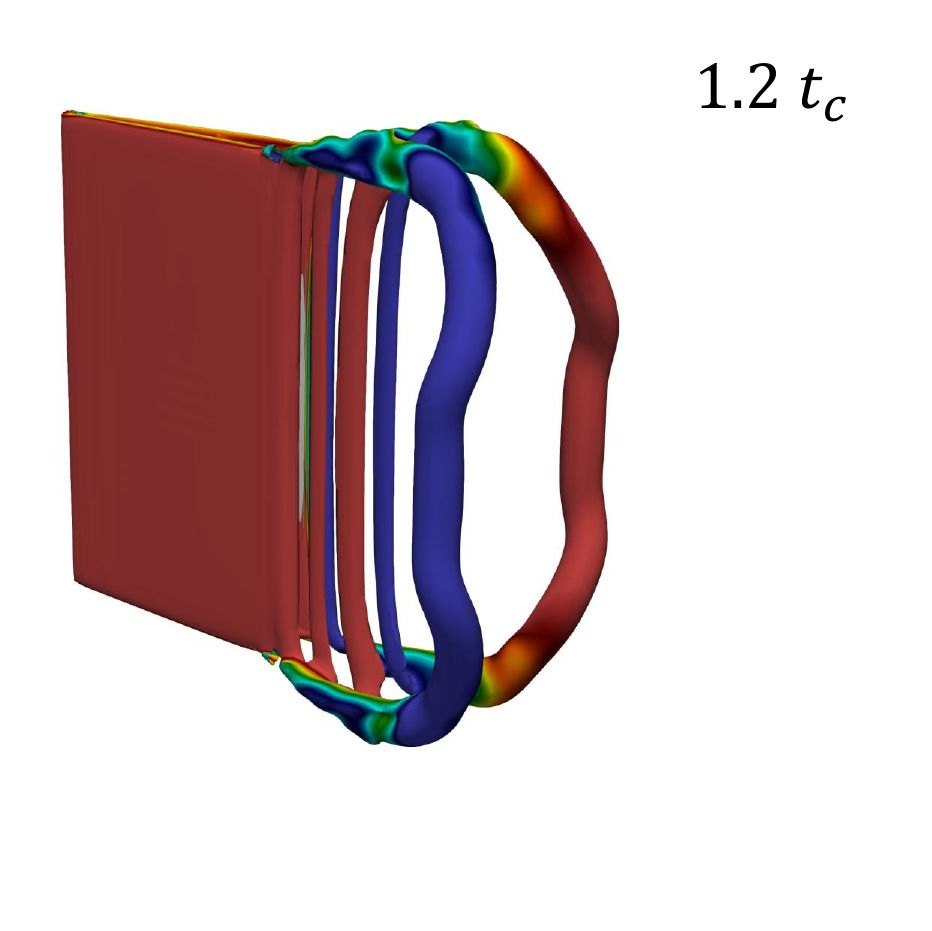}}
		\caption{}
	\end{subfigure}
	
	\begin{subfigure}[b]{0.23\textwidth}
		\framebox{\includegraphics[width=\textwidth]{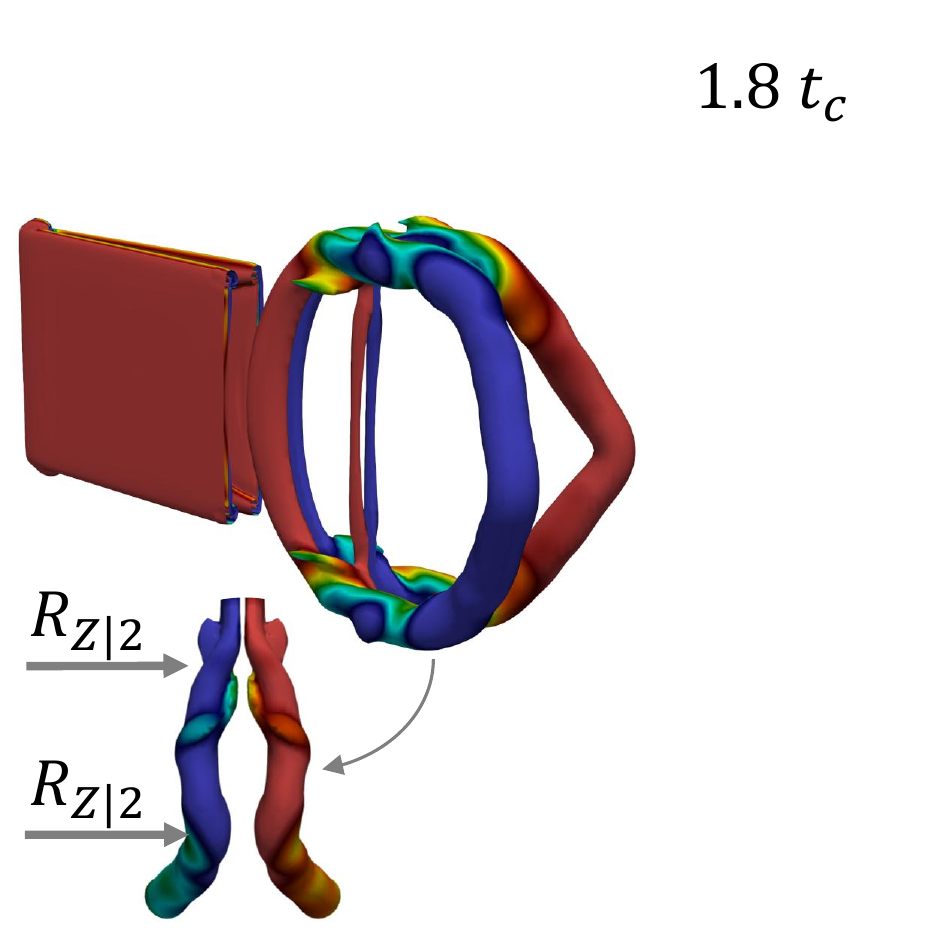}}
		\caption{}
	\end{subfigure}\hspace{20mm}
	\begin{subfigure}[b]{0.23\textwidth}
		\framebox{\includegraphics[width=\textwidth]{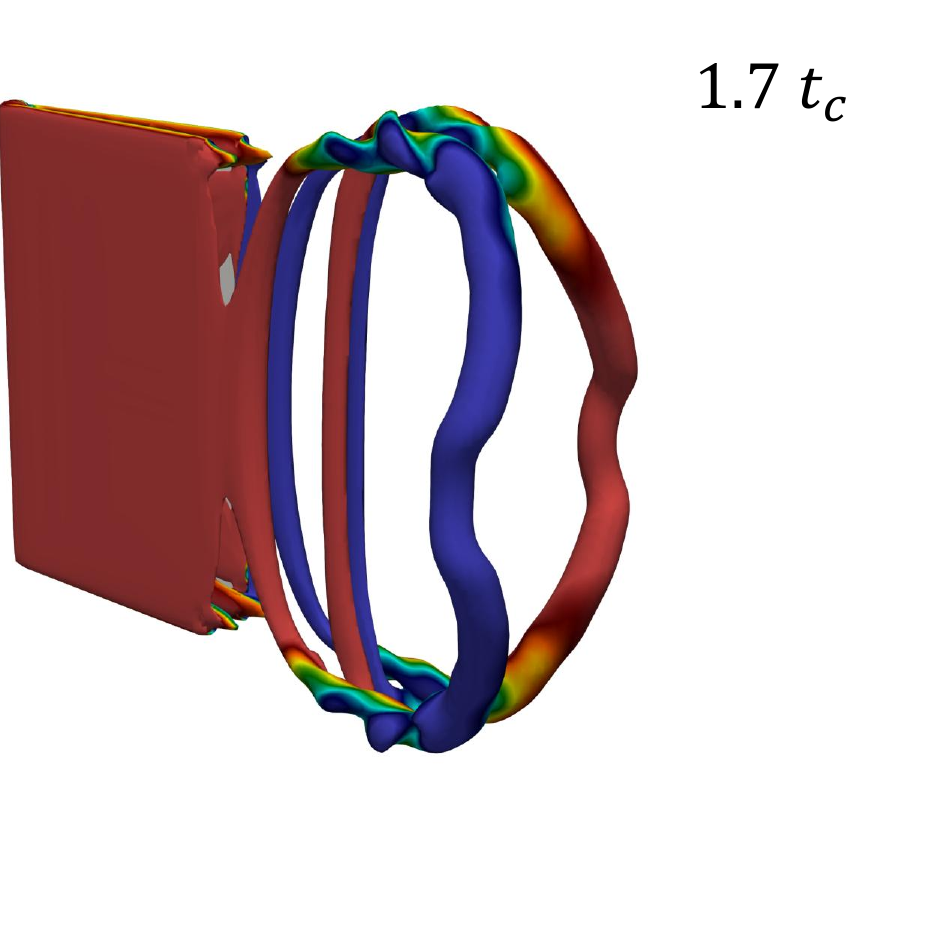}}
		\caption{}
	\end{subfigure}
	\begin{subfigure}[b]{0.60\textwidth}
		\includegraphics[width=\textwidth]{figCFD/C_IsoVort_3D}
	\end{subfigure}
	\caption{Perspective view of the vorticity iso-surface at a magnitude of 130 s$^{-1}$, colored with the Z-component of vorticity, $\omega_Z$, for the dynamic case with $d^* = 0.5$ (a, c, e, and g) and $d^* = 1.0$ (b, d, f, and h). In (c) and (g), the inset figure at the bottom left shows the top view of the vortex structure colored with streamwise vorticity, $\omega_X$, with the same color legend as for $\omega_Z$. The inset figure at the bottom right of (c) shows the wake vortices viewed from behind and colored with $\omega_Z$. The primary reconnection zone, $R_{Z|1}$, and the probable sites for secondary reconnection, $R_{Z|2}$, are identified in (c) and (g), respectively.}\label{fig:C_3D_IsoVort_Dyn}	
\end{figure}	

\begin{figure}
	\centering\
	\begin{subfigure}[b]{0.30\textwidth}
		\includegraphics[width=\textwidth]{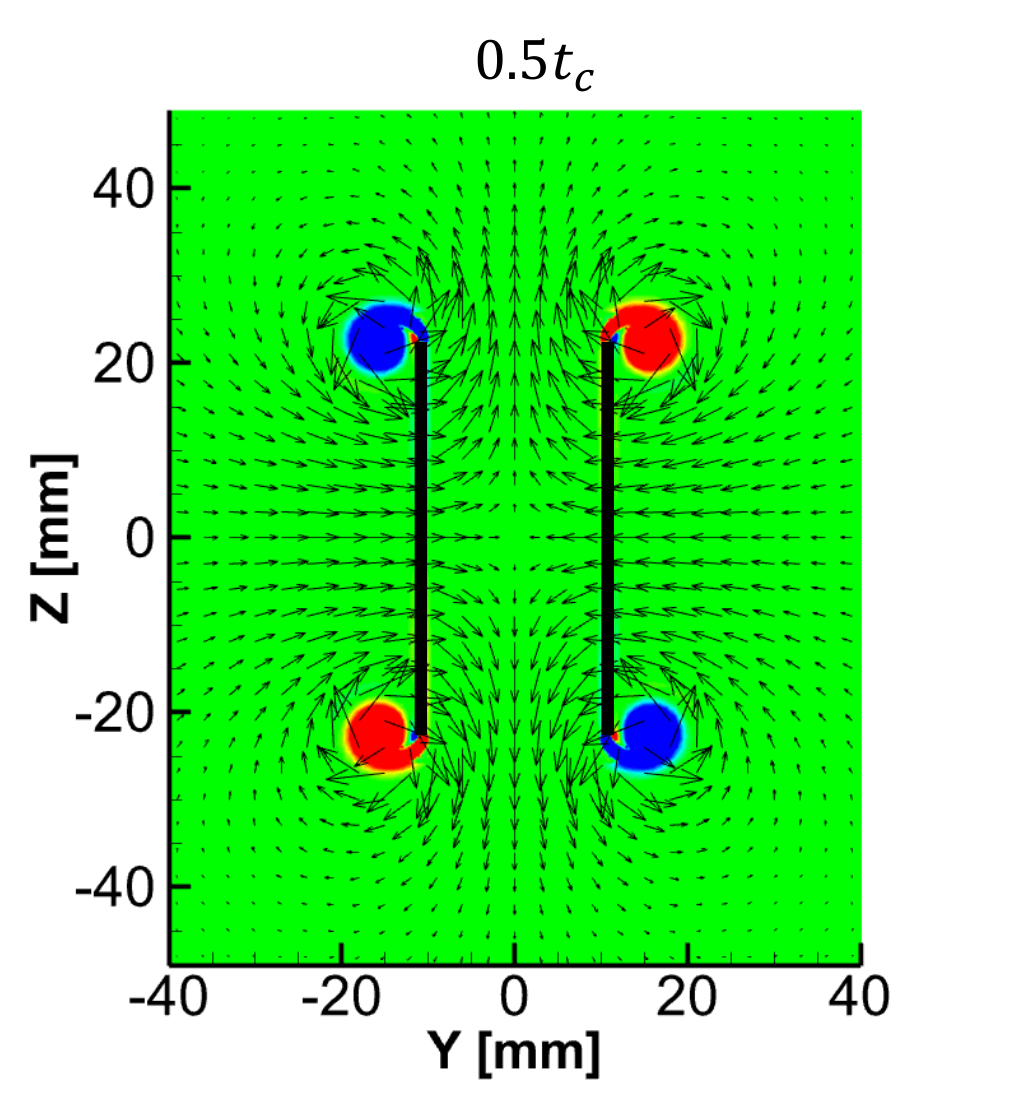}
		\caption{}
	\end{subfigure}
	\begin{subfigure}[b]{0.30\textwidth}
		\includegraphics[width=\textwidth]{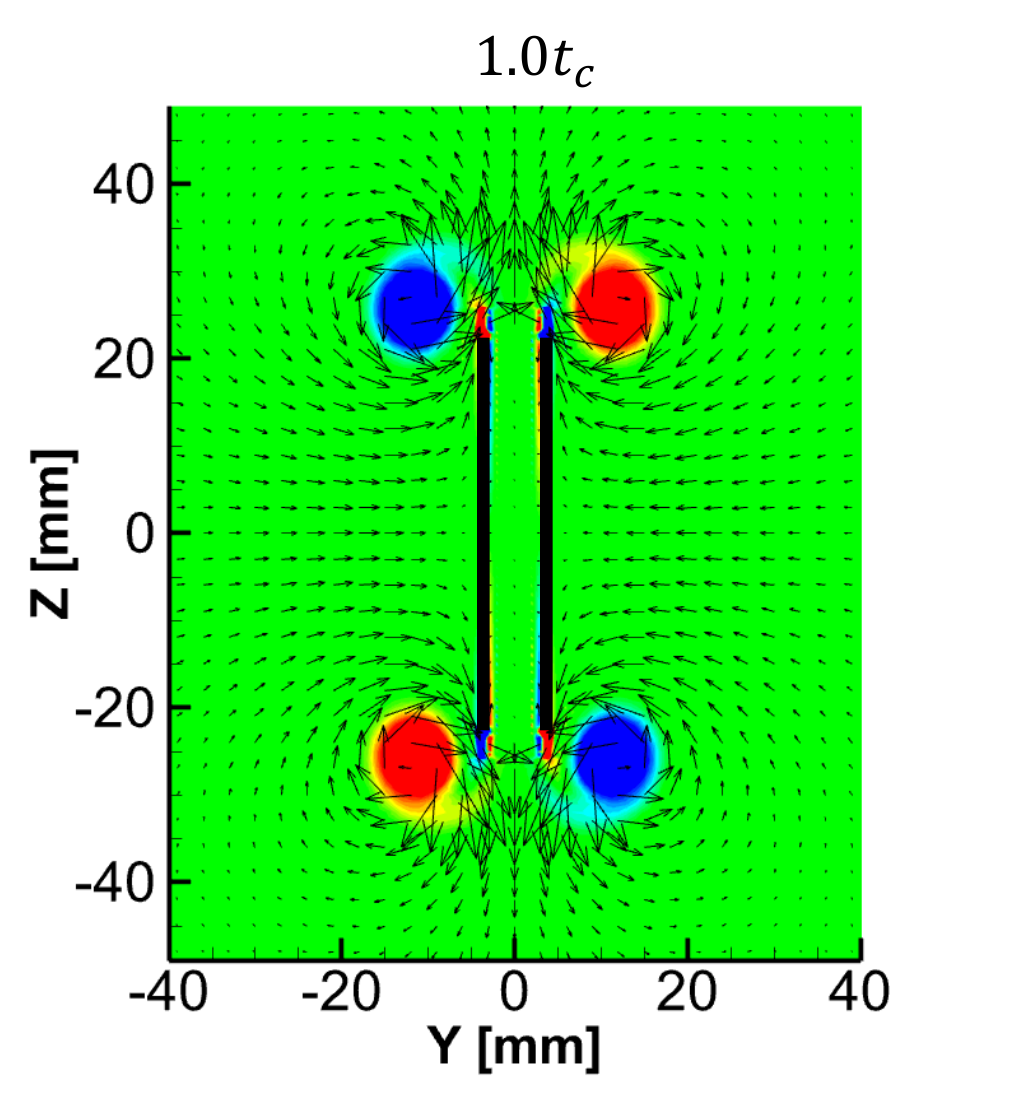}
		\caption{}
	\end{subfigure}
	\begin{subfigure}[b]{0.30\textwidth}
		\includegraphics[width=\textwidth]{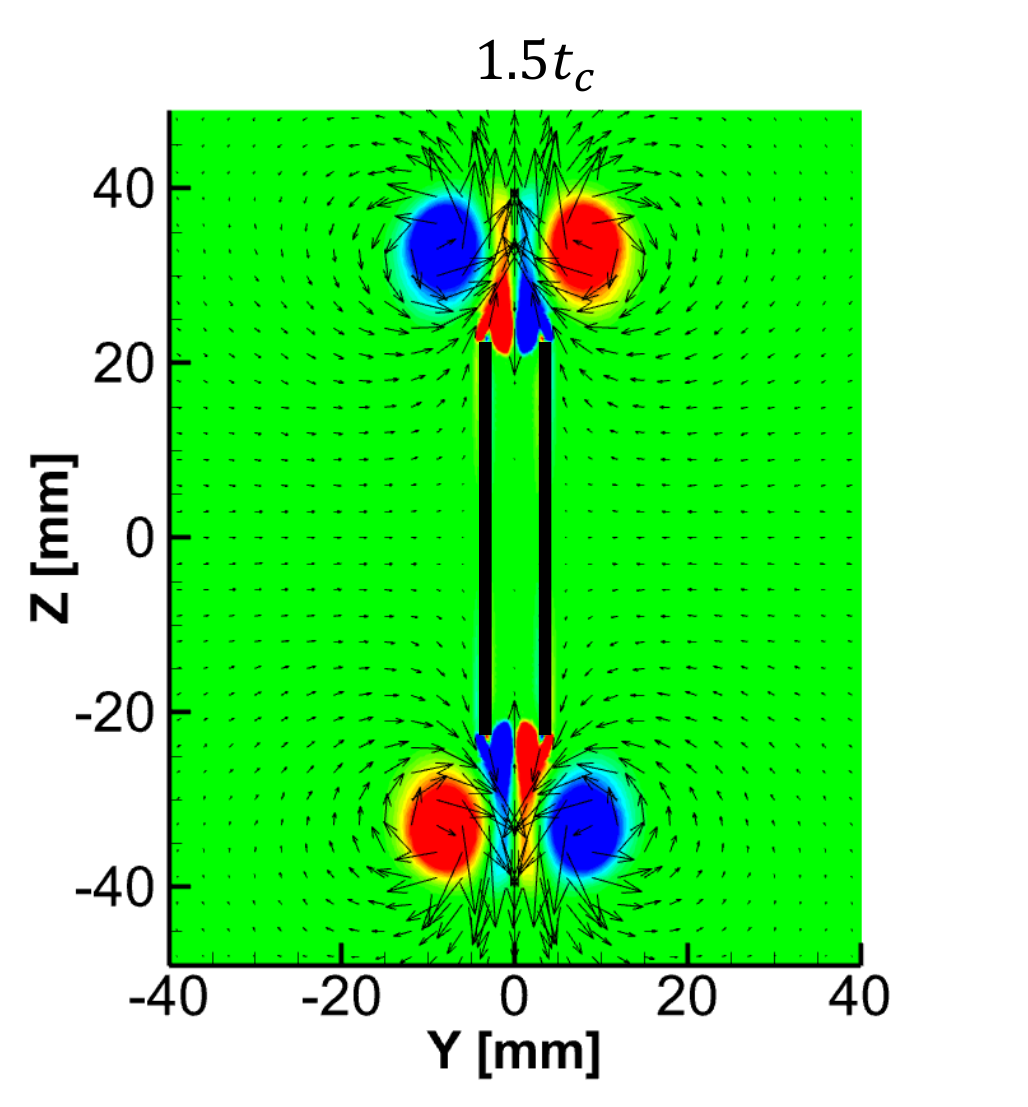}
		\caption{}
	\end{subfigure}
	\begin{subfigure}[b]{0.30\textwidth}
		\includegraphics[width=\textwidth]{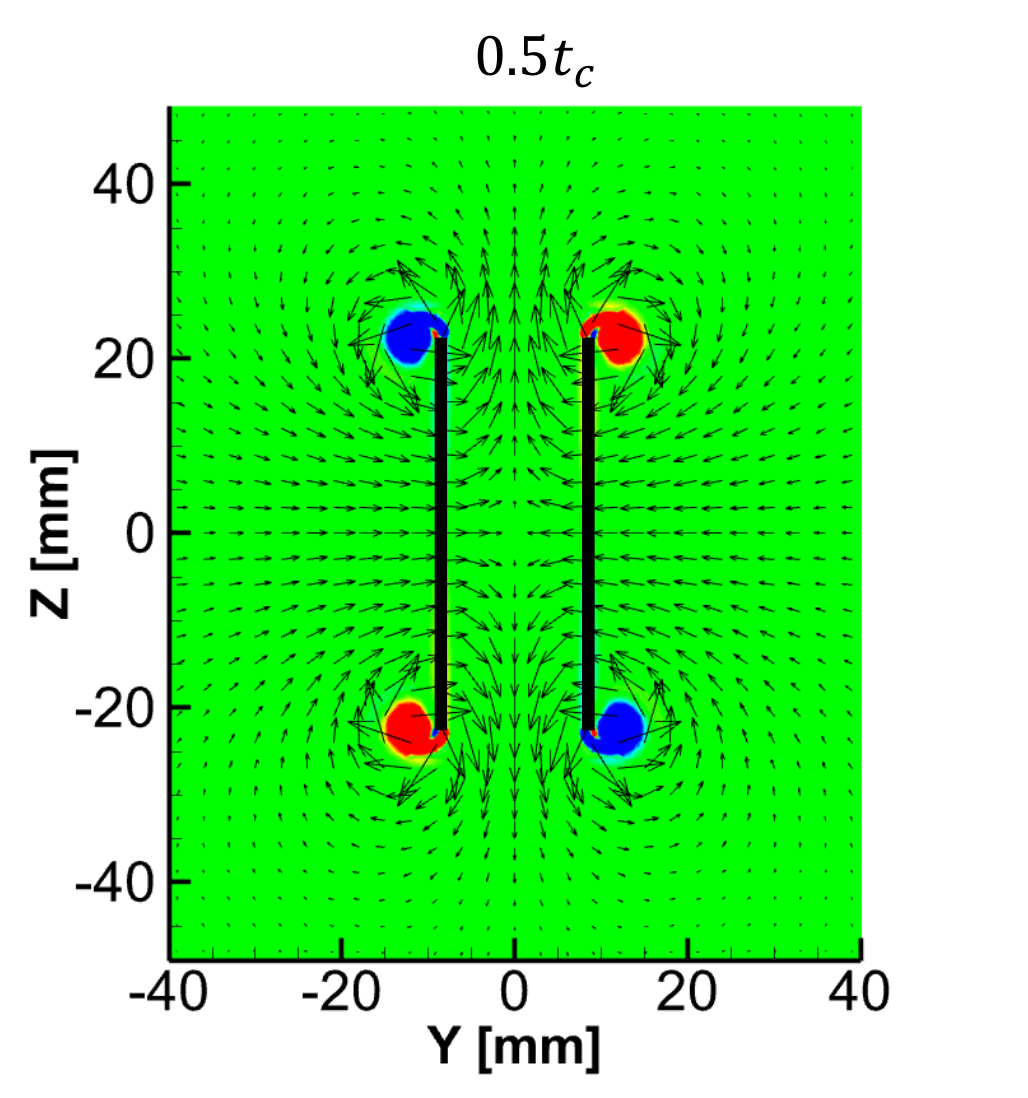}
		\caption{}
	\end{subfigure}
	\begin{subfigure}[b]{0.30\textwidth}
		\includegraphics[width=\textwidth]{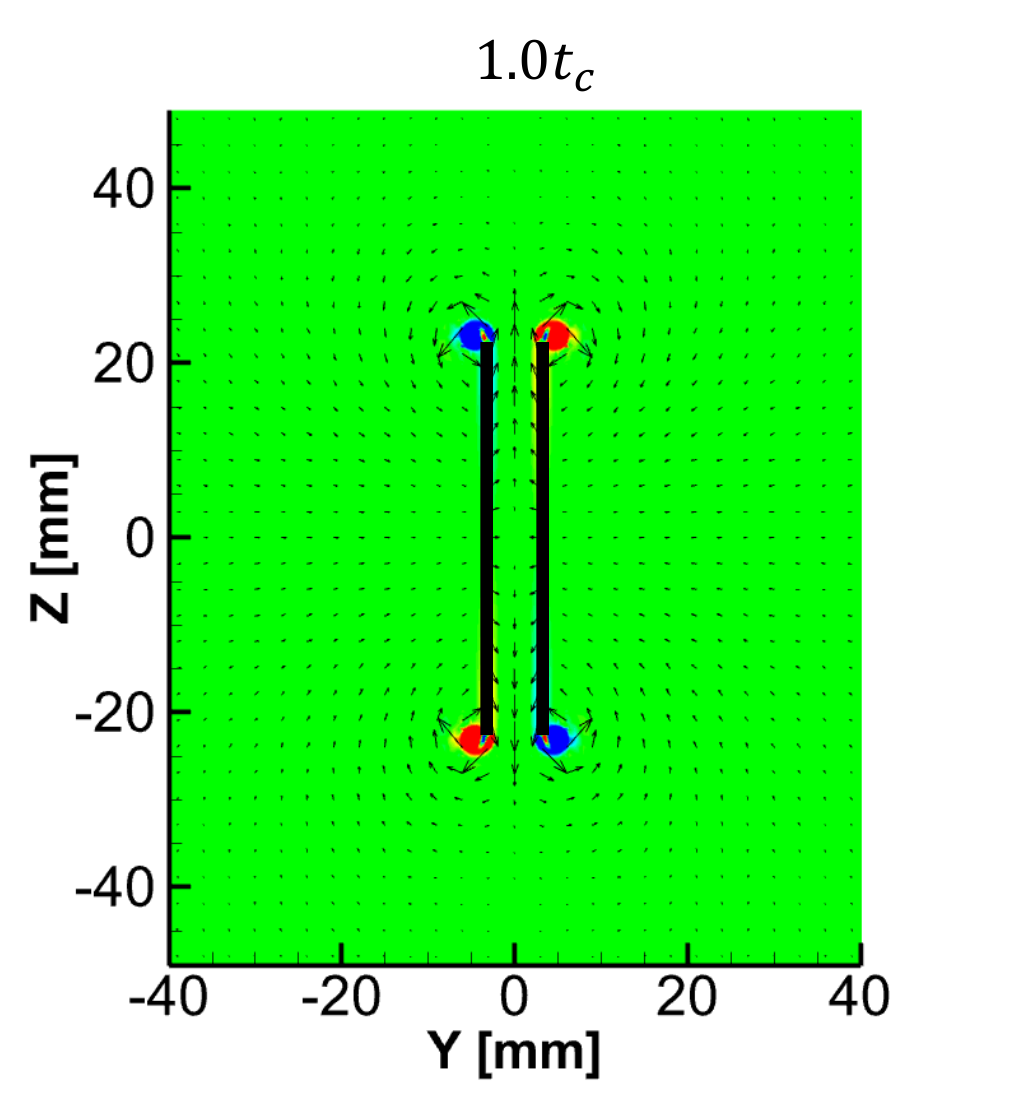}
		\caption{}
	\end{subfigure}
	\begin{subfigure}[b]{0.30\textwidth}
		\includegraphics[width=\textwidth]{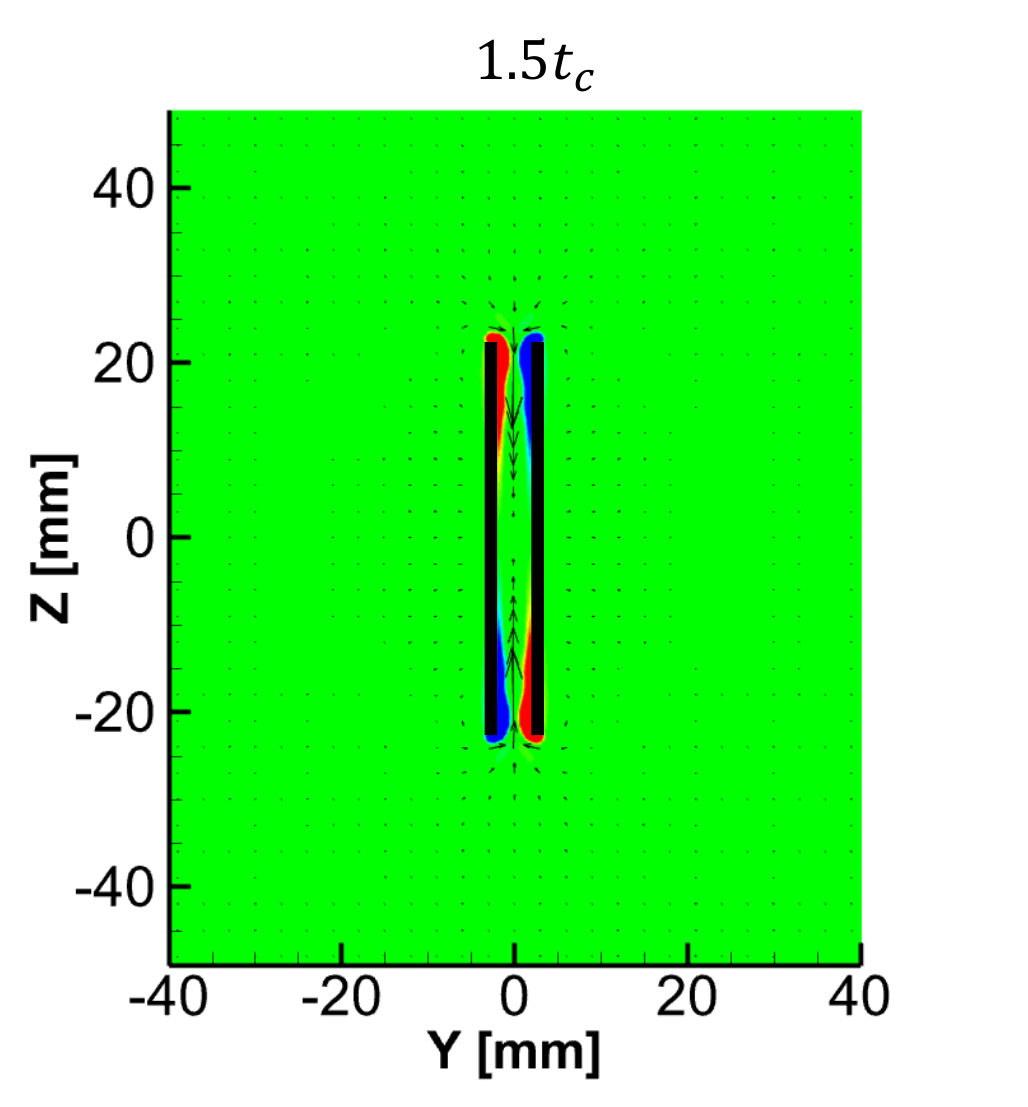}
		\caption{}
	\end{subfigure}
	
	\begin{subfigure}[b]{0.40\textwidth}
		\includegraphics[width=\textwidth]{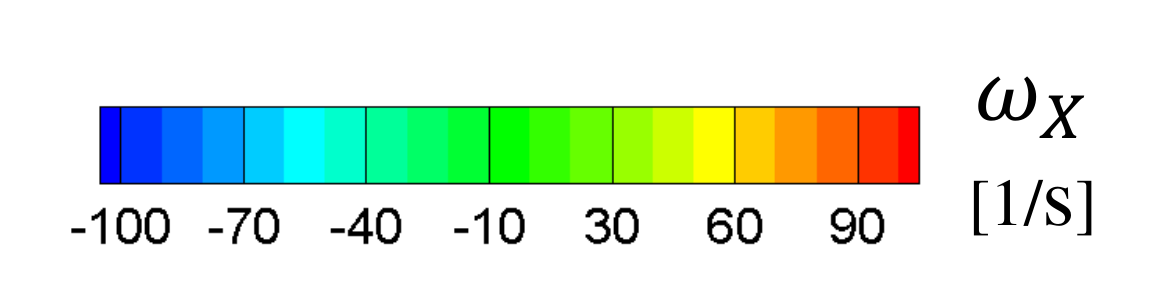}
	\end{subfigure}
	\caption{The streamwise vorticity, $\omega_X$, field and velocity vectors are shown on the YZ plane, at one-third of the plate length from the trailing edge. Black lines show the position of the plates. (a), (b), and (c) show the flow field for the stationary case of $d^* = 0.5$, at $t = 0.5 t_c$, $t = 1.0 t_c$, and $t = 1.5 t_c$, respectively, where the clapping period, $t_c = 80$ ms. (d), (e), and (f) show the flow field for the dynamic case of $d^* = 0.5$, at $t = 0.5 t_c$, $t = 1.0 t_c$, and $t = 1.5 t_c$, respectively, where $t_c = 68$ ms. The same normalized times are used to compare the stationary and dynamic cases at equivalent stages of the clapping cycle.}\label{fig:C_VortStreamWise_Stat_Dyn}	
\end{figure}

	Figure \ref{fig:C_3D_IsoVort_Stat} shows the evolution of wake vortices for the stationary case for $d^* = 0.5$ and $d^* = 1.0$. The figure presents a perspective view of iso-surfaces for vorticity magnitude of 130 $\text{s}^{-1}$ at selected time steps, chosen to highlight key stages where the vortices undergo distinct transformations. Additionally, Movies 1 and 2, provided as supplementary material, show the full temporal evolution of vortices in the stationary case for $d^* = 0.5$ and $d^* = 1.0$, respectively. In the stationary cases, the wake vortices evolve through four main phases: one, the development of the starting vortex tube; two, the formation of a weaker stopping vortex tube; three, the emergence of vortex loops due to primary reconnection between the starting and stopping vortex tubes at localized regions; and four, the bridging of vortex loops via secondary reconnection. For $d^* = 0.5$, a detailed discussion on these phases is as follows:

	\begin{enumerate}
	\item {{\it Starting vortex tube formation phase}: This phase begins with the onset of plate rotation and lasts until approximately 78 ms ($t/t_c \approx 1.0$), just before the end of rotation at $t_c =$ 80 ms. During this time, the rotation of the plates generates vorticity, which rolls up to form a starting vortex tube that envelops the top, bottom, and trailing edges of both plates (see figure~\ref{fig:C_3D_IsoVort_Stat}(a)). The vortex tube is colored by the $Z$-component of vorticity, $\omega_Z$, which clearly reveals opposite signs along the trailing edges of the two plates (red and blue). In contrast, the top and bottom regions of the tube appear green, indicating negligible $\omega_Z$, as the vorticity there is predominantly streamwise. The velocity fields associated with the vortex tubes are shown in the XY plane (figure~\ref{fig:C_VortZ_AR200_Dyn_Stat}(a--d)) and in the YZ plane (figure~\ref{fig:C_VortStreamWise_Stat_Dyn}(a--c)), at a distance one-third of the plate radius from the trailing edge. In the YZ plane, fluid jets ejected from the top and bottom sides of the interplate cavity are clearly seen in figure~\ref{fig:C_VortStreamWise_Stat_Dyn}(a), consistent with the sideways flow discussed in the previous section. The streamwise vorticity, $\omega_X$, rolls up near the top and bottom corners of the trailing edges, forming two nearly circular vortex pairs (red and blue). These pairs represent cross-sections of the starting vortex tubes along the top and bottom edges of the plates.\par
	
	Next, we examine the vortex-tube curvature, which governs its self-induced velocity. At the top and bottom corners of the leading edges (hinge region), this curvature produces a self-induced velocity, and these portions of the tube move laterally to form horseshoe-like ``U''-shaped vortices (see figure~\ref{fig:C_3D_IsoVort_Stat}(a)). The vortex tubes in the mid-depth region of the trailing edges also develop curvature, where they lag behind the rotating plates (see figures~\ref{fig:C_VortZ_AR200_Dyn_Stat}(c,d)). The resulting self-induced velocity draws the curved segments of both loops toward each other (e.g. see the vortex structure at 60 ms in Movie 1). Despite sharp curvature, the portions of the vortex tubes near the top and bottom corners of the trailing edges remain attached. These corner segments play a critical role in initiating reconnection, which is discussed below.}

	\item {{\it Stopping vortex tube formation phase}: This phase lasts for a short duration (approximately 78--95 ms; $t/t_c \approx 1.0$--$1.2$), during which a thin stopping vortex tube, associated with the deceleration of the rotating plate, forms along the three edges of each plate that face the interplate cavity. The stopping vortex formation can be observed in the streamwise vorticity field plotted on the YZ plane (figure~\ref{fig:C_VortStreamWise_Stat_Dyn}(b)), where very small vorticity patches, oriented opposite to the starting vortices, appear. Later, the starting vortices move away from the plate, while the stopping vortices slightly grow in size but remain confined in the interplate cavity (figure~\ref{fig:C_VortStreamWise_Stat_Dyn}(c)). At the end of this phase, the thicker starting vortex tube coexists with the thin stopping vortex, as shown in Movie 1.}
	
	\item {{\it Primary reconnection phase}: In this phase (approximately 95--140 ms; $t/t_c \approx 1.2$--$1.8$), the portion of the starting vortex tube attached to the top and bottom corners of the trailing edge comes in very close contact with the stopping vortex tube, initiating local vorticity cancellation. This leads to a significant reduction in the size of the starting vortex tube (see figure \ref{fig:C_3D_IsoVort_Stat}(c)). Following this interaction, away from the corner regions, the starting and stopping vortex tubes reconnect to form vortex loops on the back, top, and bottom sides of the body. The vortex reconnection is discussed in detail by Melander {\it et al.}\cite{Melander89}, who numerically studied the cross-linking of two antiparallel vortex tubes. They observed that when the tubes come into close proximity, a bridging region forms in the contact zone, where vorticity reorients perpendicular to the original tube axes. While vorticity is locally annihilated in the contact region, the overall circulation in the bridging portion remains conserved. In the present case, after reconnection, the wake evolves differently on the trailing-edge side of the body compared to the sideways wake on the top and bottom sides. On the trailing-edge side, the curved starting vortex tube reconnects with the vertical segment of the stopping vortex tube, forming a triangular vortex loop, as indicated by the dashed white line in the bottom-right inset of figure \ref{fig:C_3D_IsoVort_Stat}(c). For the two plates, the figure shows two triangular loops. In this figure, for clarity, the primary reconnection zone, $R_{Z|1}$, is marked with a white-filled circle only for one plate. At $R_{Z|1}$, a green iso-surface ($\omega_Z \approx 0$) indicates the bridging region formed during reconnection. In this portion, the vorticity is reoriented in the perpendicular ($Y$) direction from its original ($Z$) alignment in both the starting and stopping vortex tubes. Following Melander {\it et al.}\cite{Melander89}, we interpret that, despite this reorientation, the vorticity strength appears to be largely conserved.\par 
	
	In the sideways wake, the ``U''-shaped starting vortex tube interacts with a sideways-oriented portion of the stopping vortex, leading to reconnection and the formation of a horseshoe vortex resembling an $\Omega$ shape. To visualize this transformation, we include a top view of the vortex structure colored by $\omega_X$, see bottom-left inset of figure \ref{fig:C_3D_IsoVort_Stat}(c). In this figure, the dashed white line outlines the $\Omega$ shape, and the white dot marks $R_{Z|1}$, present on both sides of the loop but shown only on one side for clarity. Toward the end of this phase, despite the differing evolutions, both the $\Omega$ and triangular loops remain connected through a stretched filament of the starting vortex, as shown in figure \ref{fig:C_3D_IsoVort_Stat}(c). This stretching occurs at the top and bottom corners of the trailing edge, simultaneous with the formation of both loops.} 
	
	\item{{\it Secondary reconnection phase}: This is the final phase of wake vortex evolution, during which the vortex loops interconnect. It begins at approximately 140 ms ($t/t_c \approx 1.8$) and continues until the end. At the beginning of this phase, on the trailing-edge side of the wake, the two triangular vortex loops move closer and eventually connect at the corner locations of their bases, forming what we refer to as the secondary reconnection zone, $R_{Z|2}$, as shown for the top corner in figure \ref{fig:C_3D_IsoVort_Stat}(e). A similar reconnection zone also forms at the bottom corner, not shown. Once the secondary reconnection forms, the portions of the thin stopping vortex filaments initially located at the bases of the triangular loops come into close contact and undergo annihilation. Such merging of two triangular vortex loops into a single loop was previously observed in the stationary clapping plates study by Kim {\it et al.}\cite{Kim13}, where they used three-dimensional PIV. Eventually, the highly curved portions at the apex of both triangular loops rapidly approach each other due to their higher self-induced velocity. This transformation ultimately develops a vortex structure resembling three interconnected ringlets, identified by white dashed lines superimposed on the vortex structure in figure \ref{fig:C_3D_IsoVort_Stat}(g). A similar three-ringlet structure was observed around 200 ms in our previous experimental study on the clapping body, where a 3D representation of the imagined vortex structure was provided (see figure 15(b) in Mahulkar and Arakeri\cite{Mahulkar24}). In contrast to the complex transformation in the wake on the trailing-edge side, the sideways wake shows a simpler transformation. At the beginning of this phase, the $\Omega$-shaped horseshoe vortex filament in the sideways wake reconnects in the region $R_{Z|2}$ near the trailing edge, where the filaments were initially in close contact (see figure \ref{fig:C_3D_IsoVort_Stat}(e)). After this reconnection, one elliptical ringlet emerges from the initial horseshoe vortex in each of the top and bottom side wakes. Subsequently, these ringlets enter the coarse background mesh, where the resolution is insufficient to capture their further evolution. As a result, the ringlets appear as disconnected vorticity patches (see figure \ref{fig:C_3D_IsoVort_Stat}(g)). These ringlets convect away from the body due to their self-induced velocity, while the sideways ringlets remain connected to the trailing-edge-side ringlets through highly stretched starting vortex filaments.}
	\end{enumerate}\par

	The wake vortex evolution for the stationary case with $d^* = 1.0$, as shown in figures \ref{fig:C_3D_IsoVort_Stat}(b, d, f, and h) and Movie 2, follows the same four phases as for $d^* = 0.5$, but with a slight delay due to the later closure of the clapping motion ($t_c$ = 104 ms). The first phase of starting vortex tube formation spans $\sim$0--100 ms ($t/t_c \approx 0$--$1.0$), the second phase of stopping vortex tube formation spans $\sim$100--120 ms ($t/t_c \approx 1.0$--$1.2$), the third phase of primary reconnection lasts $\sim$120--180 ms ($t/t_c \approx 1.2$--$1.7$), and the final phase of secondary reconnection continues from $\sim$180 ms ($t/t_c \approx 1.7$) until the end of the computation. The starting and stopping vortex tubes reconnect around the top and bottom corners of the trailing edges (see figures \ref{fig:C_3D_IsoVort_Stat}(d, f)). As in the $d^* = 0.5$ case, the primary reconnection in the $d^* = 1.0$ case leads to the formation of an $\Omega$-shaped vortex loop in the sideways wake. This loop subsequently transforms into an elliptical ringlet through secondary reconnection. In contrast, the vortex evolution on the trailing-edge side for the $d^*=1.0$ case follows a slightly different path. Here, two semi-elliptical vortex loops form, with their major axes aligned along the body depth (see figure \ref{fig:C_3D_IsoVort_Stat}(d)). The main reason for the semi-elliptic shape, as opposed to the triangular-shaped loop for $d^* = 0.5$, is the higher body height, which prevents sharp curvature in the mid-depth region of the vortex loop. Subsequently, both the semi-elliptic loops undergo secondary reconnection at the top and bottom corners of their bases (see figures \ref{fig:C_3D_IsoVort_Stat}(f, h)). We expect that both semi-elliptical rings will eventually merge into a single elliptical ring, though this transformation is not captured within the given computation time. Experiments showed the formation of a single elliptical ring in the trailing-edge-side wake at around 400 ms. Its cross-section on the XZ plane (Y = 0) is shown in figure 13(e) of Mahulkar and Arakeri \cite{Mahulkar24}.\par

\subsubsection{Dynamic cases}
	When the body is allowed to move freely, the vortices in the wake evolve in a remarkably different manner compared to the stationary cases. See figures \ref{fig:C_3D_IsoVort_Dyn}(a, c, e, and g) and Movie 3 for $d^* = 0.5$, and figures \ref{fig:C_3D_IsoVort_Dyn}(b, d, f, and h) and Movie 4 for $d^* = 1.0$. The figures show vorticity structures using the iso-surface of vorticity magnitude 130 $\mathrm{s}^{-1}$ at selected time steps, chosen to highlight distinctive shape transformations. In contrast to the four-phase evolution observed in the stationary cases, the wake evolution in the dynamic cases follows three phases: first, the formation of the starting vortex tube; second, the shedding of bound vorticity, which occurs simultaneously with primary reconnection that forms vortex loops; and third, secondary reconnection between these loops. We first discuss the phase-wise evolution of the wake vortices for the dynamic case of $d^* = 0.5$.

	\begin{enumerate}
	\item{{\it Starting vortex tube formation phase}: This phase spans from 0 to 40 ms ($t/t_c \approx 0$--$0.6$). In the beginning ($\sim$20 ms), as the rotation of the plate dominates over its translation, vorticity rolls up along both side edges (top and bottom) and the trailing edge of both plates, forming the starting vortex tube, see figure~\ref{fig:C_3D_IsoVort_Dyn}(a). In this figure, the iso-surface colored by the $Z$-component of vorticity, $\omega_Z$, shows opposite orientations (red and blue) in the portions of the tube aligned along the trailing edges. The portions oriented sideways exhibit streamwise vorticity and therefore show negligible $\omega_Z$ (green). The cross-section of these streamwise vortices is shown in figure~\ref{fig:C_VortStreamWise_Stat_Dyn}(d). Compared with the stationary case, these vortices are smaller in size, and as time progresses they become even smaller (see figure \ref{fig:C_VortStreamWise_Stat_Dyn}(e)). This size reduction in the starting vortices can be understood from the observation of their 3D structural evolution towards the end of the formation phase, when the plates translate away from the vortex tubes. Because of this translation, compared to the stationary case, the time available for vorticity accumulation is reduced, resulting in thinner streamwise vortices.}
			
	\item{{\it Simultaneous bound vorticity shedding and primary reconnection phase}: This phase lasts approximately from 40 ms to 130 ms ($t/t_c \approx 0.6$--$1.9$). At the beginning of this phase, the portion of the starting vortex tube detached from the plate's trailing edge develops noticeable curvature: at the mid-depth and locations that previously rested on the top and bottom corners. See the bottom-right inset of figure \ref{fig:C_3D_IsoVort_Dyn}(c), which shows the back side view of the vortex structure. Simultaneously, the forward motion of the body develops a boundary layer on the outer surface of each plate, in contrast to the stationary case, where the absence of translation produces negligible vorticity on the outer surfaces. In this phase, where the plate’s translation dominates over its rotation, the vorticity shed from the outer surface of the plate minimally rolls along the side edges (see figure \ref{fig:C_VortStreamWise_Stat_Dyn}(f)), whereas it predominantly convects downstream from the trailing edges. At later times, the shed vorticity appears in the form of thin vertical threads just behind the trailing edges, see figure \ref{fig:C_3D_IsoVort_Dyn}(e). In contrast, in the stationary cases, around the stopping time of the plate, the vorticity rolls up along the plate’s edges in a localized zone, forming compact stopping vortices. This difference in the vorticity roll-up explains why the stopping vortex formation phase is observed only in the stationary cases. Toward the end of the plate’s rotation, primary reconnection occurs along its side edges between the thin bound vorticity tube and the remaining narrow portion of the starting vortex tube, as most of the tube has already detached by this point (figure \ref{fig:C_3D_IsoVort_Dyn}(c)). This reconnection takes place in region $R_{Z|1}$, near the top and bottom corners of the trailing edge of each plate. For clarity, only the top $R_{Z|1}$ of one plate is identified in green in the bottom-left inset of figure \ref{fig:C_3D_IsoVort_Dyn}(c), where the vortex structure is colored by the streamwise vorticity, $\omega_X$. The reconnection forms one elliptical vortex loop for each plate, consisting of the starting vortex tube on one side and the shed bound vorticity threads on the other, closer to the body, see figure \ref{fig:C_3D_IsoVort_Dyn}(g). By the end of this phase, the wake shows two elliptical vortex loops with no interconnection, as clearly illustrated in the bottom-left inset of figure \ref{fig:C_3D_IsoVort_Dyn}(g), which displays a top view of the isolated vortex loops.}
	
	\item{{\it Secondary reconnection phase}: This phase occurs from 130 ms ($t/t_c \approx 1.9$) onward, during which the isolated elliptical vortex loops interconnect. Computational data is unavailable for this phase, but based on experimental observations reported in Mahulkar and Arakeri\cite{Mahulkar24}, we discuss further probable transformations in the wake vortices. The secondary reconnection between isolated vortex loops occurs in regions of the loops oriented along the trailing edges and along the top and bottom side edges. On the trailing-edge side, the loops reconnect at three locations of sharp curvature: the mid-depth region and the top and bottom corners (figure \ref{fig:C_3D_IsoVort_Dyn}(g)). At these locations, both vortex tubes approach each other rapidly due to higher self-induced velocity, eventually reconnecting to form two ringlets through three reconnection zones. Simultaneously, in the sideways region, another reconnection occurs in the portions of the vortex loops showing close proximity, see the bottom-left inset of figure \ref{fig:C_3D_IsoVort_Dyn}(g). In this inset, two probable secondary reconnection zones, $R_{Z|2}$, are marked on the top side of the loop: one near the trailing edges of the plates and the other farther from the trailing edges. Corresponding zones on the bottom side are not shown. Reconnection in these two zones on the top side results in the formation of one elliptical ringlet, and a similar reconnection on the bottom side forms another ringlet. Ultimately, reconnection on the trailing-edge side and in the sideways regions forms a wake structure of four circumferentially connected ringlets. Such a structure was observed in the experimentally obtained flow field around 200 ms. Its cross-section on the XZ plane (Y = 0) is shown in figure 14(c), while its approximate three-dimensional structure is shown in figure 15(d) in Mahulkar and Arakeri\cite{Mahulkar24}.}
	\end{enumerate} \par
	
	\begin{figure}
	\centering\
	\begin{subfigure}[b]{0.40\textwidth}
		\includegraphics[width=\textwidth]{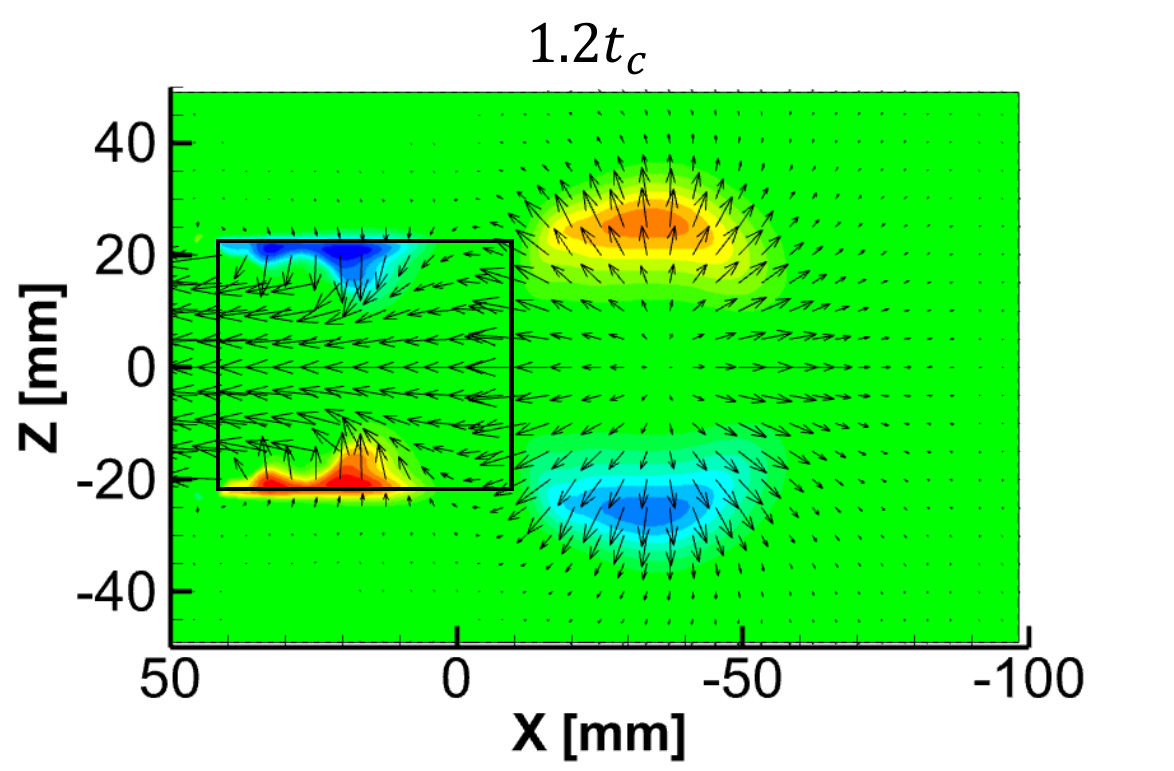}
		\caption{CFD: Dyn}
	\end{subfigure}\hspace{5mm}
	\begin{subfigure}[b]{0.40\textwidth}
		\includegraphics[width=\textwidth]{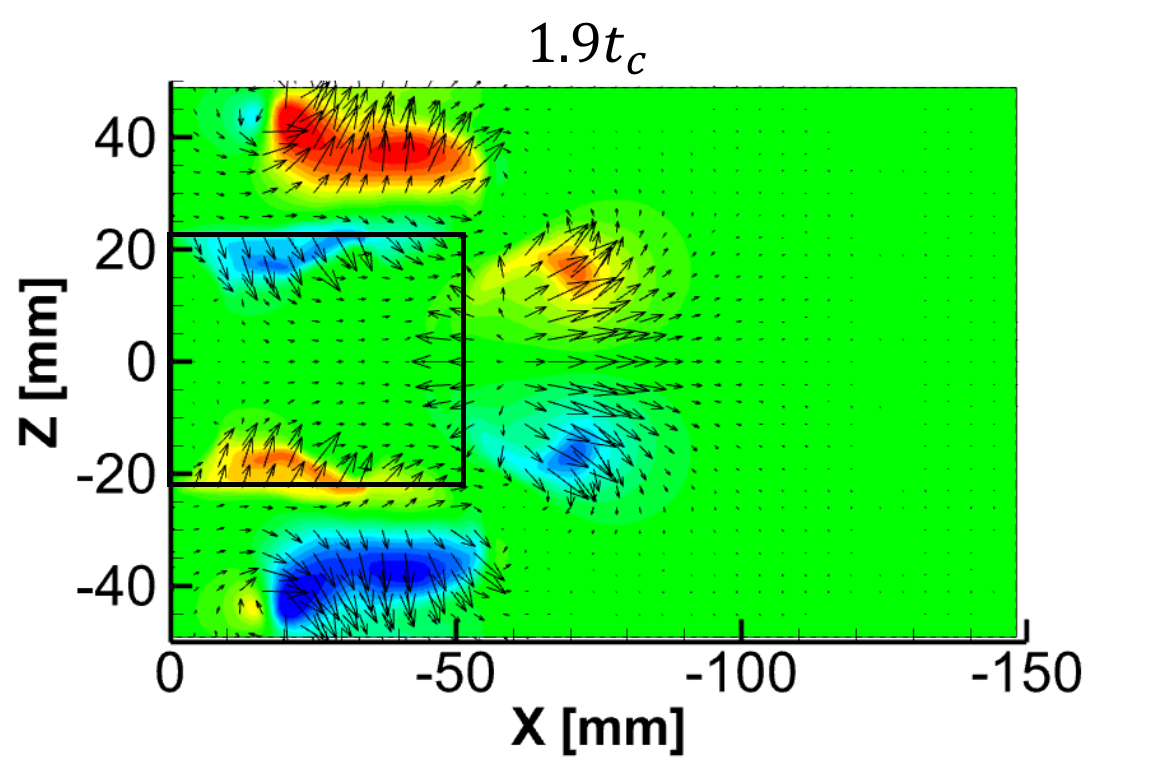}
		\caption{CFD: Stat}
	\end{subfigure}	
	\begin{subfigure}[b]{0.40\textwidth}
		\includegraphics[width=\textwidth]{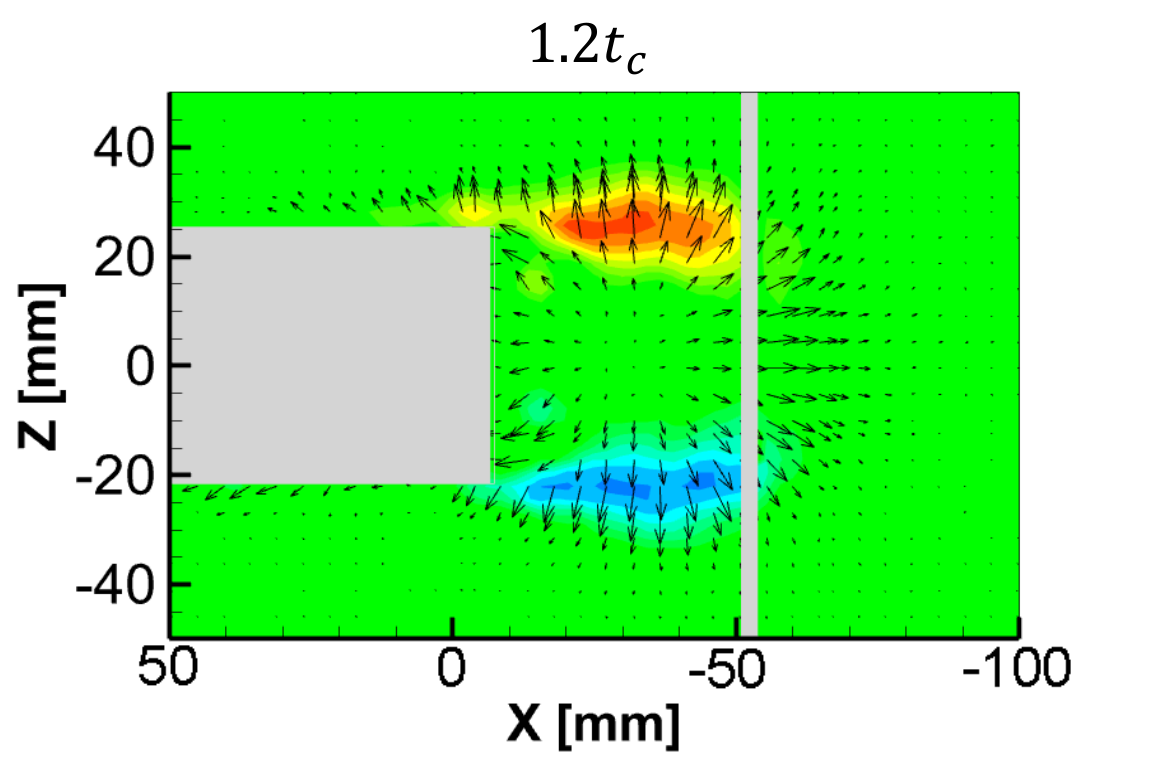}
		\caption{EXP: Dyn}
	\end{subfigure}\hspace{5mm}
	\begin{subfigure}[b]{0.40\textwidth}
		\includegraphics[width=\textwidth]{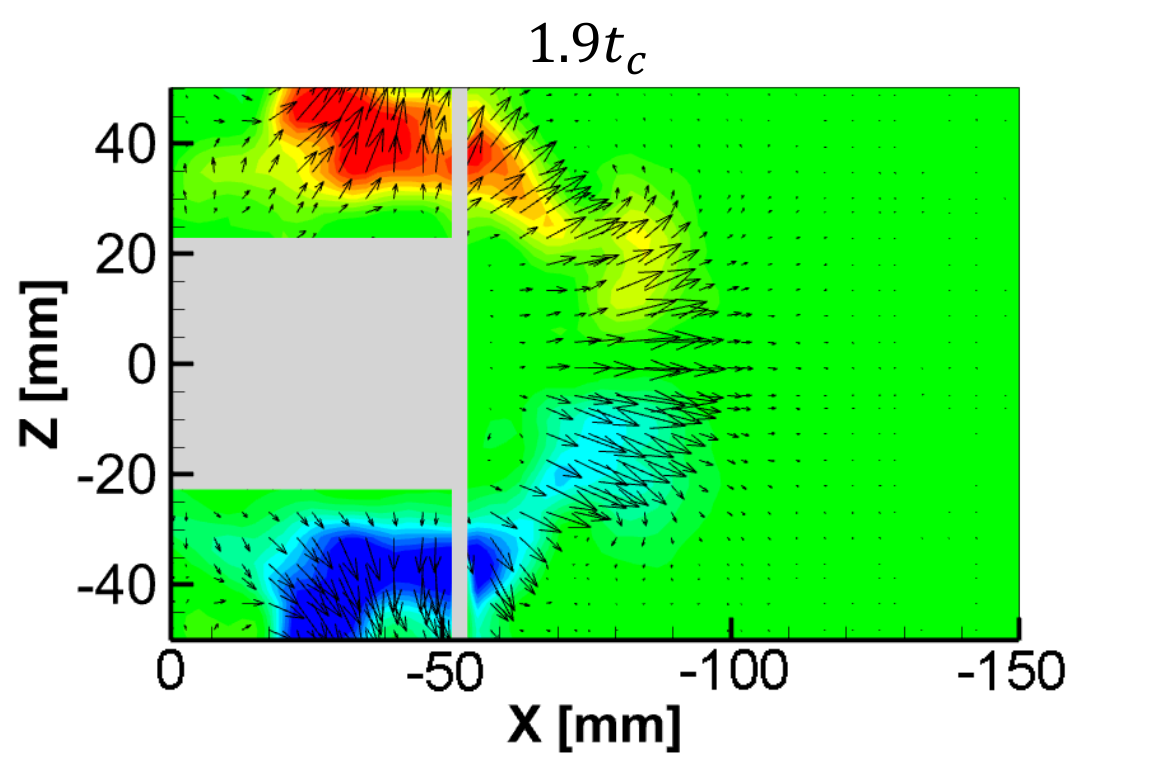}
		\caption{EXP: Stat}
	\end{subfigure}
	\begin{subfigure}[b]{0.5\textwidth}
		\includegraphics[width=\textwidth]{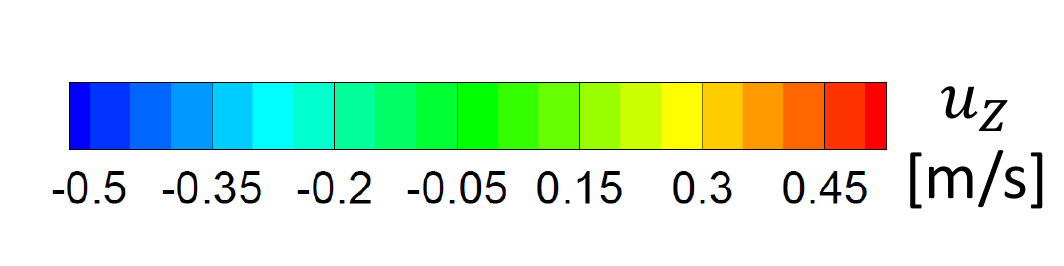}
	\end{subfigure}
	
	\begin{subfigure}[b]{0.40\textwidth}
		\includegraphics[width=\textwidth]{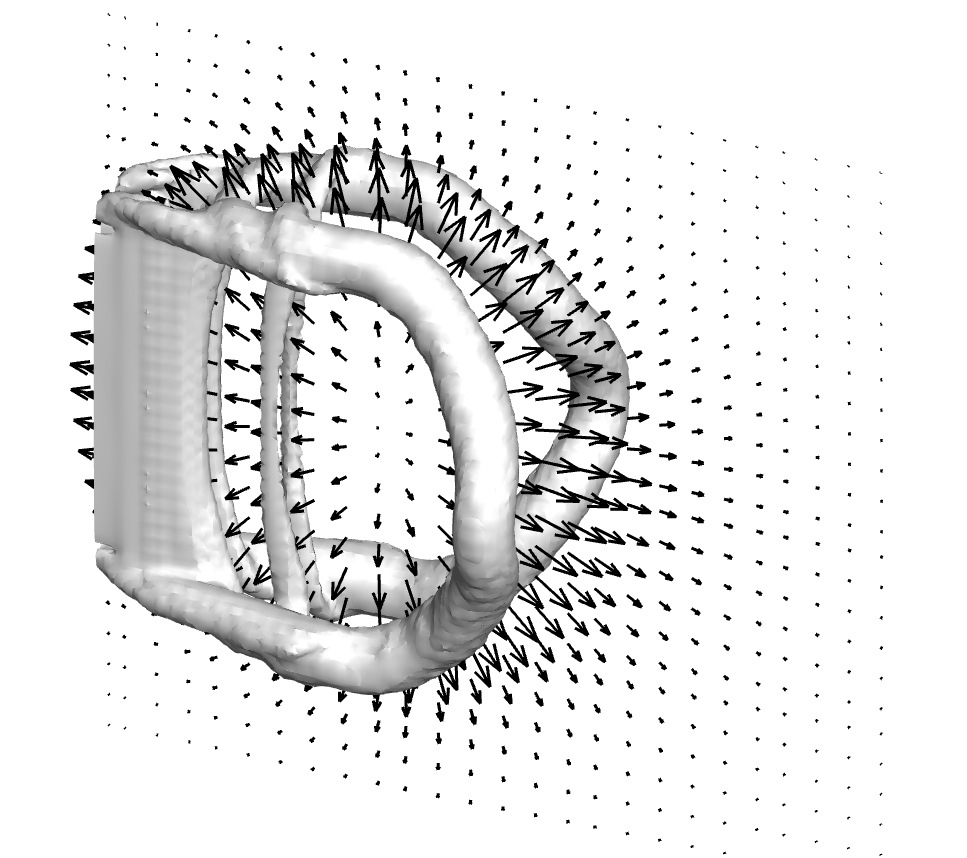}
		\caption{Dyn ($1.2t_c$)}
	\end{subfigure}
	\begin{subfigure}[b]{0.40\textwidth}
		\includegraphics[width=\textwidth]{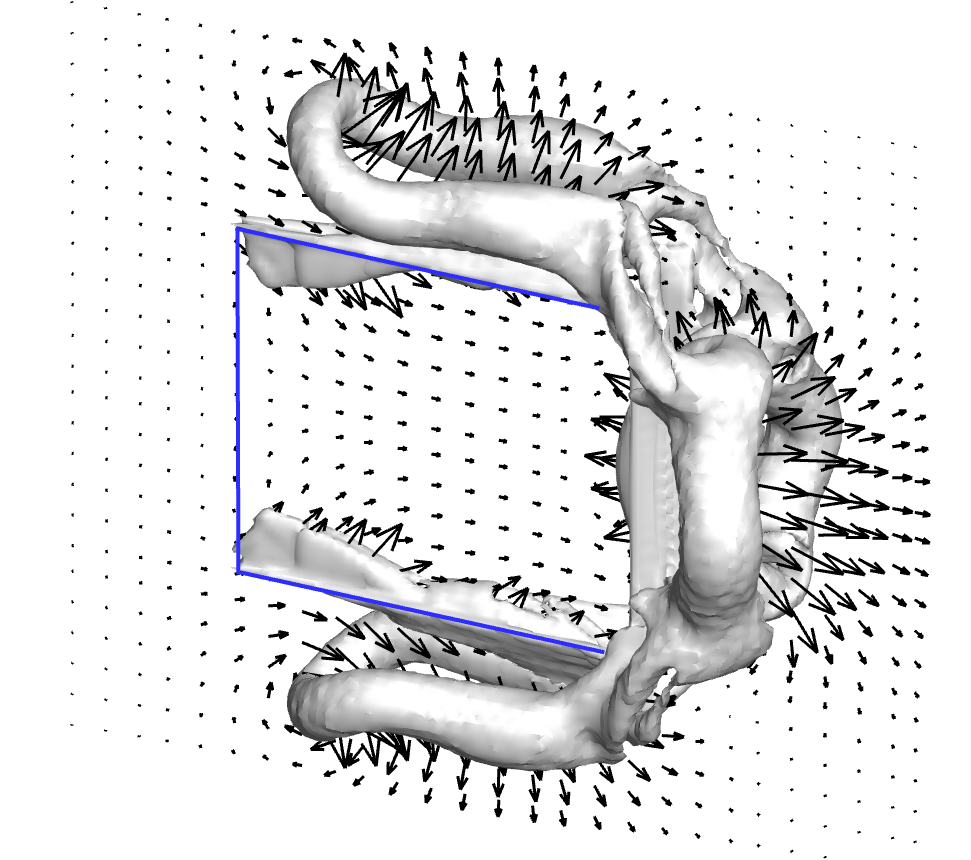}
		\caption{Stat ($1.9t_c$)}
	\end{subfigure}
	\caption{The $Z$-component of velocity, $u_Z$, plotted on the XZ plane ($Y=0$) for the clapping body with $d^*=0.5$: (a, c) dynamic case at $t \approx 1.2t_c$ and (b, d) stationary case at $t \approx 1.9t_c$. The body position is marked by a black rectangle in the computational (CFD) flow fields for (a) the dynamic (Dyn) and (b) the stationary (Stat) cases. Experimentally obtained (EXP) flow fields for (c) the dynamic and (d) the stationary cases show close agreement with the corresponding computational fields; the gray rectangle denotes the shadow region. The iso-vorticity structure at $|\omega|=130$ s$^{-1}$, superimposed with velocity vectors on the symmetry plane ($Y=0$), is shown for (e) the dynamic and (f) the stationary cases. In the dynamic case, the vorticity structure in the wake of the freely moving body is shown, whereas in the stationary case, the vorticity structure remaining near the body (blue rectangle) is shown.}\label{fig:C_sideways_flow_field}	
	\end{figure}

	In the dynamic case with increased body depth ($d^* = 1.0$), the wake-vortex evolution phases are similar to those observed for $d^* = 0.5$. A notable difference is that, for $d^* = 1.0$, the vortex loops on the trailing-edge side do not exhibit sharp curvature at mid-depth (figures \ref{fig:C_3D_IsoVort_Dyn}(d, f)). Because of the absence of such curvature, the vortex loops there do not approach each other to reconnect. However, similar to the case of $d^* = 0.5$, at the top and bottom corners of the trailing-edge side, the loops come very close to each other, as shown in figure \ref{fig:C_3D_IsoVort_Dyn}(h). Subsequent transformations in the wake vortices are inferred from experimental observations, which span longer durations than the computations. The experiments show that a large elliptical ring appears in the wake around 400 ms following secondary reconnections at the corners of the trailing-edge side, unlike the two ringlets observed for $d^* = 0.5$. The cross-section of this ring in the XZ plane (Y = 0) is shown in figure 14(b) of Mahulkar and Arakeri\cite{Mahulkar24}.

	The differences between the stationary and dynamic cases can be further understood from the sideways flow fields shown in figure \ref{fig:C_sideways_flow_field} for $d^* = 0.5$. These flow fields are shown at representative post-clapping instants, 80 ms ($t/t_c \approx 1.2$) for the dynamic case and 150 ms ($t/t_c \approx 1.9$) for the stationary case. These instants are chosen to capture the well-developed sideways-flow structures within the experimental field of view near the body. The stronger sideways flow in the stationary case compared to the dynamic case is clearly visible (figures \ref{fig:C_sideways_flow_field}(b, d)). Within the interplate cavity, the flow is nearly stagnant in the stationary case (figure \ref{fig:C_sideways_flow_field}(b)), whereas in the dynamic case, the trapped fluid is dragged forward by the moving body (figure \ref{fig:C_sideways_flow_field}(a)). In both cases, inward flow is observed at the top and bottom edges due to streamwise vortices, which are more pronounced in the stationary case (figure \ref{fig:C_VortStreamWise_Stat_Dyn}). In addition, the computationally obtained flow fields (figures \ref{fig:C_sideways_flow_field}(a, b)) show excellent agreement with the experimentally obtained fields (figures \ref{fig:C_sideways_flow_field}(c, d)), thereby validating the chosen mesh resolution in the near field of the body. A further view of the flow fields is obtained from the 3D vortex structure superimposed with velocity vectors on the plane of symmetry in figures \ref{fig:C_sideways_flow_field}(e, f). In the dynamic case, figure \ref{fig:C_sideways_flow_field}(e) shows that the 3D vortices sliding back from the moving body induce higher velocities near the top and bottom portions of the vortex loops, where the loops come closer together compared to other regions. These high-velocity regions correspond to the red and blue $u_Z$ patches in figures \ref{fig:C_sideways_flow_field}(a, c). In the stationary case, shown in figure \ref{fig:C_sideways_flow_field}(f), the 3D vortex structure reveals small vortex ringlets at the top, bottom, and on the trailing-edge side. These ringlets, formed due to the secondary reconnection described earlier, induce higher velocities in their central regions, identified by larger velocity vectors together with the red and blue $u_Z$ patches in figures \ref{fig:C_sideways_flow_field}(b, d).\par
	
\section{Thrust coefficient}
\label{sec:C_ThrustCoeff}

\begin{figure}
	\centering\
	\begin{subfigure}[b]{0.40\textwidth}
		\includegraphics[width=\textwidth]{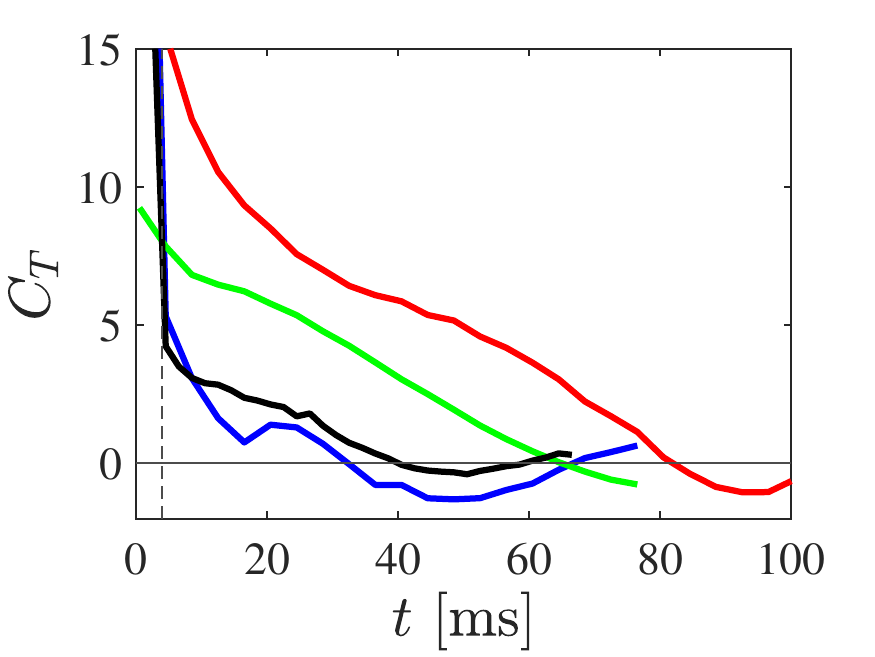}\caption{}
	\end{subfigure}
	\begin{subfigure}[b]{0.40\textwidth}
		\includegraphics[width=\textwidth]{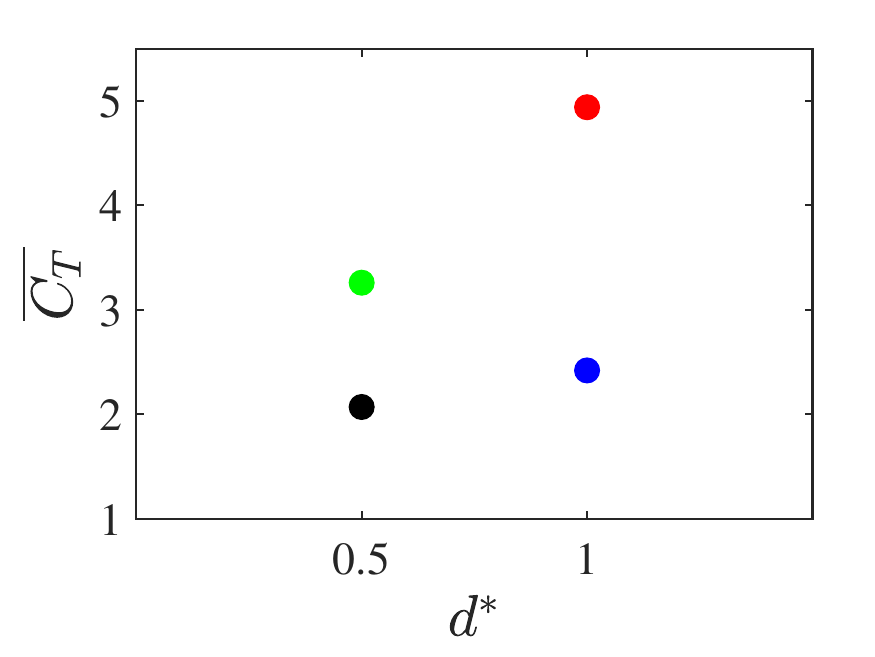}
		\caption{}
	\end{subfigure}
	\begin{subfigure}[b]{0.40\textwidth}
		\includegraphics[width=\textwidth]{figCFD/C_AngVelo_Dyn_Stat_legend}
	\end{subfigure}
	\caption{(a) Evolution of the thrust coefficient, $C_T$, with time for stationary and dynamic cases of $d^* = 0.5$ and $1.0$. The vertical dashed line indicates the time domain up to 4 ms, prior to the onset of translation in the dynamic cases. (b) Plot of the mean thrust coefficient, $\overline{C_T}$, versus $d^*$.  }\label{fig:C_Ct}
\end{figure}

As discussed in \S \ref{sec:C_Pressure distribution}, during the effective clapping period, in both stationary and dynamic cases, the pressure on each plate is predominantly positive on the front side (facing the interplate cavity) and negative on the back side. Integrating the pressure on the two faces of each plate and taking the X-component, we obtain the thrust, $F_T$, and the thrust coefficient, $C_T$, defined as:
\begin{gather}
C_T = \frac{F_T}{0.5\rho \ \overline{u}_T^2  A_c},
\end{gather} 
where $ \rho $ is the fluid density, $ \overline{u}_T(=R_{rp}\int_0^{t_c}{\dot{\theta}} \,dt/t_c) $ is the mean tip velocity obtained from the angular velocity of the plate, and $ A_c (= R_{rp} d) $ is the surface area of the plate with length $R_{rp}$ and depth, $d$. Figure \ref{fig:C_Ct}(a) shows the evolution of the thrust coefficient over the clapping period, $t_c$, for both stationary and dynamic cases. In both cases, the overall trend of $C_T$ is governed by the pressure field generated by the clapping motion. The high initial $C_T$ arises from the strong pressure associated with the acceleration of the added mass within and around the cavity, together with lower pressures developed on the outer surfaces of the plates. As the clapping motion progresses, the pressure distribution evolves and $C_T$ eventually becomes negative due to the pressure crossover occurring at $t_{\pm p}$, indicating a drag phase. In the dynamic case, however, after an initial rise during the period of pure rotation ($t \leq 4$ ms), $C_T$ shows a sharp dip when the body begins to translate impulsively. This occurs because a significant portion of the fluid’s reactive force is diverted to accelerating the body from rest, leaving less force available to generate thrust. As translation develops further, $C_T$ follows the general trend described above, becoming negative toward the end of rotation due to pressure crossover and subsequently returning to small positive values near zero due to a second pressure crossover associated with fluid dragging in the interplate cavity. \par

We next analyse the mean thrust coefficient, $\overline{C_T}$, obtained by averaging $C_T$ over the clapping period, $t_c$. Figure \ref{fig:C_Ct}(b) shows that $\overline{C_T}$ increases with $d^*$ in the stationary cases ($\overline{C_T} = 3.26$ for $d^* = 0.5$ and $4.94$ for $d^* = 1.0$), whereas it remains nearly constant in the dynamic cases ($\overline{C_T} = 2.07$ for $d^* = 0.5$ and $2.42$ for $d^* = 1.0$). This difference can be understood from the sideways-flow behaviour discussed in \S \ref{sec:C_3DWakeVortices}. In the stationary cases, increasing $d^*$ reduces the relative sideways ejection and directs a larger portion of flow momentum along the X-direction, thereby increasing the mean thrust generated per unit depth. The integral momentum conservation analysis presented in our previous study (Mahulkar and Arakeri \cite{Mahulkar24}) discusses this relation between sideways ejection and thrust per unit depth in detail. The mean tip velocity, $\overline{u}_T$, also decreases with increasing $d^*$ ($0.29$--$0.22~\mathrm{m/s}$ in the stationary cases and $0.36$--$0.28~\mathrm{m/s}$ in the dynamic cases, for $d^* = 0.5$--$1.0$). Therefore, the increase in $\overline{C_T}$ with $d^*$ in the stationary cases is consistent with greater streamwise momentum in the wake and a lower tip-velocity scale in the normalization. In the dynamic cases, since the sideways-flow contribution is negligible, the thrust generated per unit depth varies only weakly with $d^*$. This is consistent with the weak variation of $\overline{C_T}$ with $d^*$.\par

\section{Clapping wake energetics}
\label{sec:C_WakeEnergy_Clapping}
Clapping motion adds energy to the fluid domain through the work done by the unsteady rotation of the plates. During the effective clapping phase ($t \leq t_{\pm p}$), when the interplate cavity pressure exceeds the ambient pressure (see \S \ref{sec:C_Pressure distribution}), the plates perform positive work on the fluid, $WD_{T|c}$, given by
	\begin{gather}
		WD_{T|c} = 2 \int_{0}^{t_{\pm p}} T_{p|Z}\,\dot{\theta}\,dt,
		\qquad
		\boldsymbol{T_p} = \int \boldsymbol{r}_h \times \boldsymbol{dF}_p .
	\end{gather}
Here, $T_{p|Z}$ is the $Z$-component of the pressure torque vector $\boldsymbol{T_p}$, obtained by integrating the moment of the elemental pressure force $\boldsymbol{dF}_p$ acting on the plate surface, with $\boldsymbol{r}_h$ denoting the position vector measured from the hinge. The factor of two accounts for the contribution from both plates. \par
	
\begin{figure}
	\centering\
	\begin{subfigure}[b]{0.40\textwidth}
		\includegraphics[width=\textwidth]{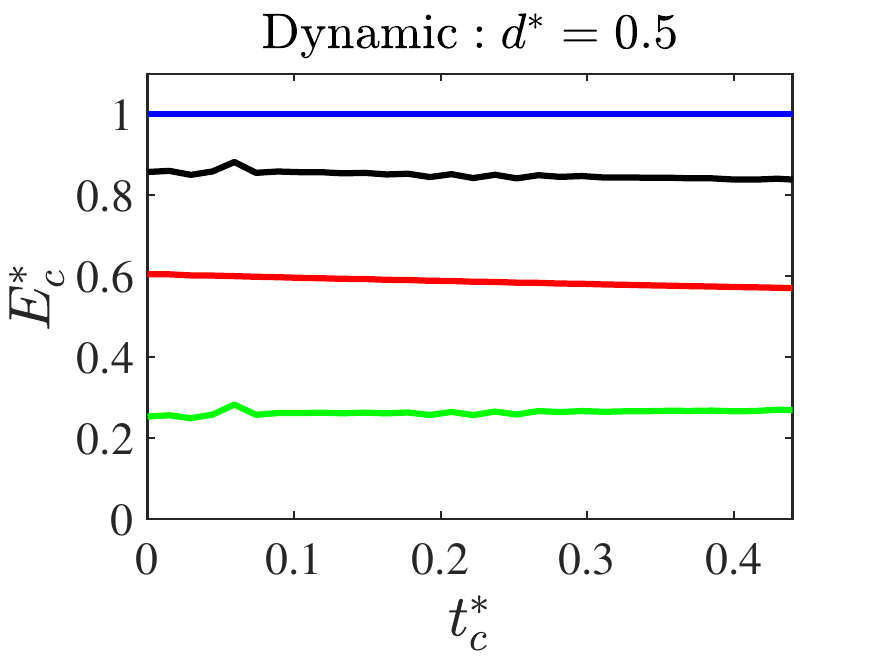}
		\caption{}
	\end{subfigure}
	\begin{subfigure}[b]{0.40\textwidth}
		\includegraphics[width=\textwidth]{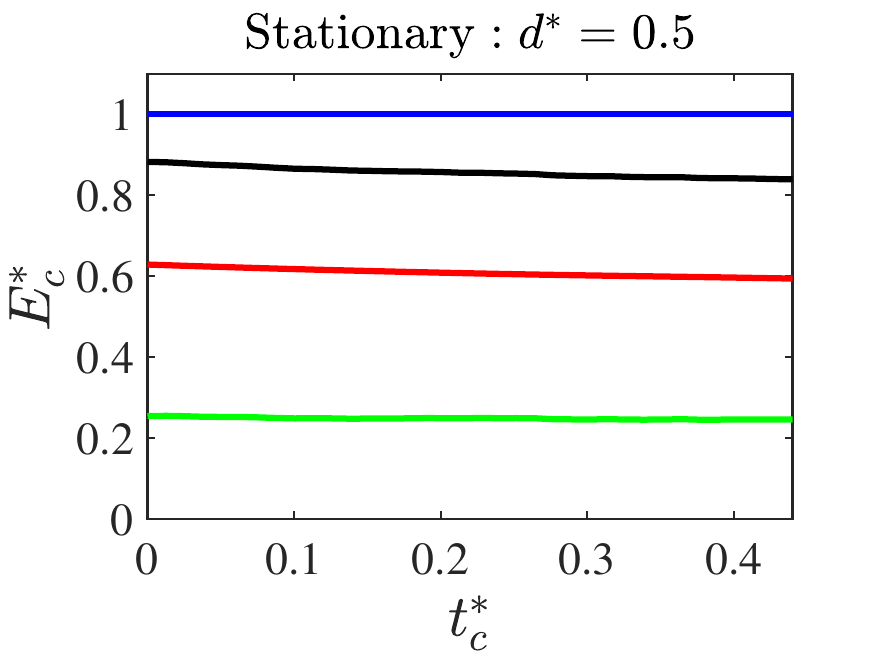}
		\caption{}
	\end{subfigure}
	\begin{subfigure}[b]{0.45\textwidth}
		\includegraphics[width=\textwidth]{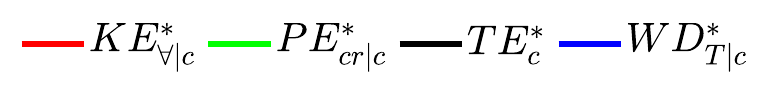}
	\end{subfigure}
	\caption{Time variation of nondimensional energies in the clapping domain after wake vortex formation for (a) the dynamic case and (b) the stationary case for $d^* = 0.5$. Here, $KE^*_{\forall|c}$, $PE^*_{cr|c}$, $TE^*_{c}$, and $WD^*_{T|c}$ denote the nondimensional kinetic energy in the fluid domain, vortex core potential energy, total energy, and work done by plate rotation, respectively.}\label{fig:C_clapping Non-dim energy}	
\end{figure}

For the clapping cases, the kinetic energy in the entire fluid domain, $KE_{\forall|c}$, is evaluated as
\begin{gather}
KE_{\forall|c} = \frac{1}{2} \rho \int |u|^2 \, d\forall_c,
\end{gather}
where $|u|$ is the velocity magnitude in the fluid domain $\forall_c$. Figure \ref{fig:C_clapping Non-dim energy} shows $KE_{\forall|c}$ nondimensionalised by the work done during the effective clapping phase, $KE^*_{\forall|c}=KE_{\forall|c}/WD_{T|c}$, plotted against the nondimensional time $t_c^*$. The nondimensional time is defined as $t_c^*=(t-t_{po})/t_c$, where $t_{po}$ is the vortex formation time and $t_c$ is the clapping period (table \ref{tab:C_BodyDyn_gamma}). The fluid energies are evaluated after the wake vortices have formed, with $t_{po}$ identified from steady circulation values and a nearly ambient pressure field around the clapping body. The corresponding values of $t_{po}$ are $79$ ms and $91$ ms for the dynamic cases of $d^* = 0.5$ and $1.0$, respectively, and $129$ ms and $161$ ms for the stationary cases of $d^* = 0.5$ and $1.0$, respectively. The chosen range, $t_c^*\leq0.45$, is used to avoid mesh interpolation errors that arise when the freely moving body transitions from the fine mesh zone $\mathrm{Z_N}$ to the coarse mesh zone $\mathrm{Z_F}$ (figure \ref{fig:C_CompModelling}(b)), ensuring that the analysis is performed while the body remains within $\mathrm{Z_N}$. Over this range, $KE^*_{\forall|c}$ remains approximately steady. Figure \ref{fig:C_clapping Non-dim energy} shows that, for the plotted $d^*=0.5$ cases, $KE^*_{\forall|c}$ is about $0.6$, indicating that the kinetic energy in the domain accounts for only about $60\%$ of the work done in both stationary and dynamic cases.\par

\begin{table}
	\centering
	\begin{tabular}{cccccccccc}
		\toprule
		\multirow{2}[2]{*}{\text{Case}} & \multirow{2}[2]{*}{\text{$d^*$}} & \text{$KE_{\forall|c}$} & \text{$PE_{cr|c}$} & \text{$TE_c$} & \text{$ WD_{T|c}$} & \multirow{2}[2]{*}{\text{$\frac{TE_c}{WD_{T|c}}$}} & \multirow{2}[2]{*}{\text{$KE^*_{\forall|c}$}} & \multirow{2}[2]{*}{\text{$PE^*_{cr|c}$}} & \multirow{2}[2]{*}{\text{$\frac{PE_{cr|c}}{KE_{\forall|c}}$}} \\
		&       & \text{[mJ]} & \text{[mJ]} & \text{[mJ]} & \text{[mJ]} &       &       &       &  \\
		\midrule
		\multirow{2}[2]{*}{\text{Stationary}} & 0.5  & 7.08  & 2.89  & 9.97  & 11.65 & 0.86  & 0.61  & 0.25  & 0.41 \\
		& 1.0  & 11.7  & 6.16  & 17.86 & 17.70 & 1.01  & 0.66  & 0.35  & 0.53 \\
		\midrule
		\multirow{2}[2]{*}{\text{Dynamic}} & 0.5  & 7.26  & 3.26  & 10.52 & 12.39 & 0.85  & 0.59  & 0.26  & 0.45 \\
		& 1.0  & 13.2  & 3.92  & 17.12 & 15.80 & 1.08  & 0.84  & 0.25  & 0.30 \\
		\bottomrule
	\end{tabular}%
	\caption{For stationary and dynamic clapping cases with $d^* = 0.5$ and 1.0, columns 3–7 list the kinetic energy in the complete domain, $KE_{\forall|c}$, vortex core potential energy, $PE_{cr|c}$, total energy $TE_c$, work done by plate rotation $WD_{T|c}$, and the ratio $TE_c/WD_{T|c}$. The last three columns give the nondimensional energies: kinetic, $KE^*_{\forall|c}$; potential, $PE^*_{cr|c}$; and the ratio $PE_{cr|c}/KE_{\forall|c}$.}
	\label{tab:Energy_clapping}%
\end{table}%

A similar kinetic energy deficit was noted by Sullivan \textit{et al.}\cite{Sullivan08} during vortex ring formation, where the missing energy was hypothesised to be associated with the vortex structure. To examine this idea in a simpler and well-defined geometry, we analyse an axisymmetric vortex ring in \S\ref{sec:C_WakeEnergy_Ring}. For that case, the additional energy associated with vortex formation is identified as the volume integral of the absolute gauge pressure over the vortex core (equation \ref{eq:PE_basic_model}). We use this definition here to quantify the vortex core potential energy of the interconnected vortices in the clapping wake:
\begin{gather}
PE_{cr|c} = \int |P|_{cr|c} \, d\forall_{cr|c},
\end{gather}
where $|P|_{cr|c}$ is the absolute gauge pressure in the vortex core volume $\forall_{cr|c}$. The vortex core is identified using the same vorticity criterion as in \S\ref{sec:C_Validation}: $\omega_Z \geq 5\%\,\omega_{Z|\max}$. Figure \ref{fig:C_clapping Non-dim energy} shows that the nondimensional potential energy, $PE^*_{cr|c}=PE_{cr|c}/WD_{T|c}$, is about $0.25$--$0.35$ over the chosen time interval. The total energy in the clapping domain, $TE_c(=KE_{\forall|c}+PE_{cr|c})$, evaluated around the formation time, nearly matches the work added to the fluid during the effective clapping phase. Table \ref{tab:Energy_clapping} lists the energies for all the cases studied and shows that $TE_c/WD_{T|c}$ varies between $0.85$ and $1.08$ (column 7) for both stationary and dynamic cases. This near closure suggests that the vortex core potential energy accounts for a substantial part of the kinetic energy deficit in the clapping wake. The table shows that, except for the dynamic case with $d^*=1.0$, where $KE^*_{\forall|c}=0.84$, the other three cases have $KE^*_{\forall|c}\approx0.6$. This deviation can be partly understood from a sudden increase in kinetic energy caused by the abrupt stopping of plate rotation in the simulations to prevent mesh intersection between the plate meshes. The sudden stop releases the pressure field generated during rotation, creating a favourable pressure gradient that accelerates the fluid and produces a kinetic energy jump immediately after rotation ceases. In contrast, in experiments the plates stop gradually, allowing smoother work--energy transfer. For the dynamic cases, the kinetic energy jump is small for $d^*=0.5$ (about $5\%$ of $KE_{\max}$), whereas for $d^*=1.0$ it reaches $1.8$ mJ (about $13\%$ of $KE_{\max}$); in the stationary cases, the jump is negligible for both $d^*$ values due to the lower plate rotation speeds.\par

Finally, since the kinetic and core potential energies require volume integration over the computational domain, the overset mesh framework may introduce double counting in overlap regions containing donor and solve cells. The resulting error was quantified by separately evaluating the energy in donor cells within the overlap zones, and since this contribution remains below $5\%$ of $KE_{\forall|c}$ and $PE_{cr|c}$ over the considered $t_c^*$ range, the associated error is considered negligible. The simpler vortex ring case discussed in the next section provides a further check on this vortex core potential energy.\par

\section{Vortex ring energetics}
\label{sec:C_WakeEnergy_Ring}

Sullivan {\it et al.}\cite{Sullivan08}, in their experimental study of piston-generated vortex rings, found that the momentum of the vortex bubble, $I_{\Omega}$, is equal to the momentum of the fluid slug, $I_p$. The slug momentum is given by $I_p = m_s u_p$, where $m_s$ is the fluid slug mass and $u_p$ is the piston velocity. The vortex bubble momentum is given by $I_{\Omega}=m_{\Omega}u_R$, where $m_{\Omega}$ is the vortex bubble mass and $u_R$ is the vortex ring velocity. In an otherwise quiescent domain, the kinetic energy in the vortex bubble, $KE_{\Omega}$, is expressed as $KE_{\Omega} = m_\Omega u_R^2/2 = I_{\Omega}^2/(2m_\Omega)$. Substituting the piston momentum into $KE_{\Omega}$ gives: 
\begin{gather}
\label{KE_deficit}
KE_{\Omega} =I^2_p/(2 m_\Omega)= (m_s/m_{\Omega}) KE_p,
\end{gather}
where $ KE_p ( = m_s u^2_p/2)$ is the kinetic energy (or slug energy) added to the fluid by the piston. Due to fluid entrainment, the mass of the vortex bubble is larger than the slug mass (Dabiri {\it et al.}\cite{Dabiri04}), $m_{\Omega} > m_s$. Consequently, the kinetic energy in the bubble is lower than the kinetic energy supplied to the fluid ($ KE_{\Omega} < KE_p $). They hypothesized that this difference in energy must be some {\it potential energy} associated with the vortex bubble. As observed in the previous section, a similar energy deficit occurs in the clapping cases, between the work done on the fluid by the rotating plates and the kinetic energy of the fluid once the wake vortices are fully developed. In the following subsections, we present an analysis of the fluid energy budget and quantify the energy deficit using CFD simulation results for the formation of a vortex ring, which is a simpler flow field than that in the clapping case.

\subsection{Computational setup and cases}	
\begin{figure}
	\centering\
	\begin{subfigure}[b]{0.45\textwidth}
	\includegraphics[width=\textwidth]{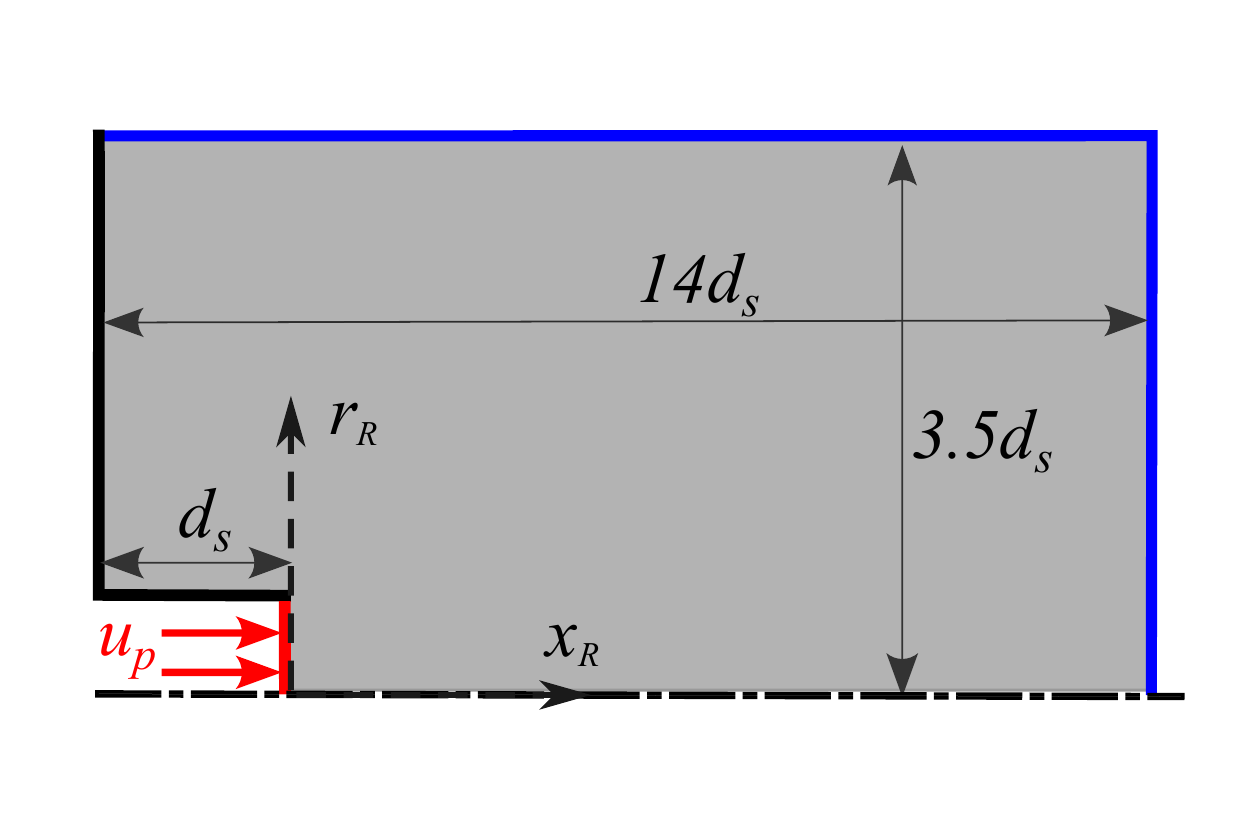}
	\caption{}
	\end{subfigure}
	\begin{subfigure}[b]{0.40\textwidth}
	\includegraphics[width=\textwidth]{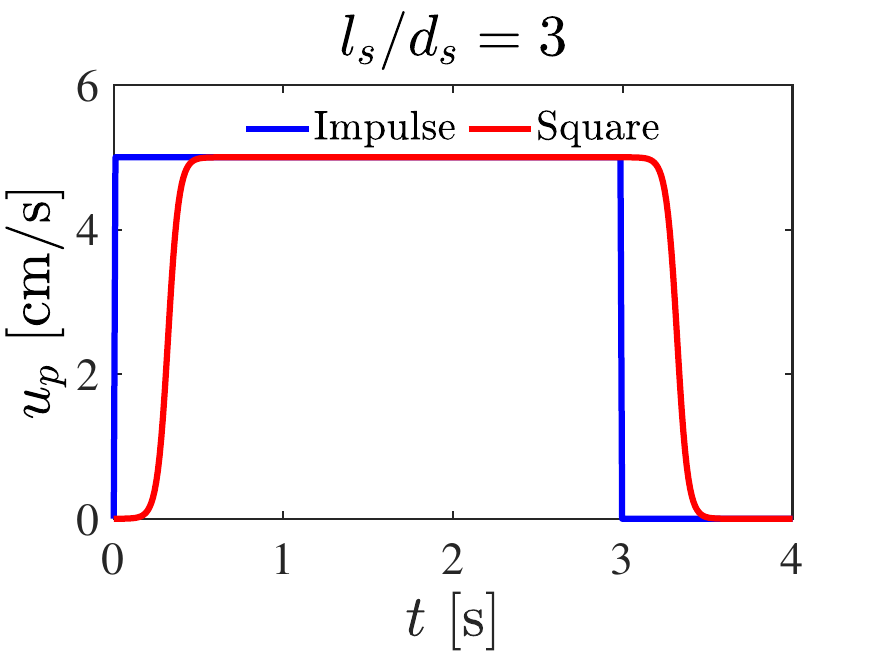}
	\caption{}
	\end{subfigure}
	\caption{Schematic of the axisymmetric computational domain used for vortex ring generation (not to scale). The domain includes a nozzle that protrudes into the interior, from which a fluid jet with velocity $u_p$ is discharged. Boundary conditions are marked by color: pressure outlet (blue), wall (black), and inlet (red). (b) Two types of time variation programs for the jet injection (or piston) velocity, $u_p$: Impulse and Square, shown for $ l_s/d_s = 3 $.
	}\label{fig:C_RingDomain}	
    \end{figure}

  The computations are performed for a vortex ring formed by unsteady flow from a nozzle. The nozzle exit, where the fluid velocity is specified, serves as the inlet to an initially quiescent fluid domain. See the axisymmetric computational domain for the vortex ring in figure \ref{fig:C_RingDomain}(a), where the nozzle exit is highlighted in red. James {\it et al.}\cite{James96} and Danaila {\it et al.}\cite{Danaila08} used a similar approach to computationally generate a vortex ring. Based on the experimental observations of Gharib {\it et al.} \cite{Gharib98}, the piston stroke-to-diameter ratio (=$ l_s/d_s $) is set to 2, 3, and 4 for generating isolated vortex rings, with the piston stroke and diameter assumed equal to the fluid slug length, $ l_s $, and slug diameter, $ d_s $, respectively. We use two types of time variation programs for the piston velocity (also referred to as the jet injection velocity), $u_p$: impulse and square, as shown in figure \ref{fig:C_RingDomain}(b). In the six cases analyzed in this study (two $ u_p $ programs for the three values of $ l_s / d_s $), the peak jet injection velocity, $ u_{pm} $, is 5 cm/s, and the nozzle diameter ($ = d_s $) is 5 cm, both kept constant. This ensures a constant Reynolds number $(Re_j = u_{pm} d_s / \nu)$ of 2500, based on the peak jet injection velocity, with water as the fluid. At each time instant of $u_p$, the jet injection velocity profile (spatial variation across the nozzle) is uniform in the impulse program, whereas, in the square program, it follows a hyperbolic tangent profile, as discussed below. James {\it et al.}\cite{James96} introduced the square program for the jet injection velocity, which we have adopted in this study, given as: 
	\begin{gather}
	\label{eq:tanh_profile}
	\frac{u_{p(r_s,t)}}{u_{p(t)}} = \frac{1}{2} \left(1 - \tanh\left(10(r_s - 1)\right)\right), \quad  r_s = 2r_R/d_s, \quad \text{and} \\\
	\frac{u_{p(t)}}{u_{pm}} = \frac{1}{2} \left[\tanh\left(\frac{5}{t_a}(t - t_a)\right) + 1.0\right], \quad \text{for } t \leq t_a + \frac{t_b}{2}, \nonumber \\
	\frac{u_{p(t)}}{u_{pm}} = \frac{1}{2} \left[\tanh\left(\frac{5}{t_a}(t_a + t_b - t)\right) + 1.0\right], \quad \text{for }  t_a + \frac{t_b}{2} < t \leq t_s  \nonumber,
	\end{gather}
	where $u_{p(r_s,t)}$ represents the transient velocity profile (hyperbolic tangent) along the normalized radius of the inlet, $r_s$, and $ u_{p(t)}$ describes its variation over time, as shown in figure \ref{fig:C_RingDomain}(b). $r_s$ is obtained by normalizing the radial distance along the nozzle, $r_R$, using half the slug diameter. In the above equations, $ t_a $ denotes the acceleration time period of the piston, which is 0.32 s for all three $ l_s/d_s $ values. $ t_b $ represents the stroke ratio, while $ t_s $ is the stopping time of the piston. $ t_s $ is the slug injection time period, calculated based on the slug length, $ l_s ( = \int_0^{t_s} u_p \, dt )$. For the square program, $ t_s $ is 2.6 s, 3.6 s, and 4.6 s for $ l_s/d_s = 2 $, 3, and 4, respectively. For the impulse program, $ t_s $ is 2 s, 3 s, and 4 s for $ l_s/d_s = 2 $, 3, and 4, respectively. \par
		
   In axisymmetric vortex ring computations, the mass and momentum conservation equations have been solved using ANSYS Fluent. Figure \ref{fig:C_RingDomain}(a) illustrates the domain used in these computations. The fluid domain extends $3.5d_s$ in the radial direction, $r_R$, and $14d_s$ in the axial direction, $x_R$. The circular nozzle, of length $d_s$, protrudes into the domain from the left, with the origin of the coordinate system positioned at the nozzle exit. In the figure, the pressure outlet boundary condition, highlighted in blue, is applied to the rightmost boundary and the boundary along the axial direction, located 3.5$ d_s $ from $ x_R $. A wall boundary condition, highlighted in black, is applied to the outer nozzle boundary and the leftmost boundary, which is attached to the nozzle. The inlet boundary condition, highlighted in red, is specified at the nozzle exit, through which fluid is discharged into the domain. For the computation of the vortex ring, we used the same fluid properties, temporal and spatial discretization schemes, and residue criteria as those used for the clapping computations. A time step of 10 ms was chosen to ensure CFL$\leq$0.5. A mesh independence study was performed using three mesh configurations: a coarse mesh with an element size of 2 mm and a total element count of 30,471; a medium mesh with an element size of 1 mm and a total element count of 1,21,250; and a fine mesh with an element size of 0.5 mm and a total element count of 4,84,997. The meshes were constructed entirely using quadrilateral elements. The mesh types are compared at the isolated vortex ring formation time, $t_R$, when the shed vorticity has rolled up into an isolated and nearly steady vortex ring structure, as shown in figure~\ref{fig:C_Ring_flowField}(a). The circulation of the ring during this phase, $ \Gamma_R $, is used as a metric for mesh comparison. $ \Gamma_R $ ($ =\int{\omega_R dA_{cr|R}} $) is calculated over the core region, $ A_{cr|R} $, where most of the azimuthal vorticity of the vortex ring, $ \omega_R $, is concentrated. To identify the core region, Danaila {\it et al.} \cite{Danaila08} provided a detailed discussion on the vorticity selection criteria. Based on their work, we defined the core region using the criterion: $ \omega_R \geq 5\% \ \omega_{R_{\text{max}}} $. For the impulse program of $ u_p $ with $ l_s/d_s = 3 $, the circulation values obtained for the coarse, medium, and fine meshes are 36.443 cm$^2$/s, 36.297 cm$^2$/s, and 36.452 cm$^2$/s, respectively. This shows that the circulation values are almost constant across mesh types. As the difference in the circulation values between the medium and fine meshes is less than 0.5\% of the medium mesh value, we used the medium mesh type for the computations of the remaining cases in the parametric space. \par

\subsection{Flowfield}	
	\begin{figure}
		\centering\
		\begin{subfigure}[b]{0.450\textwidth}
			\includegraphics[width=\textwidth]{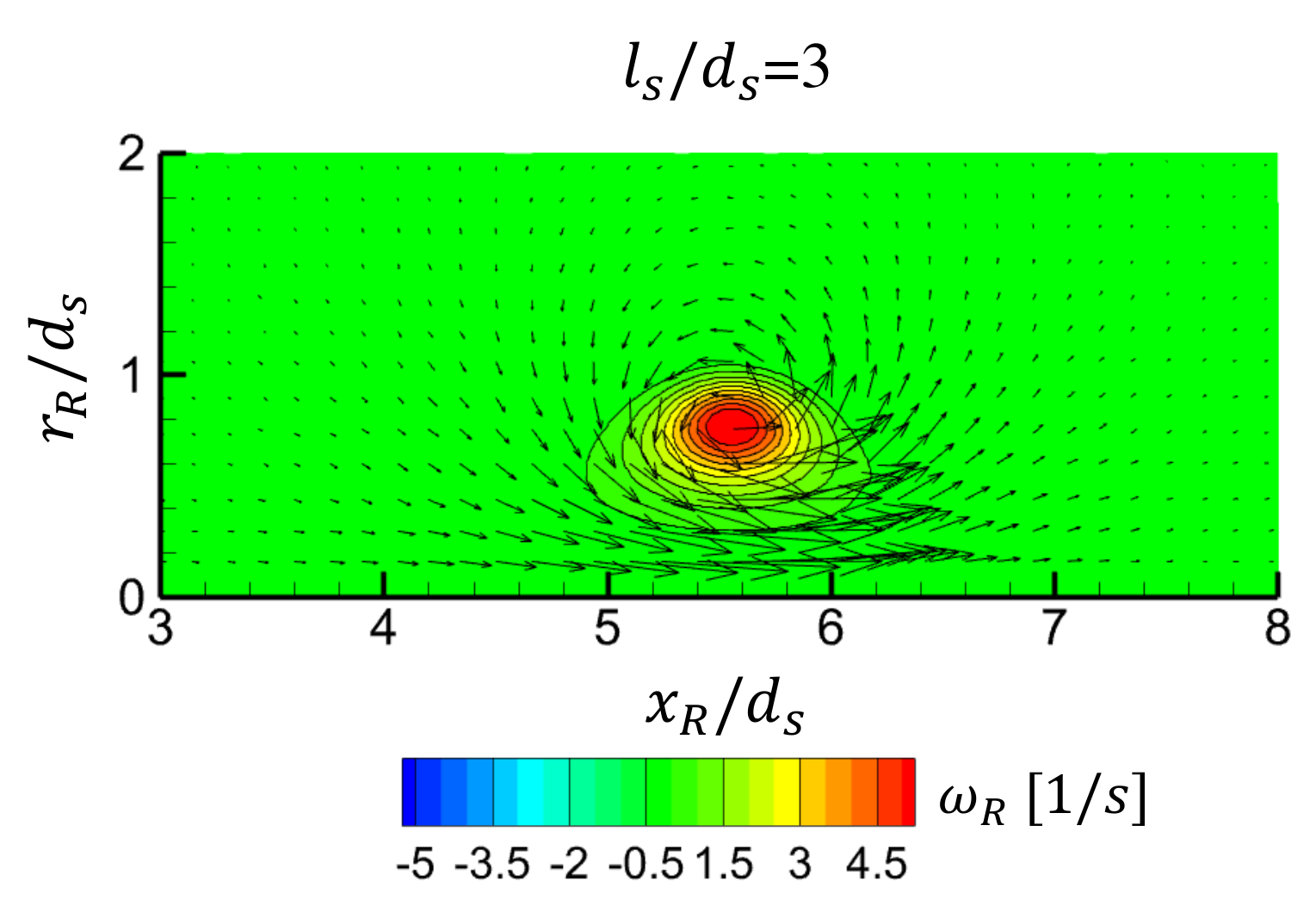}
			\caption{}
		\end{subfigure}\hspace{05mm}
		\begin{subfigure}[b]{0.40\textwidth}
			\includegraphics[width=\textwidth]{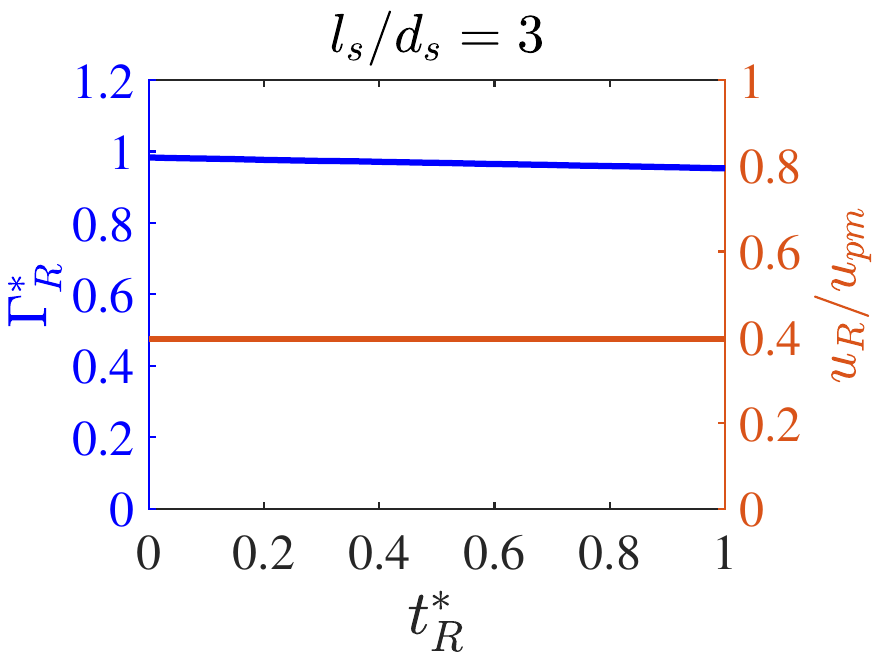}
			\caption{}
		\end{subfigure}
		\begin{subfigure}[b]{0.45\textwidth}
			\includegraphics[width=\textwidth]{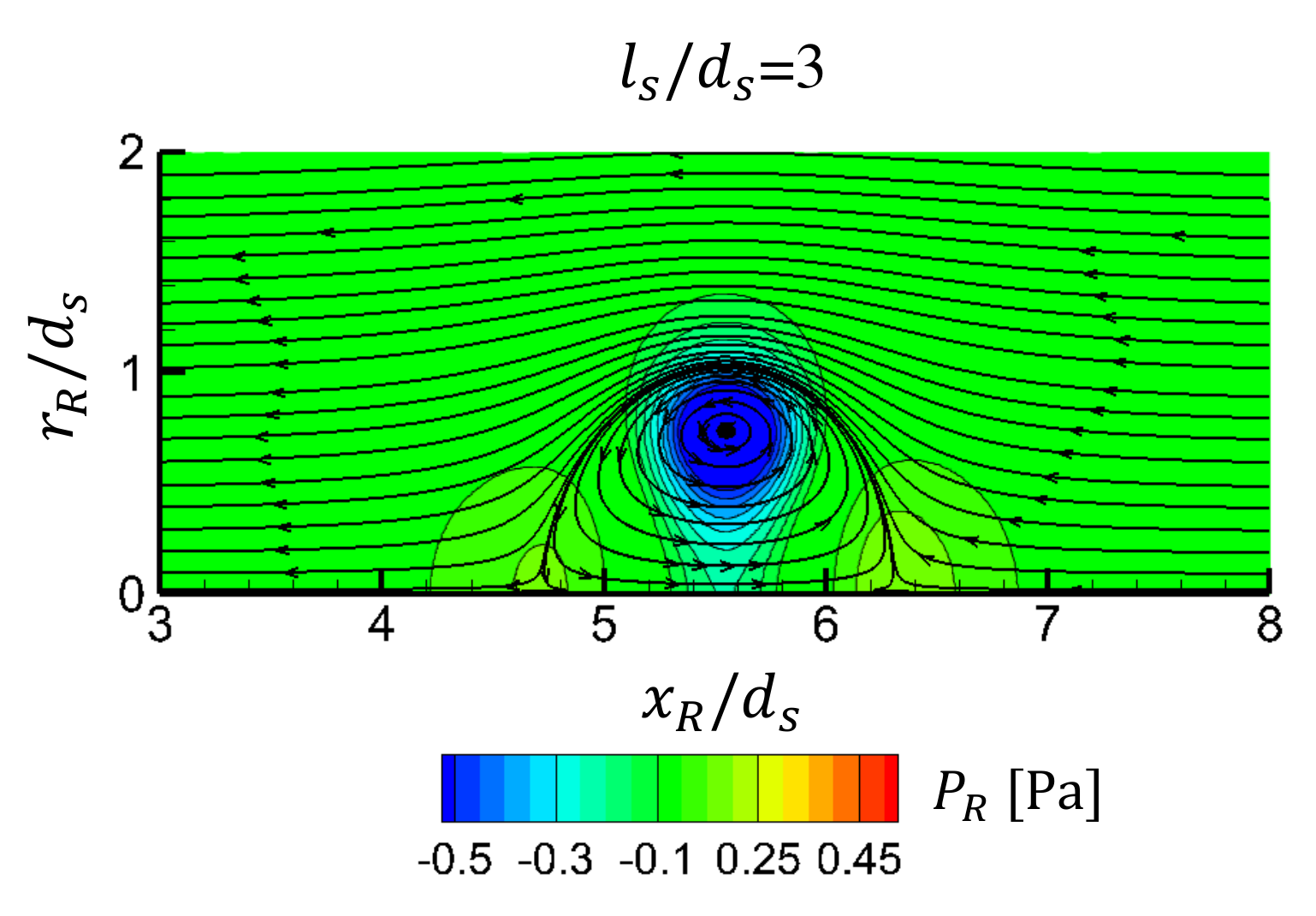}
			\caption{}
		\end{subfigure}\hspace{05mm}
		\begin{subfigure}[b]{0.40\textwidth}
			\includegraphics[width=\textwidth]{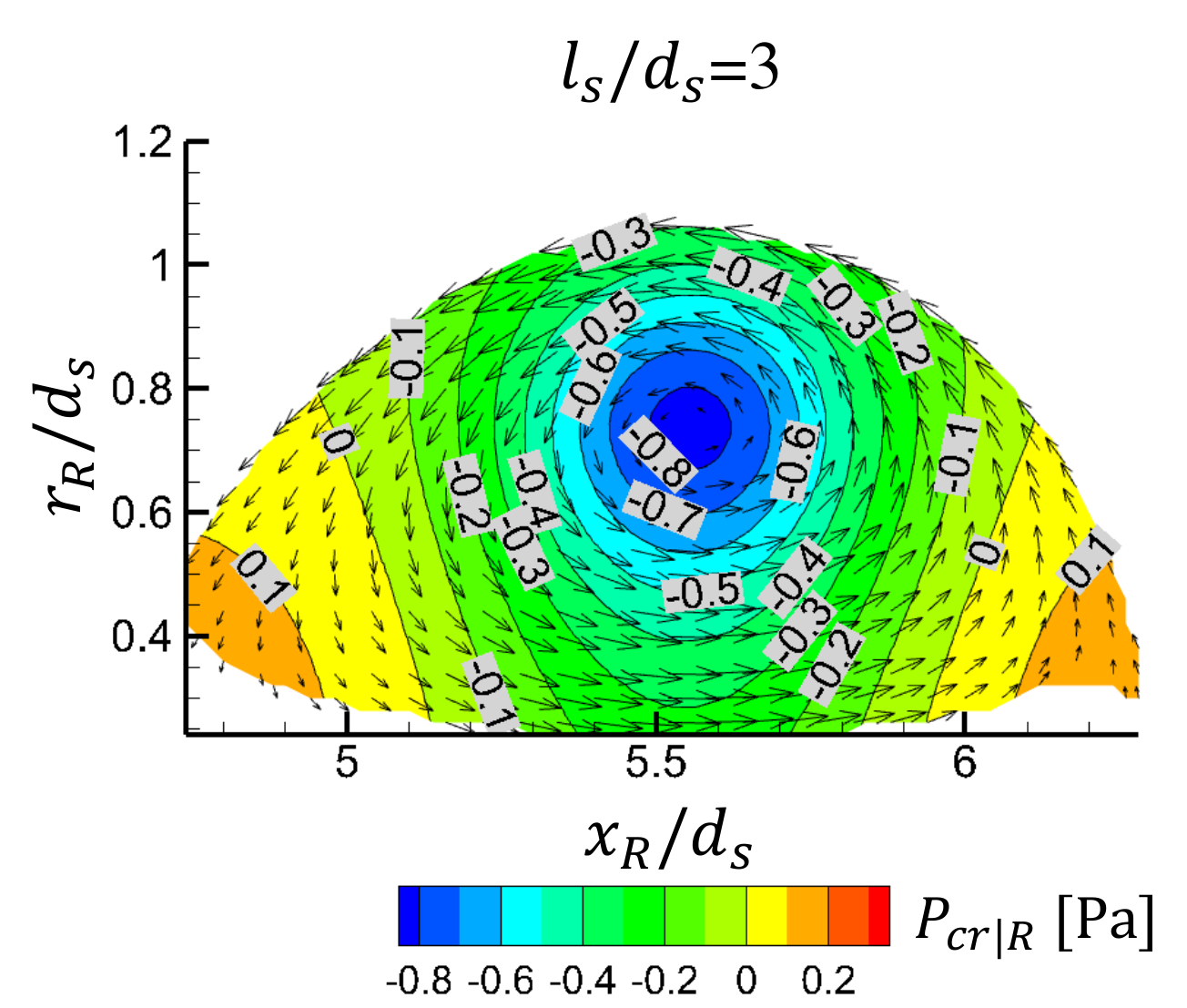}
			\caption{}
		\end{subfigure}
		\caption{The flow field of an axisymmetric vortex ring generated using the impulse program for $u_p$ with $l_s/d_s = 3$, shown at the isolated vortex ring formation time, $t_R$: (a) Azimuthal vorticity, $ \omega_R $, field and velocity vectors in the lab reference frame. (b) Plot of the variation in nondimensional circulation, $ \Gamma^*_R $, and vortex velocity ratio with nondimensional time, $ t^*_R $. The velocity ratio is obtained by normalizing the steady translational velocity of the vortex ring, $ u_R $, with the peak jet injection velocity, $ u_{pm} $. (c) Streamline plot in the reference frame moving with the steadily translating vortex and the pressure field in the domain, $P_R$. The plot shows a cross-section of the vortex bubble, with stagnation points lying on the axis of symmetry, $x_R$, at the front and back side of the bubble. (d) Pressure field in the vortex core region, $ P_{cr|R} $, and velocity vectors in the moving reference frame.}\label{fig:C_Ring_flowField}	
	\end{figure}
	Following the slug injection, the shed vorticity rolls up to form a vortex ring. In this section, we examine the axisymmetric flow field at the isolated vortex ring formation time, $t_R$, when the ring structure is isolated and nearly steady. Figure \ref{fig:C_Ring_flowField}(a) shows the azimuthal vorticity field, $\omega_R$, at $t_{R}$ for an isolated vortex ring generated using the impulse program for $u_p$ with a stroke ratio $(l_s/d_s)=3$. In the figure, the black contour enclosing the largest area marks the core region, as it captures most of the vorticity in the ring. The vector field, as shown in this figure, in the lab reference frame, shows higher velocity in the central region of the ring, predominantly aligned with the direction of propagation, $x_R$. The steady state of the vortex ring is verified by plotting the nondimensional circulation, $\Gamma_R^*$, against nondimensional time, $t_R^*$, over one slug injection time scale (see figure~\ref{fig:C_Ring_flowField}(b)). The nondimensional time, $t_R^* = (t - t_R) / t_s$, is measured starting from $t_R$ and normalized by the slug injection time, $t_s$. The values of $t_R$ are 11 s, 14 s, and 17 s for $l_s/d_s = 2, 3,$ and 4, respectively, for both the impulse and square programs. $\Gamma^*_R$ is obtained by normalizing the ring circulation $\Gamma_R$ by the slug circulation $\Gamma_s \ (= 0.5 \int_0^{t_s} u^2_p(t) \, dt)$, where $u_p(t)$ is the maximum value of the inlet velocity profile at each time instant (figure \ref{fig:C_RingDomain}(b)). Following James {\it et al.}\cite{James96}, this maximum value is used for both the uniform profile of the impulse program and the hyperbolic profile of the square program. The vortex steady velocity, $u_R$, normalized with the peak jet injection velocity, $u_{pm}(=5 \ \text{cm/s})$, also verifies the steady nature of the ring over the given duration (see figure \ref{fig:C_Ring_flowField}(b)). The flow field in a reference frame steadily moving with the vortex reveals the spheroidal geometry of the vortex bubble, as shown in figure \ref{fig:C_Ring_flowField}(c). In this figure, streamlines are plotted to identify the bubble boundaries. The distance between the two stagnation points, located on the axis of symmetry at the front and back of the bubble, gives the length of the minor axis, $2R_{\text{min}}$. The length scale along the radial direction at the bubble's centroid represents the semi-major axis, $R_{\text{max}}$. The values of $R_\mathrm{max}$ and $R_\mathrm{min}$ are provided in table \ref{tab:Ring_validation}. The streamline plot in figure \ref{fig:C_Ring_flowField}(c) is superimposed on the gauge pressure field, $P_R$, showing maximum negative pressure (blue) in the core region and low-intensity positive pressure (yellow) near the stagnation regions. To further examine the pressure in the core region, $P_{cr|R}$, we include a separate plot with a different color legend than that used for the $P_R$ range; see figure \ref{fig:C_Ring_flowField}(d). In the central region of the vortex core, the pressure field shows negative values (blue), transitioning to positive values (yellow-orange) near the axis of symmetry. The contribution of this core pressure field to the energy budget is examined later in this section.

	\begin{table}
		\centering	
		\begin{tabular}{cccccccccccc}
			\toprule
			\multirow{2}[2]{*}{\text{$l_s/d_s$}} & \multirow{2}[2]{*}{\text{Program}} & \text{$R_{\mathrm{max}}$, $R_{\mathrm{min}}$ } & \text{$\forall_{\Omega}$} & \text{$\forall_s$} & \multirow{2}[2]{*}{\text{$\eta_{R}$}} & \text{$I_{\forall|R}$} & \text{$I_s$} & \multirow{2}[2]{*}{\text{$I^*_R$}} & \text{$\Gamma_R$} & \text{$\Gamma_s$} & \multirow{2}[2]{*}{\text{$\Gamma^{*}_R$}} \\
			&       & \text{[mm]} & \text{[cc]} & \text{[cc]} &       & \text{[gm.m/s]} & \text{[gm.m/s]} &       & \text{[cm$^2$/s]} & \text{[cm$^2$/s]} &  \\
			\midrule
			\multirow{2}[2]{*}{2} & \text{Impulse} & 47, 34 & 314.60 & 196.35 & 0.38  & 9.23  & 9.82  & 0.94  & 28.55 & 25.00 & 1.14 \\
			& \text{Square} & 45, 33 & 279.92 & 196.35 & 0.30  & 7.83  & 9.50  & 0.82  & 26.84 & 24.20 & 1.11 \\
			\midrule
			\multirow{2}[2]{*}{3} & \text{Impulse} & 51,  39 & 424.91 & 294.52 & 0.31  & 14.48 & 14.73 & 0.98  & 36.30 & 37.50 & 0.97 \\
			& \text{Square} & 49.3, 38 & 386.87 & 294.52 & 0.24  & 12.50 & 14.41 & 0.87  & 34.32 & 36.70 & 0.94 \\
			\midrule
			\multirow{2}[2]{*}{4} & \text{Impulse} & 55.5, 43 & 554.81 & 392.70 & 0.29  & 19.55 & 19.63 & 1.00  & 42.18 & 50.00 & 0.84 \\
			& \text{Square} & 53.4, 42 & 501.67 & 392.70 & 0.22  & 16.99 & 19.32 & 0.88  & 39.70 & 49.20 & 0.81 \\
			\bottomrule
		\end{tabular}%
		
		\caption{The length of the semi-major axis, $R_{\text{max}}$, the semi-minor axis, $R_{\text{min}}$, the volume of the vortex bubble, $\forall_{\Omega}$, the volume of the slug injected into the domain, $\forall_s$, and the mass entrainment fraction, $\eta_R$, are listed in columns 3 to 6 for $l_s/d_s = 2, 3$, and 4, for two types of jet injection velocity programs: impulse and square. The nondimensional momentum values, $I^*_R$, are listed in column 9, obtained by normalizing the total fluid momentum in the domain, $I_{\forall|R}$, by the slug momentum, $I_s$. Column 12 shows the nondimensional circulation, $\Gamma^*_R$, obtained by normalizing the vortex ring circulation, $\Gamma_R$, by the slug circulation, $\Gamma_s$.}
		\label{tab:Ring_validation}%
	\end{table}%	
		
\subsection{Validation against published data}	
	The vortex ring simulations are validated against experimental and computational results from the literature, using three quantities:
	\begin{itemize}
		
	 \item {\it Fluid entrainment fraction}: Our simulations give entrainment fraction values of 0.38 for $l_s/d_s = 2$ and 0.29 for $l_s/d_s = 4$, as listed in table \ref{tab:Ring_validation}, closely matching the experimental results of Dabiri {\it et al.}\cite{Dabiri04}. The entrainment fraction is defined as $\eta_R = 1 - \forall_s/\forall_\Omega$, where $\forall_s(= \pi d_s^2 l_s/4)$ is the volume of the fluid slug and $\forall_\Omega (= 4 \pi R_{\text{max}}^2 R_{\text{min}}/3)$ is the volume of the spheroidal vortex bubble.
			
	\item {\it Nondimensional wake momentum}: Sullivan {\it et al.}\cite{Sullivan08} measured the vortex bubble momentum for $l_s/d_s$ values of 0.6–1.2 using a ballistic pendulum and found it closely matched the slug momentum. This led to the nondimensional momentum for the ring, $I^*_R \left(= I_{\forall|R}/I_s \right) \approx 1$, representing the ratio of the fluid domain (or bubble) momentum, $I_{\forall|R}$, to the slug momentum, $I_s$. In our axisymmetric domain, we computed the wake momentum as $I_{\forall|R} = \rho \sum u_{x_R i} \forall_{R_i}$, where $u_{x_R i}$ is the fluid velocity in the axial direction for the $i$-th cell, and $\forall_{R_i} (= 2 \pi y_i A_i)$ represents the ring volume generated by the $i$-th mesh cell revolving around the symmetry axis, $x_R$, with cell area $A_i$ at radius $y_i$. Similar to $\Gamma_s$, the slug momentum, $I_s (= \rho (\pi/4) d_s^2 \int_0^{t_s} u_p^2 \, dt)$, is computed using $u_p(t)$, the maximum value of the inlet velocity profile at each time instant from the uniform and hyperbolic tangent profiles for the impulse and square programs, respectively (figure \ref{fig:C_RingDomain}(b)). The nondimensional momentum, $I^*_R$, is close to 1 for all $l_s/d_s$ values in the impulse program (see table \ref{tab:Ring_validation}), closely matching the experimental results by Sullivan {\it et al.}\cite{Sullivan08}. For the square program, the $I^*_R$ values are slightly lower (0.82–0.88) due to an overestimation of $I_s$, which arises from using the maximum velocity of the hyperbolic tangent profile. When the slug momentum is recalculated using the mean velocity of this profile, $\bar{u}_p (= u_p / k_u)$, $I_s$ is reduced by a factor of $k_u^2$, where $k_u = 1.035$. This adjustment gives nondimensional momentum values between 0.88 and 0.93.
			
	\item {\it Nondimensional circulation}: For impulsive jet injection, the experimental nondimensional circulation values, $\Gamma^{*}_R (=\Gamma_R/\Gamma_s)$, reported by Dabiri {\it et al.}\cite{Dabiri04}, were 1.34 and 1.12 for $l_s/d_s = 2$ and 4, respectively. These values show reasonable agreement with our computational results, where $\Gamma^{*}_R$ is 1.14 for $l_s/d_s = 2$ and 0.84 for $l_s/d_s = 4$ (see table \ref{tab:Ring_validation}). Moreover, in 3D vortex ring formation simulations, James {\it et al.}\cite{James96} found a $\Gamma^{*}_R$ of 0.93 for the square program for $u_p$ with $l_s/d_s = 3.3$. This result is in close agreement with our finding of $\Gamma^{*}_R = 0.94$ for $l_s/d_s = 3$, obtained assuming the flow to be axisymmetric.
	\end{itemize}
		
\subsection{Energy budget and core potential energy}
\begin{figure}
    	\centering\
    	\begin{subfigure}[b]{0.40\textwidth}
    		\includegraphics[width=\textwidth]{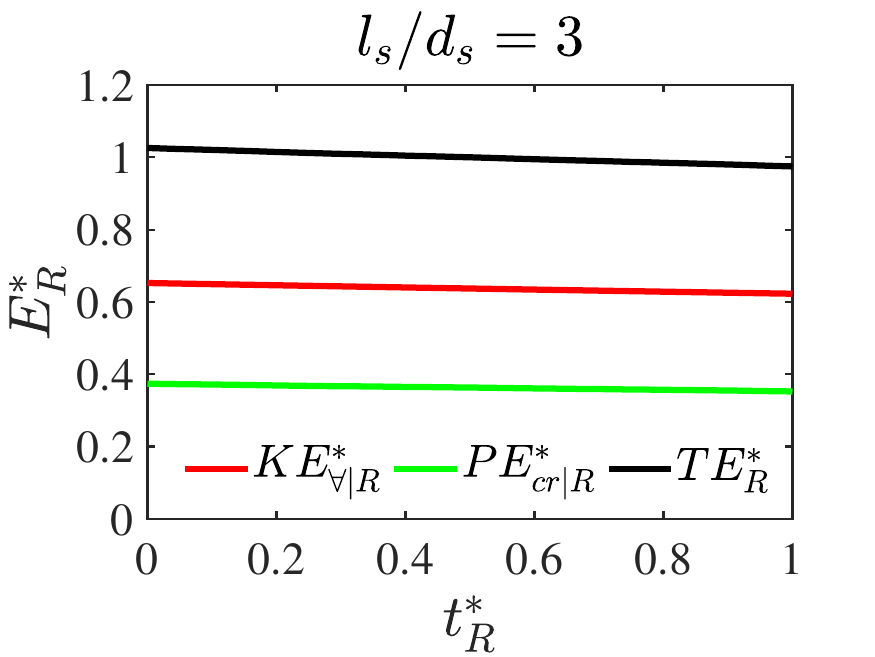}
    		\caption{}
    	\end{subfigure}
    	\begin{subfigure}[b]{0.40\textwidth}
    		\includegraphics[width=\textwidth]{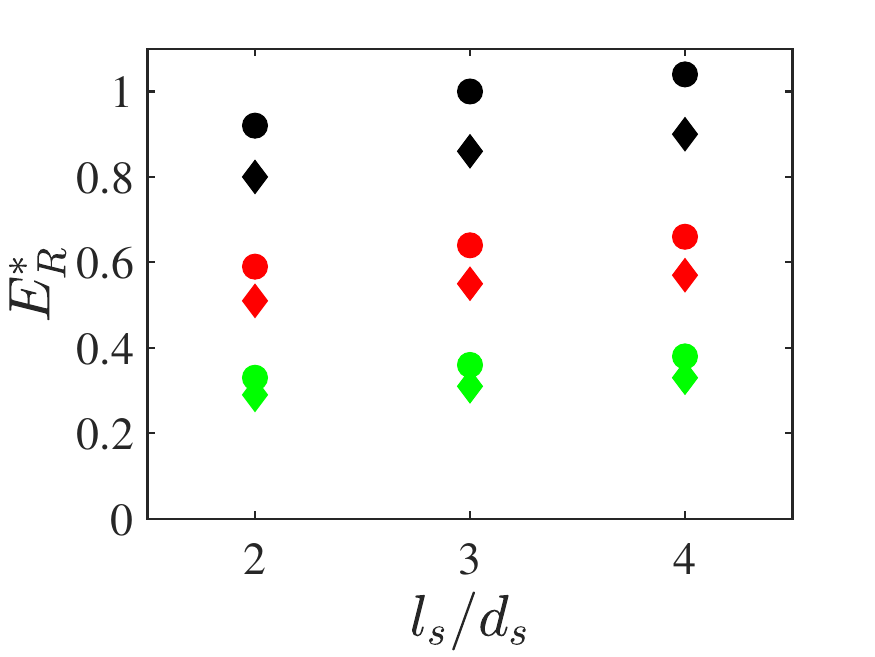}
    		\caption{}
    	\end{subfigure}
    	\begin{subfigure}[b]{0.45\textwidth}
    		\includegraphics[width=\textwidth]{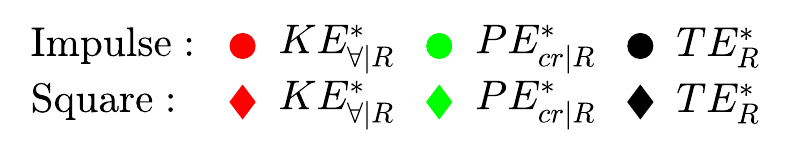}
    	\end{subfigure}
    	\caption{(a) Time variation of nondimensional energy components, collectively denoted by $ E^*_R $: kinetic energy in the full domain $ KE^*_{\forall|R} $, potential energy in the vortex core $ PE^*_{cr|R} $, and total energy $ TE^*_R = KE^*_{\forall|R} + PE^*_{cr|R} $. These are plotted against nondimensional time $ t^*_R $ for the impulse program with $ l_s/d_s = 3 $. (b) Nondimensional energies, averaged over $ t^*_R = 1 $, plotted for $ l_s/d_s$ = 2, 3, and 4 for both impulse and square programs. In both figures, the same energy symbols are used since unsteady effects are negligible.}\label{fig:C_ring_nondim_energy}	
    \end{figure}

	We now examine the evolution of kinetic energy as the vortex ring forms. The brief initial discharge of the jet into the quiescent fluid introduces kinetic energy (slug energy), which is subsequently transformed as the vortex ring evolves. We compute the kinetic energy of the domain and examine its variation starting from the isolated vortex ring formation time, $t_R$, as: $ KE_{\forall|R} = 0.5 \rho \sum |\mathbf{u}_R|_i^2 \, \forall_{R_i}, $ where $ |\mathbf{u}_R|_i $ is the velocity magnitude at the $ i $-th cell, and $ \forall_{R_i} $ represents the volume of the ring generated by the revolution of the $ i $-th cell around the axis of symmetry, as discussed earlier. The time variation of $KE_{\forall|R}$ is analyzed over one injection time scale, $t_R^* = 1$. The kinetic energy is normalized by the slug energy, $E_s = \frac{\pi}{8}\rho d_s^2 \int_0^{t_s} u_p^3(t)\, dt$, where $u_p(t)$ is the maximum value of the inlet velocity profile at each time instant. During this period, as shown in figure \ref{fig:C_ring_nondim_energy}(a), the nondimensional kinetic energy $KE^*_{\forall|R} (= KE_{\forall|R}/E_s)$ is essentially constant. For the impulse program of $ u_p $, $ KE^*_{\forall|R} $ $\approx$ 0.6 for all three $ l_s/d_s $ values, see table \ref{tab:Ring energy}. However, in the case of the square program of $ u_p $, $ KE^*_{\forall|R} $ shows slightly lower values ($ \approx 0.5 $) due to slug energy overestimation, which arises from using the maximum velocity in the hyperbolic tangent profile. Additionally, the fluid entrainment fraction $ \eta_R $ provides an alternative way to express the ratio of the kinetic energy of the vortex bubble to the kinetic energy imparted to the fluid by the piston. Using \eqref{KE_deficit}, this energy ratio is given by:  
    \begin{gather}  
	  \frac{KE_\Omega}{KE_p} = \frac{m_s}{m_\Omega} = \frac{\forall_s}{\forall_\Omega} = 1 - \eta_R.  
     \end{gather}  
    For the vortex ring cases studied in this section, $ \eta_R \approx 0.3 $ (see table \ref{tab:Ring_validation}), resulting in $ KE_\Omega/KE_p \approx 0.6 $. This kinetic energy ratio closely matches $ KE^*_{\forall|R}$, confirming that the kinetic energy in the domain, $ KE_{\forall|R} $, primarily resides in the vortex bubble, $ KE_\Omega $, as $ KE_p $ represents the slug energy $ E_s $. \par
    
    \begin{table}
    	\centering
    	
    	\begin{tabular}{cccccccccccc}
    		\toprule
    		\multirow{2}[1]{*}{\text{$l_s/d_s$}} & \multirow{2}[1]{*}{\text{Program}} & \text{$t_s$} & \text{$KE_{\forall|R}$} & \text{$PE_{cr|R}$} & \text{$E_s$} & \multirow{2}[1]{*}{$\frac{TE_R}{E_s}$} & \multirow{2}[1]{*}{\text{$\frac{PE_{cr|R}}{KE_{\forall|R}}$}} & \multirow{2}[1]{*}{\text{$KE^*_{\forall|R}$}} & \multirow{2}[1]{*}{\text{$PE^*_{cr|R}$}} & \text{${PE_{cr|+}}$} & \textbf{${PE_{cr|-}}$} \\
    		&       & \text{[s]} & \text{[mJ]} & \text{[mJ]} & \text{[mJ]} &       &       &       &       & \text{[\%]} & \text{[\%]} \\
    		\midrule
    		\multirow{2}[1]{*}{2} & \text{Impulse} & 2.00  & 0.15  & 0.08  & 0.25  & 0.93  & 0.56  & 0.59  & 0.33  & 2.17  & 97.83 \\
    		& \text{Square} & 2.60  & 0.12  & 0.07  & 0.23  & 0.80  & 0.56  & 0.51  & 0.29  & 2.41  & 97.59 \\
    		\midrule
    		\multirow{2}[2]{*}{3} & \text{Impulse} & 3.00  & 0.23  & 0.13  & 0.37  & 1.00  & 0.57  & 0.64  & 0.36  & 3.96  & 96.04 \\
    		& \text{Square} & 3.60  & 0.20  & 0.11  & 0.36  & 0.86  & 0.57  & 0.55  & 0.31  & 4.19  & 95.81 \\
    		\midrule
    		\multirow{2}[2]{*}{4} & \text{Impulse} & 4.00  & 0.32  & 0.18  & 0.49  & 1.04  & 0.57  & 0.66  & 0.38  & 5.04  & 94.96 \\
    		& \text{Square} & 4.60  & 0.27  & 0.16  & 0.48  & 0.89  & 0.57  & 0.57  & 0.33  & 5.02  & 94.98 \\
    		\bottomrule
    	\end{tabular}%
    	\caption{Columns 3 to 7 list the values for the slug discharge period, $t_s$, kinetic energy in the axisymmetric domain, $KE_{\forall|R}$, potential energy in the core, $PE_{cr|R}$, slug energy, $E_s$, and the ratio of total energy (sum of kinetic and potential) to slug energy, $TE_R/E_s$, respectively. The ratio of $PE_{cr|R}/KE_{\forall|R}$, and the nondimensional kinetic and potential energies, $KE^*_{\forall|R}$, $PE^*_{cr|R}$, are given in columns 8 to 10. The last two columns list the values for potential energy in the core region of positive and negative pressure values, $PE_{\text{cr}|+}$ and $PE_{\text{cr}|-}$, respectively, as a percentage of $PE_{cr|R}$. }
    	\label{tab:Ring energy}%
    \end{table}%
    	
	We compare the kinetic energy and steady translational velocity of the viscous vortex ring with analytical predictions for an inviscid vortex ring. These predictions are based on well-known expressions for the kinetic energy, $ KE_{R|\text{inv}} $, and steady translation velocity, $ u_{R|\text{inv}} $, of a thin inviscid ring:
	\begin{gather}  
	\label{KE_ring_inv}  
	KE_{R|\text{inv}} = \frac{\rho \Gamma^2_R \ R_R}{2} \left[ \ln \left( \frac{8R_R}{a_R} \right) - \frac{7}{4} \right], \\  
	\label{u_ring_inv}  
	u_{R|\text{inv}} = \frac{\Gamma_R}{4 \pi R_R} \left[ \ln \left( \frac{8R_R}{a_R} \right) - \frac{1}{4} \right],  
	\end{gather}  
	where $ a_R $ is the vortex core radius, $ R_R $ is the vortex ring radius, and $ \Gamma_R $ is the circulation. A detailed derivation of equations \eqref{KE_ring_inv} and \eqref{u_ring_inv} is provided in \S 2.4 of the book on vortex rings by Danaila {\it et al.}\cite{Danaila21}. The ring radius is determined by identifying the radial location of the vortex core center, where the azimuthal vorticity $ \omega_R $ reaches its maximum. To extract the vortex core radius, we analyze the tangential velocity $ u_{\theta} $ of fluid particles moving along circular streamlines within the vortex core. The variation of $ u_{\theta} $ along the axial line passing through the core center, with the origin shifted to the axial location of the core center $ x_{R|cr} $, is shown in figure \ref{fig:C_ring_KE_PE_scales}(a). In this figure, the region marked by dashed lines, where $ u_{\theta} $ varies almost linearly with $ x_{R} - x_{R|cr} $, represents the solid core. The extent of this region defines the solid core diameter, $ 2a_R $. The computed values of the core and ring radius are listed in table \ref{tab:Ring Inviscid energy}. Using these values, we calculate $ u_{R|\text{inv}} $ and $ KE_{R|\text{inv}} $. The predicted velocity $u_{R|\text{inv}}$ closely matches the velocity $u_R$ obtained from the computation (see table \ref{tab:Ring Inviscid energy}). Here, $ u_R $ is obtained by differentiating $ x_{R|cr} $ with respect to time. However, in contrast to velocity, $ KE_{R|\text{inv}} $ overestimates the kinetic energy compared to $ KE_{\forall|R} $ by approximately 25–30\%: $ KE_{R|\text{inv}} / KE_{\forall|R} = 1.25$–$1.32 $ (see table \ref{tab:Ring Inviscid energy}).\par
	
	\begin{table}
	\centering
	\begin{tabular}{ccccccccc}
		\toprule
		\multirow{2}[2]{*}{\text{$l_s/d_s$}} & \multirow{2}[2]{*}{\text{Program}} & \text{$R_R$} & \text{$a_R$} & \text{$KE_{R|\mathrm{inv}}$} & \multirow{2}[2]{*}{\text{$\frac{KE_{R|\mathrm{inv}}}{KE_{\forall|R}}$}} & \text{$u_{R|\mathrm{inv}}$} & \text{$u_R$} &  \multirow{2}[2]{*}{\text{$\frac{u_{R|\mathrm{inv}}}{u_R}$}}\\
		&       & \text{[cm]} & \text{[cm]} & \text{[mJ]} &       & \text{[cm/s]} & \text{[cm/s]} &  \\
		\midrule
		\multirow{2}[2]{*}{2} & \text{Impulse} & 3.52  & 1.38  & 0.18  & 1.25  & 1.78  & 1.79  & 1.00 \\
		& \text{Square} & 3.35  & 1.35  & 0.15  & 1.25  & 1.72  & 1.75  & 0.99 \\
		\midrule
		\multirow{2}[2]{*}{3} & \text{Impulse} & 3.87  & 1.60  & 0.31  & 1.32  & 1.98  & 2.02  & 0.98 \\
		& \text{Square} & 3.73  & 1.65  & 0.25  & 1.28  & 1.93  & 1.94  & 1.00 \\
		\midrule
		\multirow{2}[2]{*}{4} & \text{Impulse} & 4.19  & 1.90  & 0.42  & 1.29  & 2.07  & 2.10  & 0.99 \\
		& \text{Square} & 4.03  & 1.80  & 0.36  & 1.32  & 2.01  & 2.07  & 0.97 \\
		\bottomrule
	\end{tabular}%
	\caption{Columns 3 to 6 list the vortex ring radius $R_R$, core radius $a_R$, the kinetic energy for the thin inviscid vortex ring $KE_{R|\mathrm{inv}}$, and its ratio to the kinetic energy in the viscous domain, $KE_{R|\text{inv}}/KE_{\forall|R}$. Columns 7 to 9 show the steady translation velocities for the inviscid and viscous vortex rings, $u_{R|\text{inv}}$ and $u_R$, along with their ratio. The values of $KE_{R|\mathrm{inv}}$ and $u_{R|\text{inv}}$ are obtained using equations \eqref{KE_ring_inv} and \eqref{u_ring_inv}, respectively.}
	\label{tab:Ring Inviscid energy}%
	\end{table}%

	Around the isolated vortex ring formation time, $t_R$, the axial momentum in the fluid domain remains close to the slug momentum ($I_R^*\approx 1$, table \ref{tab:Ring_validation}). However, the nondimensional kinetic energy in the domain is only $KE_{\forall|R}^* \approx 0.6$ (table \ref{tab:Ring energy}), indicating that about $60\%$ of the jet's kinetic energy is retained in the flow domain. Following Sullivan's hypothesis discussed earlier\cite{Sullivan08}, the remaining energy corresponds to some potential energy associated with the isolated vortex structure, where a low-pressure region exists in the vortex core. The corresponding pressure energy in the core, obtained by integrating pressure over the core volume, is interpreted as the {\it core potential energy}, $PE_{cr|R}$. As shown in figure \ref{fig:C_Ring_flowField}(d), the pressure distribution in the core is negative in the solid core region and positive near the axis of symmetry; the corresponding contributions are therefore evaluated separately as $PE_{cr|-}$ and $PE_{cr|+}$, respectively. Our observations show that the sum of the magnitudes of $KE_{\forall|R}$, $PE_{cr|+}$, and $PE_{cr|-}$ equals the slug energy, $E_s$. For example, in an impulse program with $l_s/d_s = 3$, the values are $KE_{\forall|R} = 0.234 \, \text{mJ}$ and $PE_{cr|R} = PE_{cr|+} + PE_{cr|-} = 0.134 \, \text{mJ}$, resulting in a total energy $TE_R \ (= KE_{\forall|R} + PE_{cr|R} )= 0.368 \, \text{mJ}$, which matches the slug energy, $E_s = 0.370 \, \text{mJ}$, initially added to the domain. The core potential energy hypothesis assumes negligible viscous effects around the isolated vortex ring formation time, $t_R$, as indicated by steady circulation, near conservation of momentum, and the closure between the injected slug energy and $KE_{\forall|R}+PE_{cr|R}$.\par
	
	Across all $l_s/d_s$ values for the impulse profile, we observed an approximate match between the total energy in the domain and slug energy: $TE_{R}/E_s \approx 1$, see table \ref{tab:Ring energy}. Based on these observations, we model the total potential energy in the core as:  
		\begin{gather}  
		\label{eq:PE_basic_model}  
		PE_{cr|R} = \int |P|_{cr|R} \, d\forall_{cr|R} = \sum |P|_{cr|R_i} \, \forall_{cr|R_i},  
		\end{gather}  
	where $|P|_{cr|R}$ represents the magnitude of pressure in the core region integrated over the core volume $\forall_{cr|R}$. The subscript $i$ denotes quantities for the $i$-th computational cell, with $\forall_{cr|R_i}$ computed in the same manner as $\forall_{R_i}$. The model accounts for both negative and positive pressure contributions through the pressure magnitude. As shown in table \ref{tab:Ring energy} (columns 11–12), the core potential energy $PE_{cr|R}$ is dominated by negative pressure, with $PE_{cr|-}$ contributing $95$--$98\%$ and $PE_{cr|+}$ only $2$--$5\%$. The potential energy is approximately $60\%$ of the kinetic energy ($PE_{cr|R}/KE_{\forall|R} \approx 0.6$) and constitutes $30$--$40\%$ of the slug energy. Accordingly, the nondimensional potential energy $PE_{cr|R}^* (= PE_{cr|R}/E_s)$ ranges between $0.3$ and $0.4$ (see table \ref{tab:Ring energy}). Similar to $KE_{\forall|R}^*$, both $PE_{cr|R}^*$ and the total nondimensional energy $TE_R^* (= KE_{\forall|R}^* + PE_{cr|R}^*)$ remain nearly steady over a period of one $t_R^*$ (see figure \ref{fig:C_ring_nondim_energy}(a)). Around the isolated vortex ring formation time, $t_R$, $TE_R^*(=TE_R/E_s) \approx 1$ for all $l_s/d_s$ values in the impulse program, whereas for the square program $TE_R^*$ is slightly less than one (see figure \ref{fig:C_ring_nondim_energy}(b) and table \ref{tab:Ring energy}). This discrepancy arises because $E_s$ is overestimated when the maximum velocity in the hyperbolic tangent profile is used; using the mean velocity instead reduces $E_s$ by a factor of $k_u^3$, leading to $TE_R^* \approx 1$. These results support the core pressure energy hypothesis and demonstrate its role in resolving the energy deficit. Physically, this potential energy represents a portion of the initial jet kinetic energy redistributed to form a stable rotational core region that maintains relatively low pressure within an otherwise quiescent fluid domain.\par

	To further examine how the core potential energy depends on circulation and ring radius, a simplified analytical model based on force balance within the vortex core is presented in Appendix~\ref{sec:Analytical_PE_cr}. The model shows reasonable agreement with the values obtained from computations and offers useful physical insight into the scaling of $PE_{cr|R}$.

\section{Concluding remarks}
\label{sec:C_conclusion}

We report a computational study of clapping propulsion using a body consisting primarily of two thin rigid plates, with plate kinematics prescribed from experiments reported in our previous study (Mahulkar and Arakeri \cite{Mahulkar24}). The study considers two configurations: the dynamic case, where the body translates while the interplate cavity closes, and the stationary case, where forward motion is constrained. Simulations are performed for two aspect ratios, $d^*=0.5$ and $1.0$, using an overset mesh to capture both body translation and contraction of the interplate cavity.

For both the dynamic and stationary cases, a high pressure develops within the interplate cavity during the effective clapping phase, and starting vortices form at the two side edges and the trailing edge of each plate. Differences between the two cases emerge quickly once the body begins to translate forward in the dynamic case. Compared with the stationary case, the dynamic case shows lower interplate cavity pressure and more rapid plate closure, while the starting vortices are left behind. The unsteady Bernoulli equation applied in a non-inertial frame (see Appendix~\ref{sec:Pr_ub_dot}) provides further insight into the reduced pressure levels in the dynamic case. Another major difference is the strong sideways flow in the stationary case, which is almost absent in the dynamic case and partly explains the depth dependence of circulation and thrust per unit depth in the former. In the stationary case, the starting and stopping vortices interact and reconnect to form triangular loops on the back side and $\Omega$-shaped loops in the sideways region, followed by secondary reconnection that produces interconnected structures. In the dynamic case, the starting vortices detach earlier and reconnect with vorticity shed from the translating plates, forming one elliptical vortex loop for each plate, followed by secondary reconnection.

We next examine the energy budgets for the two cases, focusing on how the work done by the plates is transformed into fluid energy. The work input is obtained by integrating the product of pressure torque and angular velocity over the clapping period. The fluid energies are evaluated after clapping ends and the vortices are formed, when the plate-induced pressure field becomes negligible. Calculations show that the kinetic energy in the flow is substantially lower than the work input for most clapping cases. Sullivan {\it et al.}\cite{Sullivan08}, in their experimental study of vortex rings, also found that the kinetic energy was substantially lower than the input slug energy. They hypothesized that this energy deficit was associated with vortex formation. To further examine this hypothesis, we performed axisymmetric simulations of vortex ring formation. When the ring reaches a nearly steady state, the momentum in the flow domain matches the injected slug momentum, while the kinetic energy is about $60\%$ of the slug energy. We found that, for both the clapping wake and the axisymmetric vortex ring wake, the energy deficit closely matched what we term core potential energy. We define the core potential energy as the integral of the absolute gauge pressure over the vortex core volume (see equation~\ref{eq:PE_basic_model}), evaluated over the cores of the vortex structures in the clapping case and over the vortex core in the vortex ring case. This observed energy closure requires further analysis for a complete explanation. It appears to be related to vortex formation and to the development of pressure differentials in a fluid domain that was initially at uniform ambient pressure.

\section*{Acknowledgements}

The authors are grateful to Prof. Gaurav Tomar (IISc, Bangalore) for insightful discussions during the initial phase of this work. We sincerely thank Prof. Pramod Kumar and Prof. Susmita Dash (IISc, Bangalore) for providing access to the Ansys-Fluent software.

Portions of the manuscript text were edited for English clarity using ChatGPT (OpenAI), accessed via chatgpt.com in July 2026. The authors reviewed and verified all edits and take full responsibility for the content.	
	
\appendix

\section{Unsteady Bernoulli equation in a non-inertial frame}
\label{sec:Unsteady_Bernoulli}

We begin with the Euler momentum equation for incompressible and inviscid flow in a non-inertial reference frame:
\begin{gather}
\label{diff_mom}
\frac{\partial \bm{u}_r}{\partial t} + (\bm{u}_r \cdot \boldsymbol{\nabla}) \bm{u}_r
= -\frac{\boldsymbol{\nabla} P}{\rho} - \bm{a}_F,
\end{gather}
where $\bm{u}_r$ is the fluid velocity in the moving frame, $\rho$ is the fluid density, and $P$ is the static pressure. The term $\bm{a}_F$ is the apparent acceleration in the non-inertial frame and accounts for the corresponding fictitious body forces.

For irrotational flow, the velocity can be expressed as $\bm{u}_r = \boldsymbol{\nabla}\phi$, where $\phi$ is the velocity potential. Using the identity
$\boldsymbol{\nabla}(\bm{u}_r \cdot \bm{u}_r)
= 2(\bm{u}_r \cdot \boldsymbol{\nabla})\bm{u}_r
+ 2\bm{u}_r \times (\boldsymbol{\nabla} \times \bm{u}_r)$,
and noting that $\boldsymbol{\nabla} \times \bm{u}_r = \bm{0}$, equation \eqref{diff_mom} simplifies to
\begin{gather}
\label{diff_mom_simplified}
\frac{\partial}{\partial t}\boldsymbol{\nabla}\phi
+ \frac{1}{2}\boldsymbol{\nabla}(\boldsymbol{\nabla}\phi \cdot \boldsymbol{\nabla}\phi)
+ \frac{\boldsymbol{\nabla}P}{\rho}
+ \bm{a}_F = 0.
\end{gather}

Using $\bm{u}_r=\boldsymbol{\nabla}\phi$, the first term can be written as
$\frac{\partial}{\partial t}\boldsymbol{\nabla}\phi
= \frac{\partial\bm{u}_r}{\partial t}$.
Integrating equation \eqref{diff_mom_simplified} between two points along a streamline gives
\begin{gather}
\label{prefinal_unsteady_bernoulli}
\int_{1}^{2} \frac{\partial \bm{u}_r}{\partial t} \cdot d\bm{s}
+ \frac{u_{r2}^2}{2} - \frac{u_{r1}^2}{2}
+ \frac{P_2-P_1}{\rho}
+ \int_{1}^{2} \bm{a}_F \cdot d\bm{s} = 0,
\end{gather}
where $d\bm{s}$ is the differential displacement vector along the streamline. For the freely moving clapping body, the translational acceleration of the reference frame is considered, $\bm{a}_F=\dot{\bm{u}}_b$. Substituting this into equation \eqref{prefinal_unsteady_bernoulli}, we obtain the unsteady Bernoulli equation in the non-inertial frame:
\begin{gather}
\boxed{
	\int_{1}^{2} \frac{\partial \bm{u}_r}{\partial t} \cdot d\bm{s}
	+ \int_{1}^{2} \dot{\bm{u}}_b \cdot d\bm{s}
	+ \frac{P_2}{\rho} + \frac{u_{r2}^2}{2}
	= \frac{P_1}{\rho} + \frac{u_{r1}^2}{2}.
}
\end{gather}

\section{Pressure reduction due to forward translation}
\label{sec:Pr_ub_dot}
\begin{figure}
	\centering\
	\begin{subfigure}[b]{0.40\textwidth}
		\includegraphics[width=\textwidth]{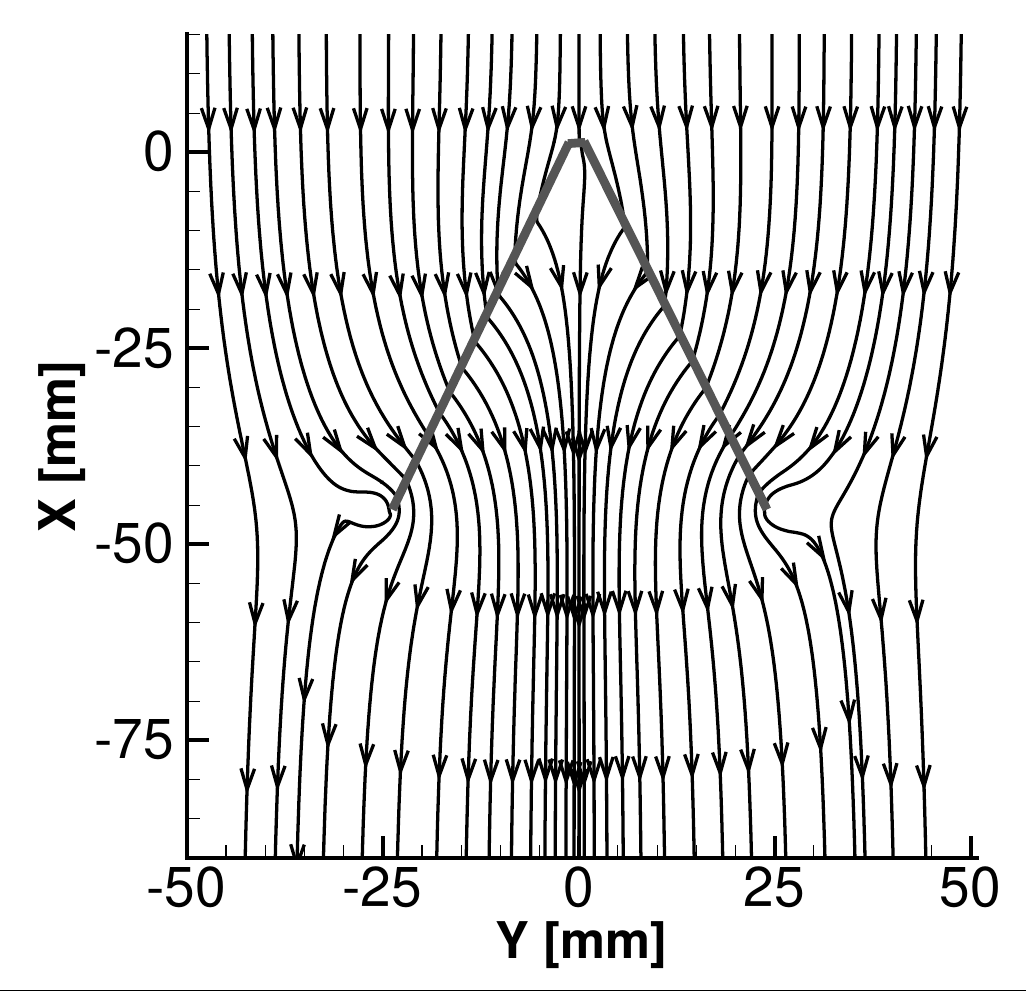}
		\caption{}
	\end{subfigure}\hspace{05 mm}
	\begin{subfigure}[b]{0.40\textwidth}
		\includegraphics[width=\textwidth]{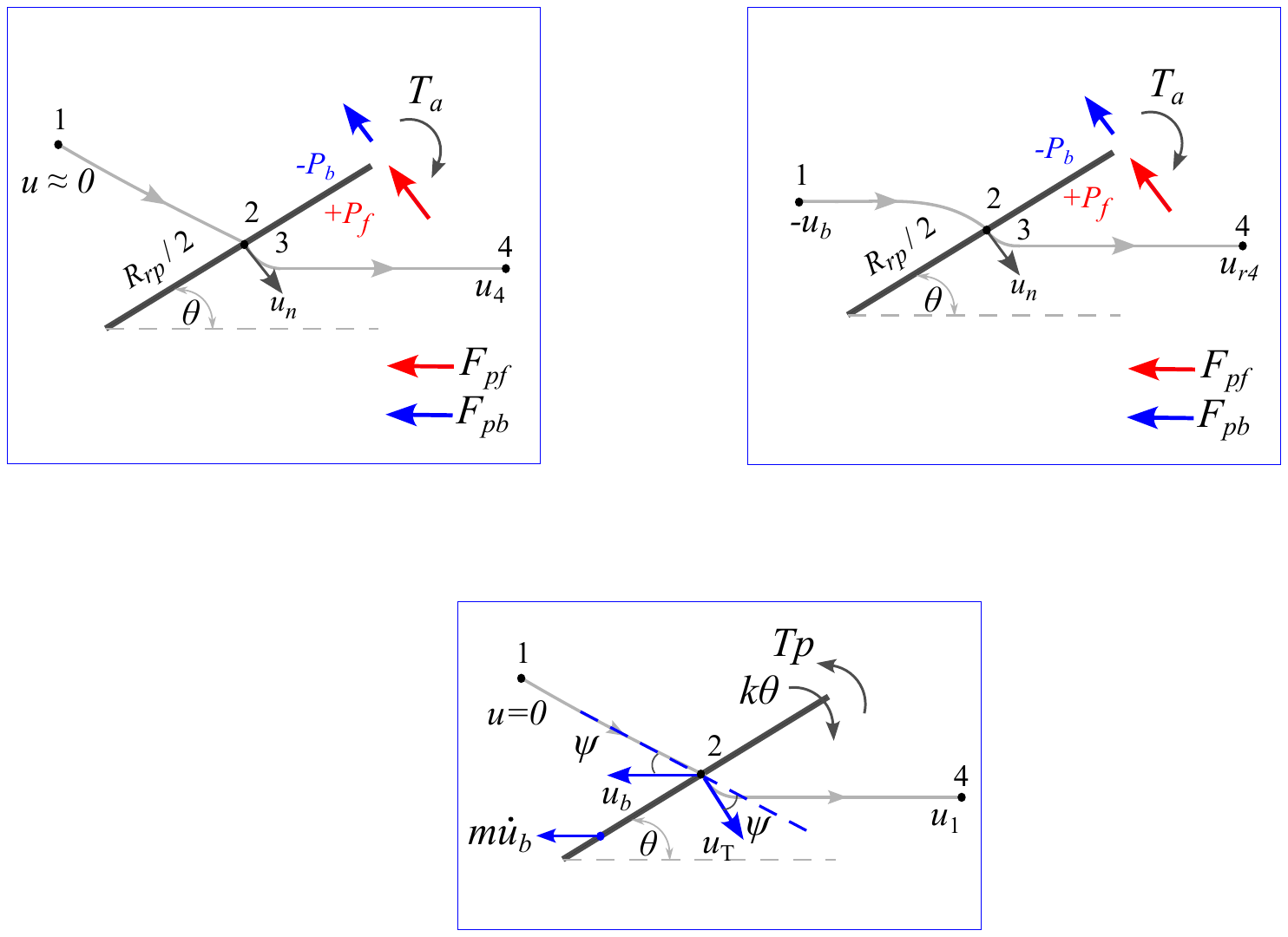}
		\caption{}
	\end{subfigure}
	\caption{(a) Streamline plot on the XY plane ($Z=0$) in the reference frame moving with the body for $d^* = 0.5$, just after the start of clapping ($t=10$ ms), where thick gray lines indicate the body position. (b) Schematic of the streamlines at the mid-radius of the plate (gray), oriented at a semi-clapping angle, $ \theta $. These streamlines are used to estimate the pressures on the plate’s back side, $ -P_b $, and front side, $ +P_f $, by applying the unsteady Bernoulli equation in a non-inertial frame between points 1--2 and 3--4, respectively. The resulting pressure forces on the back side, $ F_{pb} $, and front side, $ F_{pf} $, due to $ -P_b $ and $ +P_f $, respectively, act in the same direction, generating a torque that opposes the applied spring torque, $ T_a $, on the plate.
	}\label{fig:C_StrmLin_Dyn}
\end{figure}

\begin{figure}
	\centering\
	\begin{subfigure}[b]{0.40\textwidth}
		\includegraphics[width=\textwidth]{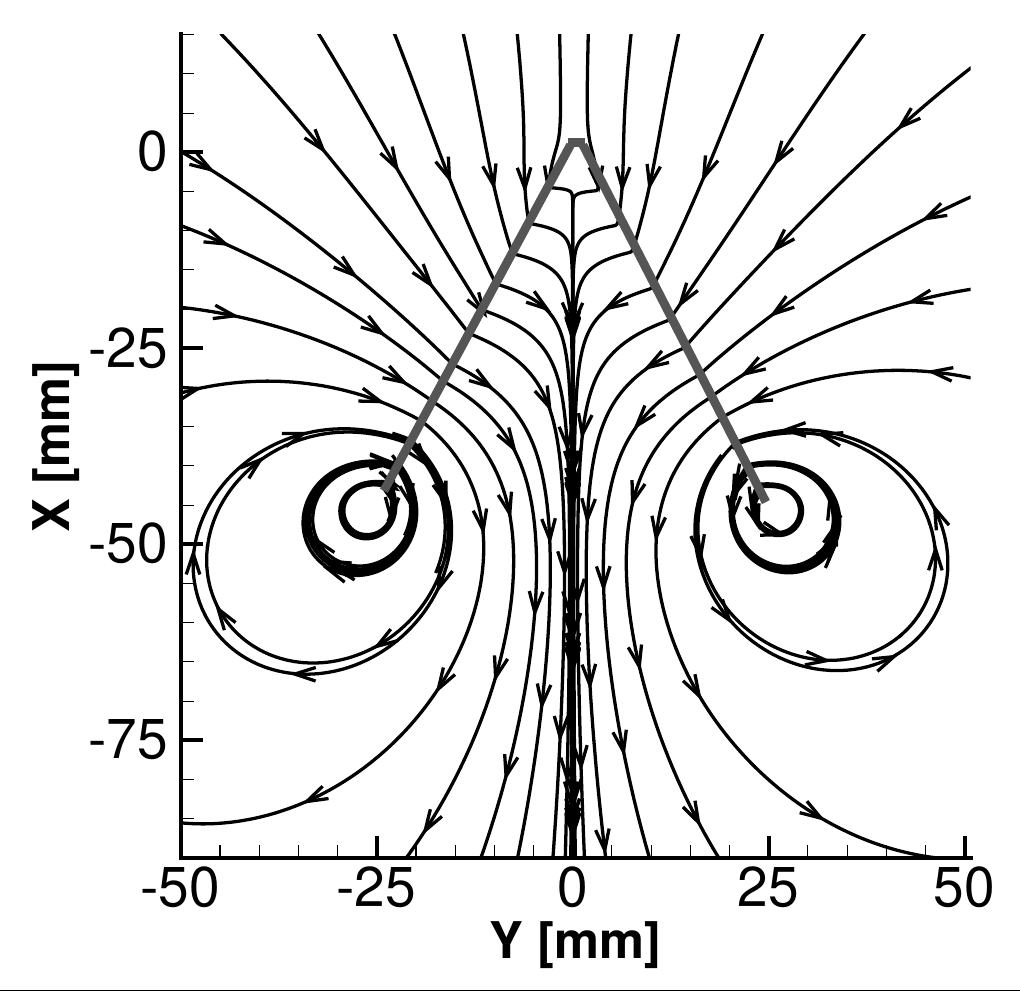}
		\caption{}
	\end{subfigure}\hspace{05 mm}
	\begin{subfigure}[b]{0.40\textwidth}
		\includegraphics[width=\textwidth]{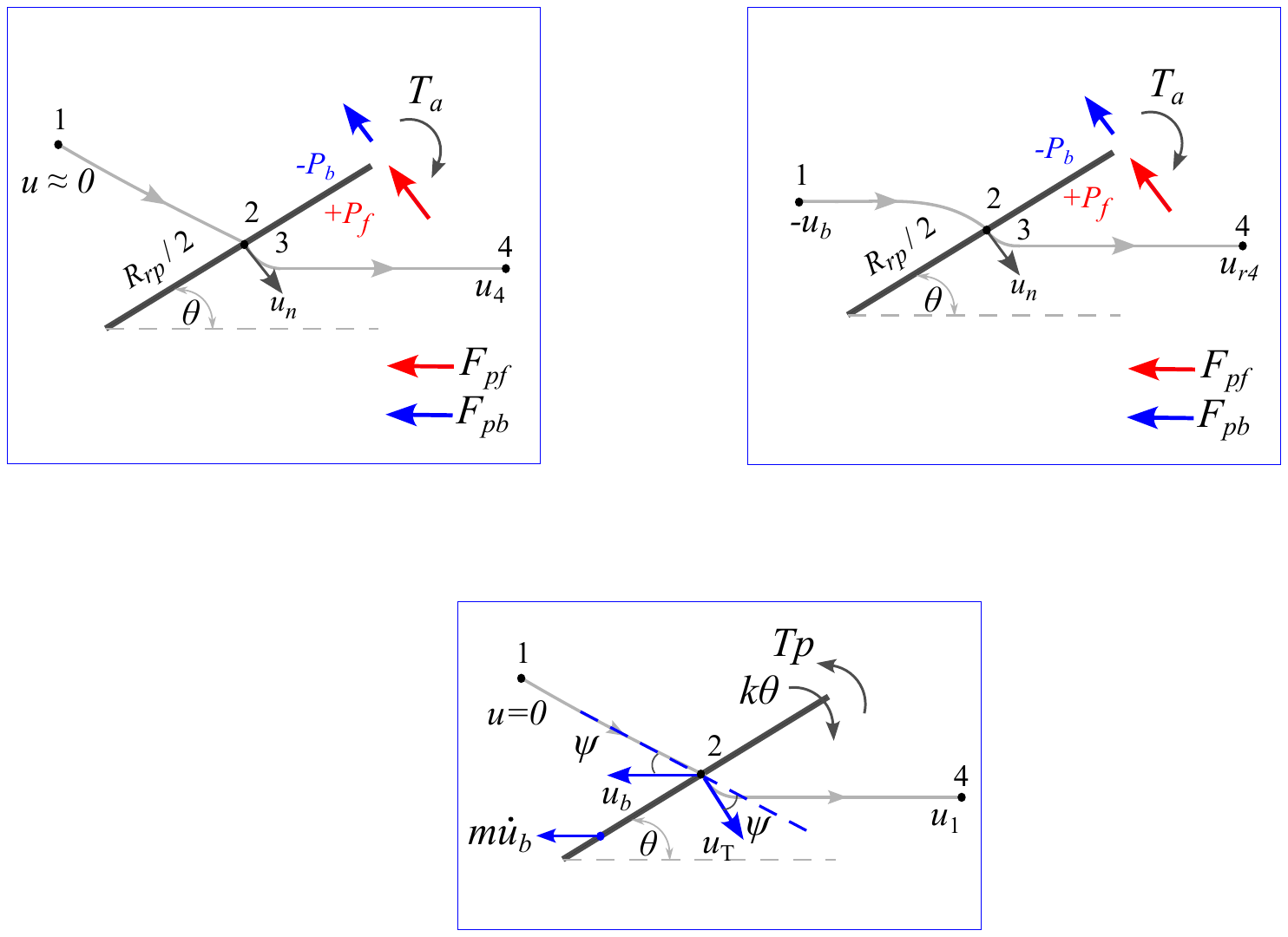}
		\caption{}
	\end{subfigure}
	\caption{(a) Streamlines in the lab reference frame just after the start of clapping ($t=10$ ms) on the XY plane ($Z=0$) for the stationary case of $d^* = 0.5$. (b) Schematic of the streamlines at the mid-radius of the plate, used for an approximate estimation of pressures on the plate. The schematic in (b) follows the same description as figure \ref{fig:C_StrmLin_Dyn}(b). }\label{fig:C_StrmLin_Stat}
\end{figure}

To understand how the forward translation of the body influences the pressure distribution across the rotating plate, we analyze the pressure difference using the Bernoulli equation. In both cases, until the end of effective clapping, mean positive gauge pressure, $P_f$, on the front side and mean negative gauge pressure, $-P_b$, on the back side produce unidirectional pressure forces on the plate (see red and blue arrows in figures \ref{fig:C_StrmLin_Dyn}(b) and \ref{fig:C_StrmLin_Stat}(b)). The resultant pressure force on the plate allows us to express the net pressure difference across the plate, $P_{f+b}$, as the sum of the magnitudes of the pressures on both sides:
\begin{gather}
\label{eq:mean net P}
P_{f+b} = P_f + P_b .
\end{gather}

Now, to determine $P_f$  and $P_b$ in the dynamic case, we apply the unsteady Bernoulli equation in a non-inertial frame. Here, representative streamlines near the mid-radius on the back and front sides of the plate are selected for the Bernoulli analysis because they are relatively unaffected by the starting vortices and have simpler geometries (see figure \ref{fig:C_StrmLin_Dyn}(a)). A schematic of these streamlines is provided separately in figure \ref{fig:C_StrmLin_Dyn}(b), showing that points 1 and 2 lie on the back side streamline, while points 3 and 4 lie on the front side streamline. The unsteady Bernoulli equation in the accelerating reference frame, applied between points 1 and 2, is  
\begin{gather}
\label{eq:UnsteadyBenoulli}
\int_{1}^{2} \frac{\partial \bm{u_r}}{\partial t} \cdot d{\bm{s}} + \int_{1}^{2} \bm{a_F} \cdot d{\bm{s}} + \frac{P_2}{\rho} + \frac{u^2_{r2}}{2} = \frac{P_1}{\rho} + \frac{u^2_{r1}}{2},
\end{gather} 
where $ \bm{u_r} $ is the fluid velocity relative to the translating body, $ d\bm{s} $ is a differential segment along the streamline, $ \bm{a_F} $ is the acceleration of the reference frame, and the pressures and magnitudes of relative velocity at points 1 and 2 are denoted by $ P_1, u_{r1} $ and $ P_2, u_{r2} $, respectively. For the derivation of this equation, see Appendix \ref{sec:Unsteady_Bernoulli}. The first term on the LHS of \eqref{eq:UnsteadyBenoulli}, the \textit{unsteady term}, requires variation in fluid acceleration along the streamline, which is difficult to determine. To address this, we use a scaling approach, assuming two scales: first, the characteristic scale for fluid acceleration is taken to be the acceleration normal to the plate at point 2, given as $ R_{rp} \ddot{\theta} / 2 $; second, the length scale of the streamline is taken as the plate radius, $ R_{rp} $, since the flow field variations are largely captured within this length. Substituting these scales, we obtain the scaling expression for the unsteady term as $ \int_{1}^{2} {\partial \bm{u_r}}/{\partial t} \cdot d\bm{s} \sim R^2_{rp} \ddot{\theta} / 2 $. We now solve for the second term on the LHS of \eqref{eq:UnsteadyBenoulli}, the \textit{accelerated-frame term}, where the reference frame accelerates with the body acceleration, $ \dot{u}_b $, in the direction opposite to the streamline. Substituting $ R_{rp} $ as the length scale in this term gives $ \int_{1}^{2} \bm{a_F} \cdot d{\bm{s}} \sim - \dot{u}_b R_{rp} $. With the scales for both the unsteady and accelerated-frame terms established, the remaining terms can be evaluated using the pressure–velocity description at points 1 and 2. At the beginning of translation of the body, at point 1, the pressure is assumed to be ambient ($P_1 \approx 0$), and the velocity is $u_{r1} \approx -u_{b}$ since the fluid is nearly quiescent in the lab reference frame. At point 2, the pressure is $P_2=-P_b$, and the relative fluid velocity is approximated by the local plate-normal velocity at the mid-radius, $u_n$: $u_{r2} \approx u_n = R_{rp}\dot{\theta}/2$. Substituting these values into \eqref{eq:UnsteadyBenoulli}, we obtain an expression for the pressure on the back side of the plate in the dynamic case:
\begin{gather}
\label{eq:pb_dyn}
P_{b|D} \sim \rho \left( \frac{R^2_{rp}}{2}\ddot{\theta} - \dot{u}_b R_{rp}  + \frac{u^2_n}{2} - \frac{u^2_{b}}{2}\right). 
\end{gather} 
Similarly, an expression for $P_f$ can be obtained by applying the Bernoulli equation between points 3 and 4 as shown in figure \ref{fig:C_StrmLin_Dyn}(b). Before deriving $P_f$, we obtain the mean fluid velocity ejected relative to the body, $u_{re}$, using integral mass conservation applied to the 2D triangular interplate cavity, under the small-angle approximation, which gives
\begin{gather}
\label{eq:u_re_dyn}
u_{re} = \frac{R_{rp}\dot{\theta}}{2\theta}.
\end{gather} 
We now apply the Bernoulli equation between points 3 and 4, where the pressure at point 3 is $ P_3 = +P_f $, and the pressure at point 4 is assumed to be ambient, $ P_4 \approx 0 $. The velocities at these points are taken as $ u_{r3} \approx u_n $ and $ u_{r4} = u_{re} $, respectively. Using the previously derived scaling estimates for the unsteady and accelerated-frame terms, we obtain the pressure on the front side of the plate in the dynamic case:
\begin{gather}
\label{eq:pf_dyn}
P_{f|D} \sim \rho \left( \frac{R^2_{rp}}{2}\ddot{\theta} - \dot{u}_b R_{rp} + \frac{u^2_{re}}{2} - \frac{u^2_n}{2} \right).  
\end{gather} 

At the onset of translation, both angular and linear velocities of the plate are negligible: $ u_n \approx 0 $, $ u_{re} \approx 0 $, and $ u_b \approx 0 $. Under these assumptions, substituting \eqref{eq:pb_dyn} and \eqref{eq:pf_dyn} into \eqref{eq:mean net P} gives an approximate expression for the mean net pressure difference across the plate during the initial stage of motion in the dynamic case:
\begin{gather}
\label{eq:pfb_dyn}
P_{f+b|D} \sim 2\rho R_{rp}\left( \frac{R_{rp}}{2}\ddot{\theta} -\dot{u}_b \right).  
\end{gather} \par

Similarly, the mean pressure across the plate in the stationary case, $ P_{f+b|S} $, is obtained using the unsteady Bernoulli equation in the laboratory reference frame. As in the dynamic case, we select the streamline at the mid-radius of the plate (see figure \ref{fig:C_StrmLin_Stat}(a)). We now apply the unsteady Bernoulli equation between points 1 and 2, as shown in figure \ref{fig:C_StrmLin_Stat}(b), and substitute the scaling for the unsteady term, as derived in the dynamic case. This gives an expression for the pressure on the back side of the plate as:
\begin{gather}  
\label{eq:pb_stat}
P_{b|S} \sim \rho \left( \frac{R^2_{rp}}{2}\ddot{\theta} + \frac{u^2_n}{2} \right),  
\end{gather}  
where the fluid at point 1 is nearly quiescent, and the pressure is assumed to be ambient ($ u_1 \approx 0 $, $ P_1 \approx 0 $). At point 2, $P_2=-P_b$, and the fluid velocity is approximated by the local plate-normal velocity at the mid-radius, $u_2 \approx u_n$. Next, the pressure on the front side of the plate is obtained by applying the Bernoulli equation between points 3 and 4 (see figure \ref{fig:C_StrmLin_Stat}(b)), expressed as:  
\begin{gather} 
\label{eq:pf_stat} 
P_{f|S} \sim \rho \left( \frac{R^2_{rp}}{2}\ddot{\theta} + \frac{u^2_e}{2} - \frac{u^2_n}{2} \right),  
\end{gather}  
where the unsteady term is replaced by its scale and at point 3, $ P_3 = +P_f $ and $ u_3 \approx u_n $, while at point 4, $ P_4 \approx 0 $ and $ u_4 = u_e $. The jet ejection velocity, $u_e$, is obtained using integral mass conservation for the triangular cavity in the lab reference frame as $u_e = R_{rp}\dot{\theta}/2\theta$. At the initial times, $u_n \approx 0$ and $u_e \approx 0$. Substituting \eqref{eq:pb_stat} and \eqref{eq:pf_stat} into \eqref{eq:mean net P} then gives the net mean pressure across the plate in the stationary case as:
\begin{gather}
\label{eq:pfb_stat}
P_{f+b|S} \sim \rho R^2_{rp} \ddot{\theta}.
\end{gather} \par
Now, by comparing the expressions for the net mean pressure force in the stationary and dynamic cases (\eqref{eq:pfb_stat} and \eqref{eq:pfb_dyn}, respectively), we observe that the initial translational acceleration of the body reduces the pressure difference across the rotating plate in the dynamic case. At the start of rotation, for the same initial torque applied in both the stationary and dynamic cases, the reduced pressure in the dynamic case lowers the opposing hydrodynamic torque on the plate, resulting in a higher initial rotational acceleration of the plates, as shown in figure \ref{fig:C_VeloInput_Stat_Dyn}(a).

\section{Analytical model for the core potential energy}
\label{sec:Analytical_PE_cr}
	\begin{figure}
	\centering\
	\begin{subfigure}[b]{0.40\textwidth}
		\includegraphics[width=\textwidth]{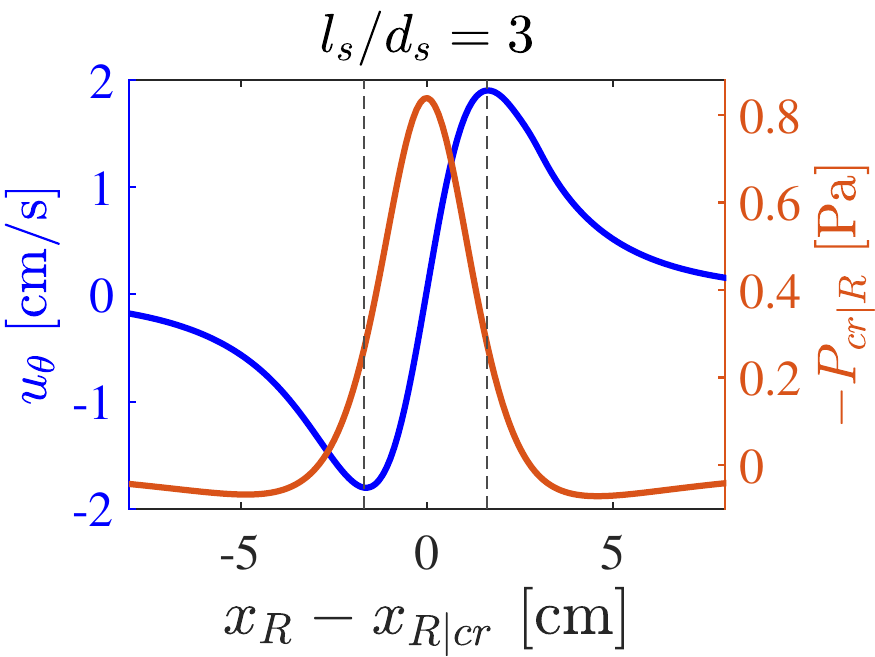}
		\caption{}
	\end{subfigure}
	\begin{subfigure}[b]{0.40\textwidth}
		\includegraphics[width=\textwidth]{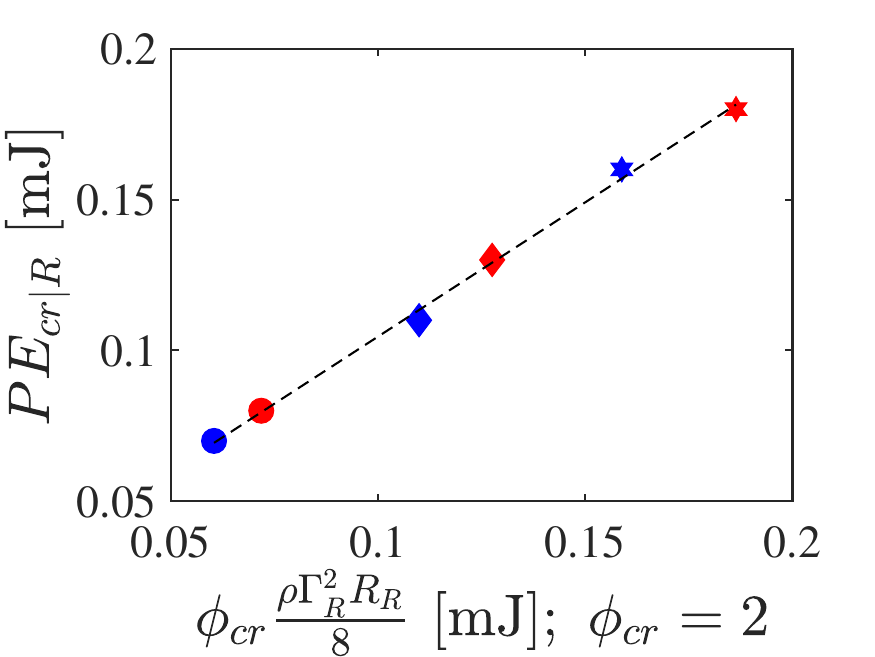}
		\caption{}
	\end{subfigure}
	\begin{subfigure}[b]{0.40\textwidth}
		\includegraphics[width=\textwidth]{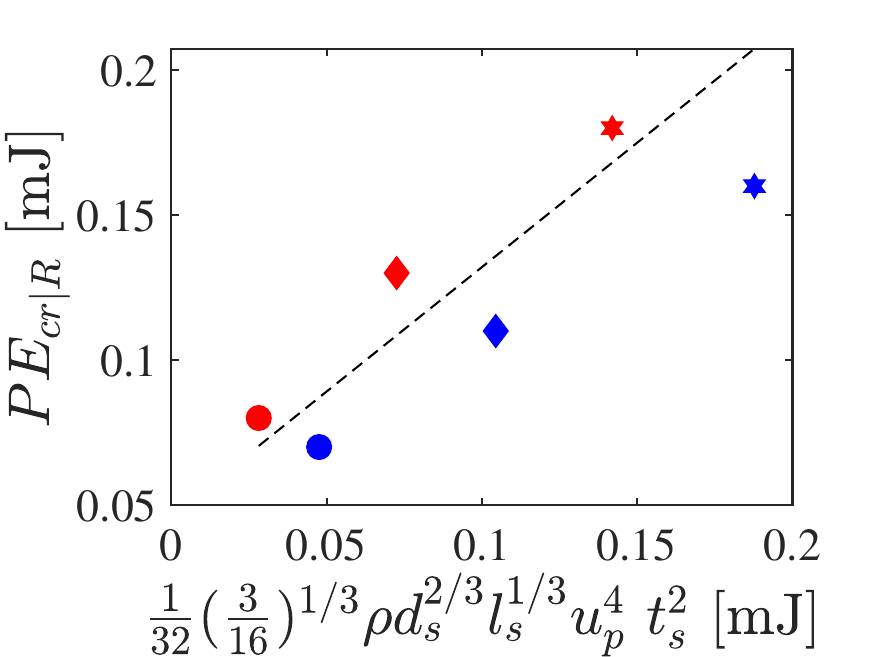}
		\caption{}
	\end{subfigure}
	\begin{subfigure}[b]{0.40\textwidth}
		\includegraphics[width=\textwidth]{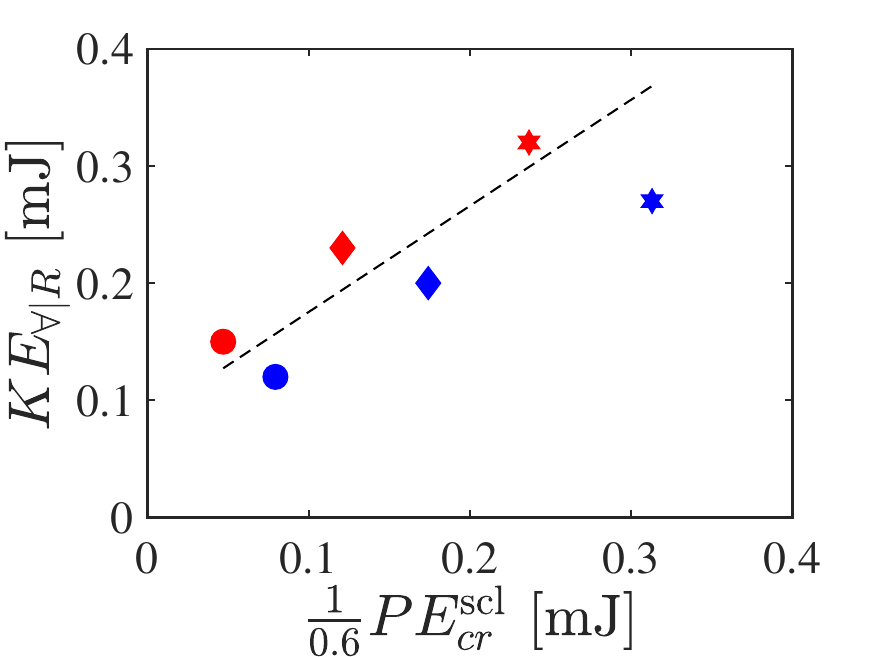}
		\caption{}
	\end{subfigure}
	\begin{subfigure}[b]{0.50\textwidth}
		\includegraphics[width=\textwidth]{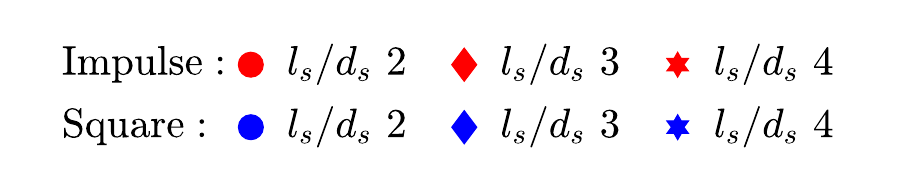}
	\end{subfigure}
	\caption{(a) Plot of the tangential velocity in the reference frame moving with the vortex, $u_{\theta}$ (blue), of fluid particles traveling along curved streamlines in the vortex core, and the pressure in the core region, $P_{cr|R}$ (orange), along the axial line passing through the center of the core at the isolated vortex ring formation phase ($t_R$).  For this line, the origin of the axial coordinate, $x_R$, is shifted to the core center, $x_{R|cr}$. Dotted lines identify the extent of the solid core of the vortex. (b) Potential energy in the core, $ PE_{cr|R} $, versus the analytically estimated potential energy, $ PE^a_{cr} $ (equation \eqref{eq:ring_PE_expression}), scaled by the core area correction factor, $ \phi_{cr} $. (c) Potential energy in the core region versus the scale for potential energy, given by \eqref{eq:ring_PE_scale}. (d) Kinetic energy in the ring domain versus the scale for kinetic energy, given by \eqref{eq:ring_KE_scale}.}\label{fig:C_ring_KE_PE_scales}	
\end{figure}

\begin{table}
	\centering
	\begin{tabular}{ccccccccccc}
		\toprule
		\multirow{2}[2]{*}{\text{$l_s/d_s$}} & \multirow{2}[2]{*}{\text{Program}} & \text{$u^a_{\theta m}$} & \text{$u_{\theta m|R}$} & \multirow{2}[2]{*}{\text{$\frac{u^a_{\theta m}}{u_{\theta m|R}}$}} & \text{$\overline{P}^a_{cr}$} & \text{$\overline{P}_{cr|R}$} & \multirow{2}[2]{*}{\text{$\frac{\overline{P}^a_{cr}}{\overline{P}_{cr|R}}$}} & \text{$2PE^a_{cr}$} & \text{$PE_{cr|R}$} & \multirow{2}[2]{*}{\text{$\frac{2PE^a_{cr}}{PE_{cr|R}}$}} \\
		&       & \text{[cm/s]} & \text{[cm/s]} &       & \text{[Pa]} & \text{[Pa]} &       & \text{[mJ]} & \text{[mJ]} &  \\
		\midrule
		\multirow{2}[2]{*}{2} & \text{Impulse} & 3.29  & 1.85  & 1.78  & -0.27  & -0.25  & 1.07  & 0.07  & 0.08  & 0.90 \\
		& \text{Square} & 3.16  & 1.77  & 1.79  & -0.25  & -0.23  & 1.08  & 0.06  & 0.07  & 0.86 \\
		\midrule
		\multirow{2}[2]{*}{3} & \text{Impulse} & 3.61  & 1.90  & 1.90  & -0.33  & -0.27  & 1.21  & 0.13  & 0.13  & 0.98 \\
		& \text{Square} & 3.31  & 1.83  & 1.81  & -0.27  & -0.25  & 1.08  & 0.11  & 0.11  & 1.00 \\
		\midrule
		\multirow{2}[2]{*}{4} & \text{Impulse} & 3.53  & 1.93  & 1.83  & -0.31  & -0.28  & 1.12  & 0.19  & 0.18  & 1.04 \\
		& \text{Square} & 3.51  & 1.87  & 1.88  & -0.31  & -0.27  & 1.16  & 0.16  & 0.16  & 0.99 \\
		\bottomrule
	\end{tabular}%
	
	\caption{Columns 3 to 5 list the maximum tangential velocity of fluid particles in the vortex solid core, in the reference frame moving with the vortex, estimated analytically, $u^a_{\theta m}$, extracted from the computational flow field, $u_{\theta m|R}$, and their ratio. Columns 6 to 8 list the analytically estimated area-averaged pressure in the core region, $\overline{P}^a_{cr}$, the corresponding value obtained from the simulation, $\overline{P}_{cr|R}$, and their ratio. Columns 9 to 11 show twice the analytically estimated core potential energy, $2PE^a_{cr}$, the corresponding value obtained from the simulation, $PE_{cr|R}$, and their ratio.}\label{tab:ring_analytical}%
\end{table}%

We derive an analytical estimate for the core potential energy at the isolated vortex ring formation phase ($t_R$), when the shed vorticity has rolled up into a distinct, nearly steady vortex ring. For this simplified analysis, vorticity is assumed to be uniformly distributed within a solid circular core. The pressure inside the core is assumed negative and nearly zero outside, consistent with the pressure profile in figure \ref{fig:C_ring_KE_PE_scales}(a). The tangential velocity, $u_\theta^a$, is defined in the reference frame moving with the vortex. The figure also shows that, within the core, $u_\theta^a$ varies linearly with the radial coordinate $r_{cr}$,
\begin{gather}
u_\theta^a = C_u r_{cr},
\label{eq:linear_ring_velo}
\end{gather}
where $C_u$ is a constant and the superscript $a$ denotes analytical quantities. At the core boundary ($r_{cr}=a_R$), the velocity reaches its maximum value $u_{\theta m}$. Using the circulation relation $\Gamma_R = u_{\theta m} (2\pi a_R)$ gives
\begin{gather}
u_\theta^a = \frac{\Gamma_R}{2\pi a_R^2} r_{cr}.
\label{eq:linear_ring_velo_Gamma}
\end{gather}
Using known values of $\Gamma_R$ and $a_R$, this expression predicts the maximum tangential velocity $u_{\theta m}^a$. The predicted values are nearly twice the values extracted from the $u_\theta$ profiles, $u_{\theta m|R}$ (see column 5 in table \ref{tab:ring_analytical}). This difference arises because the viscous core velocity profile is not perfectly linear. Nevertheless, this velocity expression provides a useful estimate for the pressure distribution in the core. The pressure field follows from the force balance along a curved streamline in the reference frame moving with the vortex,
\begin{gather}
\frac{\partial P^a_{cr}}{\partial r_{cr}} = \rho \frac{(u_\theta^a)^2}{r_{cr}} .
\label{eq:ring_pr_balance}
\end{gather}
Integrating equation \eqref{eq:ring_pr_balance} with the boundary condition $P^a_{cr}=0$ at $r_{cr}=a_R$ gives
\begin{gather}
P^a_{cr} = \frac{1}{2}\rho C_u^2 (r_{cr}^2 - a_R^2).
\label{eq:ring_pr_expression}
\end{gather}
Averaging over the core area yields the mean core pressure
\begin{gather}
\overline{P}^a_{cr} = - \frac{1}{16}\frac{\rho \Gamma_R^2}{\pi^2 a_R^2}.
\label{eq:ring_mean_pr_expression}
\end{gather}
The analytically estimated mean pressure agrees well with the pressure averaged over the core area from the simulations (see column 8 in table \ref{tab:ring_analytical}). Using the mean core pressure, the vortex core potential energy from equation \eqref{eq:PE_basic_model} is
\begin{gather}
PE^a_{cr} = |\overline{P}^a_{cr}|\,\forall_{cr|R},
\end{gather}
where $\forall_{cr|R}=2\pi R_R A_{cr}$ and $A_{cr}=\pi a_R^2$. Substitution gives
\begin{gather}
PE^a_{cr} = \frac{1}{8}\rho \Gamma_R^2 R_R.
\label{eq:ring_PE_expression}
\end{gather}
The potential energy obtained from simulations is approximately twice the analytical estimate (see column 11 in table \ref{tab:ring_analytical}). This difference arises because the viscous core occupies a larger effective area than the idealized circular core assumed in the model. To account for this effect, equation \eqref{eq:ring_PE_expression} is multiplied by a core-area correction factor $\phi_{cr}=2$, giving
\begin{gather}
\boxed{PE^a_{cr|\text{vis}} = \frac{1}{4}\rho \Gamma_R^2 R_R}.
\label{eq:ring_PE_visc_expression}
\end{gather}
Figure \ref{fig:C_ring_KE_PE_scales}(b) shows good agreement between $PE^a_{cr|\text{vis}}$ and the computed core potential energy $PE_{cr|R}$.

Next, we obtain a scaling for the core potential energy using the mass conservation relation introduced by Sullivan {\it et al.} \cite{Sullivan08},
\begin{gather}
\rho\frac{\pi}{4} d_s^2 l_s \approx \rho\frac{4\pi}{3} R_R^3 ,
\end{gather}
where the LHS represents the injected slug mass and the RHS the mass of the vortex bubble's inner spheroid.
This gives the radius scale $R_R^{\text{scl}} \approx \left(\frac{3}{16} d_s^2 l_s\right)^{1/3}$. Substituting this and the slug circulation scale $\Gamma_R^{\text{scl}} \approx \frac{1}{2}u_p^2 t_s$ into equation \eqref{eq:ring_PE_expression} gives the potential energy scale as
\begin{gather}
PE_{cr}^{\text{scl}} \sim \frac{1}{32}\left(\frac{3}{16}\right)^{1/3}\rho\, d_s^{2/3} l_s^{1/3} u_p^4 t_s^2.
\label{eq:ring_PE_scale}
\end{gather}
Using the observed relation $PE_{cr|R}\approx0.6\,KE_{\forall|R}$ (column 8 in table \ref{tab:Ring energy}) gives the kinetic energy scale,
\begin{gather}
KE_{\forall}^{\text{scl}} \sim \frac{1}{0.6}PE_{cr}^{\text{scl}} .
\label{eq:ring_KE_scale}
\end{gather}
Figures \ref{fig:C_ring_KE_PE_scales}(c, d) show good agreement between the predicted potential and kinetic energy scales and the computational data. These relations estimate vortex ring energetics using the parameters $d_s$, $l_s$, $u_p$, and $t_s$, and hold for both impulse and square programs with $l_s/d_s=$ 2 to 4 for isolated vortex rings.

	\bibliographystyle{jfm}
	%\bibliography{jfm2esam}

\end{document}